\documentclass[aps,prx,twocolumn,superscriptaddress,nofootinbib,nolongbibliography,nobalancelastpage,10pt,floatfix]{revtex4-2}

\usepackage{xcolor,amsthm,amsmath,amssymb,amsxtra,amsfonts,dsfont,graphicx,bm,bbm,bbold,braket,natbib,booktabs,mathtools}
\usepackage[colorlinks,citecolor=blue,linkcolor=blue,urlcolor=blue]{hyperref}

\usepackage[normalem]{ulem}
\usepackage[utf8]{inputenc}
\usepackage{microtype}
\usepackage{comment}
\usepackage{multirow}

\usepackage{float}
\usepackage[percent]{overpic}
\usepackage{placeins}
\usepackage[caption=false]{subfig}
\usepackage[capitalize]{cleveref}
\crefname{appendix}{App.}{App.}

\def\Id{{\openone}}

\newcommand*\dd{\mathop{}\!\mathrm{d}}
\DeclareMathOperator{\tr}{tr}

\newif\ifshowcomments
    \showcommentstrue 

\usepackage{tikz}
\usetikzlibrary{quantikz2,backgrounds,fit,positioning,shapes.geometric,calc,decorations.pathreplacing}

\pdfoutput=1

\begin{document}

\title{Quantum algorithms for cooling: a simple case study}

\author{Daniel Molpeceres}
\email{daniel.molpeceres@tum.de}
\affiliation{Max-Planck-Institut f\"ur Quantenoptik, Hans-Kopfermann-Strasse 1, D-85748 Garching, Germany}
\affiliation{Munich Center for Quantum Science and Technology (MCQST), Schellingstrasse 4, D-80799 M\"unchen, Germany}
\affiliation{Technical University of Munich, TUM School of Natural Sciences, Physics Department, 85748 Garching, Germany}

\author{Sirui Lu}
\affiliation{Max-Planck-Institut f\"ur Quantenoptik, Hans-Kopfermann-Strasse 1, D-85748 Garching, Germany}
\affiliation{Munich Center for Quantum Science and Technology (MCQST), Schellingstrasse 4, D-80799 M\"unchen, Germany}

\author{J. Ignacio Cirac}
\affiliation{Max-Planck-Institut f\"ur Quantenoptik, Hans-Kopfermann-Strasse 1, D-85748 Garching, Germany}
\affiliation{Munich Center for Quantum Science and Technology (MCQST), Schellingstrasse 4, D-80799 M\"unchen, Germany}

\author{Barbara Kraus}
\affiliation{Munich Center for Quantum Science and Technology (MCQST), Schellingstrasse 4, D-80799 M\"unchen, Germany}
\affiliation{Technical University of Munich, TUM School of Natural Sciences, Physics Department, 85748 Garching, Germany}

\date{\today}

\begin{abstract}
    Preparation of low-energy quantum many-body states has a wide range of applications in quantum information processing and condensed matter physics. Quantum cooling algorithms offer a promising alternative to other methods based, for instance, on variational and adiabatic principles, or on dissipative state preparation. In this work, we investigate a set of cooling algorithms in a simple, solvable fermionic model which allows us to identify the mechanisms which underlie the cooling process and, also, those which prevent it. We derive analytical expressions for the cooling dynamics, steady states, and cooling rates in the weak coupling limit. We find that multi-frequency and randomized cycle strategies can significantly enhance the performance of the quantum algorithm and circumvent some of the obstacles. We also analyze the effects of noise and evaluate the conditions under which cooling remains feasible. Furthermore, we present optimized cooling protocols that can significantly enhance cooling performance in the presence of noise. Additionally, we compare cooling and dissipative state preparation and show that, in the model analyzed here, cooling generally achieves lower energies and is more resilient to noise.
\end{abstract}

\maketitle

\section{Introduction}
\label{sec:intro}

The preparation of the ground state of a Hamiltonian is a central problem in quantum many-body physics, in quantum computing and in quantum simulation. Although the general problem is computationally hard~\cite{Kempe2006Complexity}, there exist a variety of heuristic methods to prepare low energy states that work well in different situations. They include adiabatic~\cite{Albash2018Adiabatic} and variational quantum algorithms~\cite{Cerezo2021Variational}, and filtering methods based on quantum phase estimation~\cite{Poulin2009Preparing,Ge2019Faster,Lin2020Nearoptimal}.
Each of them, however, faces some limitations. For instance, adiabatic methods require devising a path in the set of Hamiltonians with a sufficiently large energy gap, since the computational time scales inversely with a power of the minimum energy gap along the path~\cite{Jansen2007Bounds}. Variational methods may encounter barren plateaus~\cite{McClean2018Barren} and those based on phase estimation demand the preparation of a state with a significant overlap with the ground state as well as deep circuits that may require fault-tolerant quantum hardware.

Other approaches use the coupling to an external bath, composed by a set of auxiliary qubits in a pure state, to extract energy from the system and thus approach the ground state. This is the case of dissipative state preparation (DSP)~\cite{Poyatos1996Quantum,Kraus2008Preparation,Diehl2008Quantum,Verstraete2009Quantuma} which engineers the coupling so that the unique stationary state of the process is the desired state. Similar to DSP, quantum cooling algorithms~\cite{Terhal2000Problem} leverage engineered system-bath interactions to approach the ground state. However, unlike DSP, while coupled to the bath, the system is also subjected to the action of the original Hamiltonian, which avoids the need for precisely designing a dissipative process to steer the system to a predetermined state. An advantage of both methods operating with an external bath is that the dissipative or cooling processes converge to a fixed point, which may provide additional robustness against noise.
A schematic representation of the cooling setup is shown in \cref{fig:cooling_strategies}.

\begin{figure}
    \centering
    \begin{tikzpicture}[
            scale=0.9,
            >=Stealth,
            sys_atom/.style={circle, fill=red!80!black, inner sep=0pt, minimum size=5pt},
            anc_atom/.style={circle, fill=blue!80!black, inner sep=0pt, minimum size=5pt},
            sys_pot/.style={thick, color=black!70},
            anc_pot/.style={thick, color=black!60},
            interaction/.style={<->, thick, shorten <=1pt, shorten >=1pt},
            recycle/.style={->, thick, blue!60!black, looseness=1.5},
            evolution/.style={->, thick, red!60!black, decorate, decoration={snake, amplitude=0.4mm, segment length=2mm}},
            label_node/.style={font=\bfseries}
        ]

        \coordinate (S1) at (0, 0.6);
        \coordinate (S2) at (2, 0.6);
        \coordinate (S3) at (4, 0.6);
        \coordinate (S4) at (6, 0.6);

        \coordinate (A1) at (0, 2.5);
        \coordinate (A2) at (2, 2.5);
        \coordinate (A3) at (4, 2.5);
        \coordinate (A4) at (6, 2.5);

        \begin{scope}[yshift=0cm]
            \node[font=\small] at (-1.5, 0.7) {System ($a_n$)};

            \draw[sys_pot] plot[domain=-1:7, samples=100, smooth]
            (\x, {0.8 - 0.4*cos(\x*180/1)});

            \node[sys_atom] at (S1) {};
            \node[sys_atom] at (S2) {};
            \node[sys_atom] at (S3) {};
            \node[sys_atom] at (S4) {};

            \node[label_node, red!60!black] at (6.6, 0.5) {(0)};
        \end{scope}

        \begin{scope}[yshift=0cm]
            \node[font=\small] at (-1.5, 2.7) {Bath ($b_n$)};

            \def\trapwidth{0.7}
            \def\trapdepth{0.8}
            \draw[anc_pot] plot[domain=-\trapwidth:\trapwidth, samples=20, smooth] (\x+0, {(\x)^2 * (\trapdepth/(\trapwidth)^2) + 2.2});
            \draw[anc_pot] plot[domain=-\trapwidth:\trapwidth, samples=20, smooth] (\x+2, {(\x)^2 * (\trapdepth/(\trapwidth)^2) + 2.2});
            \draw[anc_pot] plot[domain=-\trapwidth:\trapwidth, samples=20, smooth] (\x+4, {(\x)^2 * (\trapdepth/(\trapwidth)^2) + 2.2});
            \draw[anc_pot] plot[domain=-\trapwidth:\trapwidth, samples=20, smooth] (\x+6, {(\x)^2 * (\trapdepth/(\trapwidth)^2) + 2.2});

            \node[anc_atom] at (A1) {};
            \node[anc_atom] at (A2) {};
            \node[anc_atom] at (A3) {};
            \node[anc_atom] at (A4) {};

            \draw[recycle] ($(A1)+(0.2, 0.6)$) arc (0:270:0.3);
            \draw[recycle] ($(A2)+(0.2, 0.6)$) arc (0:270:0.3);
            \draw[recycle] ($(A3)+(0.2, 0.6)$) arc (0:270:0.3);
            \draw[recycle] ($(A4)+(0.2, 0.6)$) arc (0:270:0.3);
            \node[label_node, blue!60!black] at (6.5, 3.2) {(2)};
        \end{scope}

        \begin{scope}
            \draw[interaction, blue!70!black] (S1) -- (A1);
            \draw[interaction, blue!70!black] (S2) -- (A2);
            \draw[interaction, blue!70!black] (S3) -- (A3);
            \draw[interaction, blue!70!black] (S4) -- (A4);

            \draw[interaction, green!60!black] ([xshift=1pt]S1.north) -- ([xshift=-1pt]A2.south);
            \draw[interaction, green!60!black] ([xshift=1pt]S2.north) -- ([xshift=-1pt]A3.south);
            \draw[interaction, green!60!black] ([xshift=1pt]S3.north) -- ([xshift=-1pt]A4.south);

            \draw[interaction, green!60!black] ([xshift=-1pt]S2.north) -- ([xshift=1pt]A1.south);
            \draw[interaction, green!60!black] ([xshift=-1pt]S3.north) -- ([xshift=1pt]A2.south);
            \draw[interaction, green!60!black] ([xshift=-1pt]S4.north) -- ([xshift=1pt]A3.south);

            \node[label_node, black] at (-0.5, 1.5) {(1)};
        \end{scope}

    \end{tikzpicture}
    \caption{Schematic representation of the quantum cooling algorithm. The system consists of interacting fermionic modes $a_n$ (red dots) evolving under the system Hamiltonian $H_S$, indicated by the connected potential and label (0). The system is coupled via interaction $V_{SB}$ (1) to a bath of independent fermionic modes $b_n$ (blue dots), each in its own potential well representing $H_B$. The interaction (1) facilitates energy exchange, shown here with local ($a_n \leftrightarrow b_n$, blue arrows, $nn=0$) and nearest-neighbor ($a_n \leftrightarrow b_{n\pm 1}$, green arrows, $nn=1$) couplings. After an interaction time $t$, the bath modes are reset to their ground state (2). A cooling cycle involves the joint evolution under $H_{SB} = H_S + H_B + V_{SB}$ (steps (0) and (1) occur concurrently) for time $t$, followed by the bath reset (2). This paper investigates several protocols based on repeating this cycle: (i) single-frequency cooling with fixed $t$ and bath frequency $\Delta$ (\cref{subsec:single_time}), (ii) randomized-time cooling where $t$ varies randomly (\cref{subsubsec:randomized_times}), (iii) multi-frequency cooling using several $\Delta_r$ (\cref{subsec:multi_freq}), and (iv) protocols with optimized, potentially longer-range couplings (\cref{sec:nn_cooling}).}
    \label{fig:cooling_strategies}
\end{figure}

Some of the most advanced cooling techniques have their roots in atomic physics~\cite{Hansch1975Cooling, Wineland1975Storage,Chu1998Nobel}. The theoretical foundations for quantum cooling of many-body systems were established by Davies~\cite{Davies1974Markovian}, who proved that a quantum system coupled to an infinite-dimensional thermal bath approaches its Gibbs state, which at zero temperature is the ground state. Terhal and DiVincenzo~\cite{Terhal2000Problem} considered a finite-dimensional bath and proposed using cooling for quantum many-body applications.

Recent years have seen renewed interest in quantum cooling algorithms for many-body state preparation. Mi et al.~\cite{Mi2024Stable} experimentally demonstrated the preparation of quantum-correlated states in superconducting circuits using engineered dissipation. The theoretical framework for this protocol was later developed by Lloyd et al.~\cite{Lloyd2025Quasiparticle}, who specifically targeted Floquet systems. Their method employs time-dependent coupling between the system and bath qubits and is designed to satisfy a detailed balance condition. Moreover, Marti et al.~\cite{Marti2025Efficient} recently proposed a cooling algorithm for fermionic systems, initially at very low temperatures, where an auxiliary qubit is used to decrease the energy further by first detecting resonant frequencies and then addressing them in sequences of cooling cycles. Other approaches have focused on cooling with minimal resources, such as the single-ancilla cooling methods proposed by Polla et al.~\cite{Polla2021Quantum} and Ding et al.~\cite{Ding2024Singleancilla}. Other developments include coherent cooling for quantum computation~\cite{Feng2022Quantuma} and classical optimization problems~\cite{Feng2024Escaping,Arisoy2021Fewqubit} and adiabatic demagnetization protocols~\cite{Matthies2024Programmable}. Despite all these advances, a detailed understanding of cooling algorithms, their performance and limitations, is still incomplete. However, such an understanding is indispensable to improve the efficiency and robustness of cooling algorithms in the many-body context.

The aim of this work is to contribute to such an understanding by concentrating on a simple, exactly solvable fermionic model. This allows us to derive analytical formulae for the cooling rates, the final energies and fidelities, as well as to investigate the limitations posed by the presence of noise. With that knowledge, we devise some refinements that improve the cooling algorithm and also develop re-optimization techniques that improve the performance, especially in noisy environments. We also compare the efficiency between cooling and DSP and show that the former generally achieves lower energies. While all these results are specific for the simple model we analyze, we believe that some of the techniques developed here may become relevant in other, more general, situations.

The outline of the remainder of the paper is the following. In \cref{sec:prelim}, we introduce the fundamental concepts and most relevant figures of merit used throughout this work. We begin by presenting the two state preparation algorithms under consideration: the cooling algorithm and the DSP algorithm and highlight the fundamental differences between them.

In \cref{sec:model}, we describe the main model studied here. Specifically, we consider a system of $N$ fermionic modes governed by a fixed system Hamiltonian, which is present throughout the cooling process. These $N$ system modes are coupled to $N$ fermionic bath modes, with both the system and bath Hamiltonians, as well as their coupling, being quadratic in the fermionic creation and annihilation operators. The translational invariance of the model significantly simplifies the derivations. To cool the system modes, the bath, prepared in a pure state, is coupled to the system and gets reset after a cycle time $t$. This process is repeated to extract energy or entropy from the system to obtain a state with a low energy with respect to the system Hamiltonian. Additionally, we analyze the impact of noise on the cooling process and present the considered noise models also in this section.

In \cref{sec:cooling_times_and_ss}, we derive the cooling time required to bring the system sufficiently close to its ground state and analyze how long it takes to reach a state with a sufficiently low energy density. In \cref{sec:multifreq_longtime}, we focus on the weak-coupling limit and analyze in depth three different cooling protocols:
\begin{itemize}
    \item Single-frequency cooling---where a single bath frequency $\Delta$ and a fixed cycle time $t$ are used (see \cref{subsec:single_time}).

    \item Randomized cycle times---where the same bath frequency $\Delta$ is used, but the cycle times $t_m$ are randomly chosen around an average value $t$ (see \cref{subsubsec:randomized_times}).

    \item Multi-frequency cooling---where each cycle has one of $R$ different bath frequencies $\Delta_r$ combined with randomized cycle times averaging to $t$ (see \cref{subsec:multi_freq}).
\end{itemize}
In fact, the insight that we obtain from the analytic investigation of cooling teaches us that using randomized times and more frequencies lead to more effective cooling algorithms. We also analyze the effect of noise on the cooling process.
In \cref{sec:nn_cooling}, we extend our analysis to various coupling scenarios with variable coupling strengths. By optimizing over all free parameters, we demonstrate that efficient cooling is achievable even in the presence of noise. Finally, we compare our optimized cooling protocol to the optimal DSP approach and show that cooling continues to outperform DSP in multiple scenarios.

\section{Quantum algorithms for cooling and state preparation}
\label{sec:prelim}

\subsection{Statement of the problem}

We consider a quantum system composed of $N$ subsystems (e.g., fermionic modes) and described by a Hamiltonian $H_S$. The goal is to prepare a state, $\rho$, with low energy, ideally close to the ground state. As a figure of merit, we will mostly use the relative energy with respect to that of the ground state, i.e.,
\begin{equation}
    e = \left|\frac{E-E_{GS}}{E_{GS}}\right|,
    \label{eq:relative_energy}
\end{equation}
where $E=\tr(H_S \rho)$ and $E_{GS}$ denotes the ground-state energy. In some cases, we will consider the fidelity with the ground state,
\begin{equation}
    \mathcal{F}= \bra{\Psi_{GS}}\rho\ket{\Psi_{GS}},
    \label{eq:fidelity}
\end{equation}
where we have assumed that the ground state, $\Psi_{GS}$, is non-degenerate. We emphasize that, in physical systems, one is typically interested in the large-$N$ limit, and thus intensive variables, like the relative energy, are better suited. In addition, we will be interested in the time, $T$, it takes to reach the state $\rho$.

\subsection{Simple algorithm}

In order to prepare the state $\rho$, we will use the following cooling algorithm~\cite{Terhal2000Problem}. We couple the system to a bath, consisting of $N$ subsystems, and initially prepared in a pure state $\rho_B$. At each cooling cycle, the coupling between the system and the bath is activated (assumed to be instantaneous), and the composite system evolves under the many-body Hamiltonian:
\begin{equation}
    H_{SB} = H_S + H_B + V_{SB},
    \label{eq:HSB}
\end{equation}
where $H_B$ is the bath Hamiltonian and $V_{SB}$ represents the system-bath interaction. This evolution occurs for a duration $t$, referred to as the ``cycle time''. During this joint evolution, energy is transferred between the system and the bath. Then, the system and bath are decoupled, the bath is reset to its original state, $\rho_B$ (assumed to be instantaneous), and the process repeats. This process is depicted schematically in \cref{fig:cooling_strategies}. Each cycle is represented mathematically by the completely positive trace-preserving (CPTP) map
\begin{equation}
    \mathcal{E}(\rho_S) = \tr_B\left[e^{-i H_{SB}t}(\rho_S \otimes \rho_B)e^{i H_{SB}t}\right],
    \label{eq:cooling_map0}
\end{equation}
where $\rho_S$ is the density matrix of the system, and $\tr_B$ denotes the partial trace over the bath degrees of freedom. Starting from an initial state of the system (e.g., the vacuum), repeated application of the cooling cycle will drive the system into the stationary state, $\rho_{S}^\text{ss}$, which we assume here to be unique. This stationary state satisfies
\begin{equation}
    \mathcal{E}(\rho_{S}^\text{ss}) = \rho_{S}^\text{ss}.
\end{equation}
Our objective is to design a simple system-bath coupling such that the resulting steady state exhibits a low relative energy $e$ as defined in \cref{eq:relative_energy}. We will also be interested in the number of cycles, $n^c$, which is directly proportional to the total time required to reach the stationary state, $T=n^c t$.

\subsection{Composed algorithm}
\label{subsec:composed_alg}
As we will demonstrate, it can be advantageous to vary the coupling Hamiltonian across different cycles. This leads to a composed algorithm, where a global cycle, built from several individual cycles with different parameters, is repeatedly applied until the stationary state is reached. Mathematically, each global cycle is described by the CP map
\begin{equation}
    \mathcal{E}(\rho_S) = \left[\circ_{m=1}^L \mathcal{E}_m\right](\rho_S),
    \label{eq:cooling_map2}
\end{equation}
where $\circ$ denotes the composition of maps, and each $\mathcal{E}_m$ is given by \cref{eq:cooling_map0} but potentially with different $V_{SB}$ and cycle times $t_m$. Here $L$ represents the number of elementary subcycles with potentially different parameters that are concatenated within each global cooling cycle, while $n^c$ quantifies the number of global cycles required to reach the steady state.

\subsection{Cooling versus dissipative state preparation}
\label{subsec:cooling_vs_dsp}

The cooling scheme is closely related to dissipative state preparation~\cite{Poyatos1996Quantum,Kraus2008Preparation,Diehl2008Quantum,Verstraete2009Quantuma},
where the system-bath coupling is engineered so that the system ends up in some desired steady state.
This state could have the property of having a low energy with respect to some system Hamiltonian. In the present setup, this would correspond to taking in the cooling cycle $H_{SB} = H_B + V_{SB}$, i.e., $H_S=0$. A key difference is that in cooling, the fact that the system Hamiltonian $H_S$ is present during the evolution is essential, whereas in a DSP protocol it is not. This may be much simpler to implement in practice, since it does not require the quantum simulation of the system Hamiltonian, but only that of the bath and its coupling. However, as we will show, in some situations the attained energies are significantly lower in the case of cooling, and thus using the action of the system Hamiltonian in the procedure may provide a big advantage. In particular, in the weak-coupling regime (to be defined below), this is the case and has a simple interpretation: by choosing $\rho_B$ to be the ground state of $H_B$, the coupling with the bath may transfer some excitations from the system, so that the energy of the bath increases, whereas the one of the system must then decrease due to energy conservation (ignoring the energy of the interaction, which is assumed to be small).

In many physical systems, such as cold atoms or ions, cooling rather than dissipative state preparation is widely used. Furthermore, it is well known that with the action of the system Hamiltonian, by coupling the system to an infinite bath, it is possible to prepare a state with arbitrarily low energy~\cite{Davies1974Markovian}. Note, however, that this usually comes at the expense of a long time, typically scaling exponentially with the system size.

\section{Model}
\label{sec:model}
In this section, we describe the textbook model we consider in this work. It is deliberately chosen to be very simple so that we can analyze cooling and dissipative state preparation algorithms in depth, and understand how they operate.
Throughout this work, we will also use dimensionless units for all energy scales and coupling constants, so that it will be clear how they scale.

\subsection{System}
\label{subsec:sys_model}

The system will be composed of an even number, $N$, of fermionic modes, with the system Hamiltonian
\begin{align}
    H_S &= \frac{1}{2} \sin\theta \sum_{n=1}^N (a_n^\dagger a_n - a_n a_n^\dagger) \nonumber\\
        &\quad + \frac{1}{2} \cos\theta \sum_{n=1}^N \left[a_n^\dagger (a_{n+1} + i a_{n+1}^\dagger) + \text{h.c.}\right].
    \label{eq:system_hamiltonian_realspace}
\end{align}
Here, $a_n$ ($a_n^\dag$) denotes the fermionic annihilation (creation) operator at site $n$ fulfilling $\{a_n,a_m\}=\{a_n^\dagger,a_m^\dagger\}=0, \{a_n,a_m^\dagger\}=\delta_{n,m}$. We consider periodic boundary conditions, i.e., $a_{N+1} = a_{1}$. With this definition, the system Hamiltonian is traceless and possesses several symmetries: For instance, it is invariant under the change $a_{2n}\to -a_{2n}$ and $\theta\to \pi-\theta$, or under the particle-hole transformation $a_n\to i a_n^\dagger$ and $\theta\to \pi+\theta$. These symmetries allow us to restrict ourselves to $\theta\in [0,\pi/2]$, as any value of $\theta$ outside this interval can be mapped back into it using these transformations (see~\cref{app:theta_mapping}).

This model can be exactly solved by using momentum representation and a Bogoliubov transformation~\cite{Sachdev2011Quantum}, as detailed in~\cref{app:Hk_derivation}. We define the momentum mode operators
\begin{align}
    \tilde{a}_k &= \frac{1}{\sqrt{N}}\sum_{n=1}^N e^{i 2\pi k n/N} a_n \mbox{ and }\nonumber\\
    \hat{a}_k   &= \cos(\varphi_k) \tilde{a}_k + \sin(\varphi_k) \tilde{a}_{-k}^\dagger,
    \label{eq:Bogolibov}
\end{align}
where $k=-N/2+1,\ldots,N/2$. In this basis, the Hamiltonian can be written as
\begin{equation}
    H_S = \frac{1}{2} \sum_{k=-N/2+1}^{N/2} \epsilon_k\,(\hat{a}^\dagger_k \hat{a}_k-\hat{a}_k \hat{a}_k^\dagger),
    \label{eq:Hs}
\end{equation}
with mode energies $\epsilon_k$ and Bogoliubov angles $\varphi_k$ given by
\begin{align}
    \epsilon_k &= \sqrt{1 + \sin2\theta \cos\tfrac{2\pi k}{N}},\label{eq:epsilonk}\\
    \varphi_k  &= \frac{1}{2}\tan^{-1}\left(\frac{\sin\tfrac{2\pi k}{N}}{\tan\theta + \cos\tfrac{2\pi k}{N}}\right).
    \label{eq:varphik}
\end{align}
Note that $\epsilon_k=\epsilon_{-k}$, and that for $k=0$, $\hat{a}_0=\tilde{a}_0$, whereas for $k=N/2$, $\hat{a}_{N/2}=\tilde{a}_{N/2}^\dagger$.

The ground state $\ket{\Psi_{GS}}$ of $H_S$ is the vacuum state with respect to the Bogoliubov modes $\hat{a}_k$, i.e., $\hat{a}_k\ket{\Psi_{GS}}=0$ for all $k$. The ground state energy is
\begin{equation}
    E_{GS} = - \frac{1}{2}\sum_{k=-N/2+1}^{N/2} \epsilon_k.
    \label{eq:EnergyGS}
\end{equation}
This model exhibits two distinct phases separated by a quantum critical point at $\theta = \pi/4$~\cite{Sachdev2011Quantum}. At this critical point, the mode with $k=N/2$ has zero energy, $\epsilon_{N/2}=0$, leading to a degenerate ground state, spanned by $\ket{\Psi_{GS}}$ and $\hat{a}_{N/2}^\dagger \ket{\Psi_{GS}}$. Unless otherwise stated, we initialize the system in its highest-energy Gaussian state, defined by $\hat{a}_k^\dagger |\Psi_1\rangle = 0$ for all $k$.

\subsection{Coupling to the bath}
\label{subsec:coupling_bath}

The system is coupled to a bath consisting of $N$ independent fermionic sites with Hamiltonian:
\begin{equation}
    H_B = \frac{\Delta}{2}\sum_{n=1}^N \left(b_n^\dag b_n - b_n b_n^\dag\right),
    \label{eq:bath-hamiltonian-realspace}
\end{equation}
where $b_n$ ($b_n^\dag$) are bath fermionic annihilation (creation) operators at site $n$, and $\Delta$ is the bath frequency. The bath is initialized in its ground state $\rho_B=\ket{\Omega}\bra{\Omega}$ at the beginning of each cooling cycle, where $b_n\ket{\Omega}=0$ for all $n$.

The system-bath coupling is assumed to be translation-invariant and extends up to a range of $nn$ nearest neighbors. It is given by
\begin{equation}
    {V}_{SB} = g \sum_{n=1}^N \sum_{j=-nn}^{nn} \left[\left(\lambda_j a_n^\dag {b}_{n+j} + i \mu_j a_n {b}_{n+j}\right) + \text{h.c.}\right],
    \label{eq:system-bath-coupling_realspace}
\end{equation}
where $g$ is the coupling strength, and $\lambda_j,\mu_j\le 1$ are dimensionless coupling constants to modes for different neighbor distances $j$. For $nn=0$, both $H_B$ and $V_{SB}$ are invariant under the transformations: (i) $a_{2n}\to -a_{2n}$ and $\theta\to \pi-\theta$, provided we also transform $b_{2n}\to -b_{2n}$; and (ii) $a_{n}\to i a_{n}^\dagger$ and $\theta\to \pi+\theta$, provided we take $\lambda_{j}\leftrightarrow \mu_{j}$. In the first part of our work, we will choose $nn=0$, which corresponds to a local coupling in order to obtain simple analytical formulas. Furthermore, for simplicity, we chose $\lambda_0=\mu_0=1$.

The effect of the bath on the system can be better understood by moving to the interaction picture with respect to the bath and the system Hamiltonian. For the bath, this will induce oscillating terms in \cref{eq:system-bath-coupling_realspace}
at a frequency given by $\Delta$. For the system, this will create terms in the same equation at frequencies given by $\epsilon_k$. As long as the coupling constant is sufficiently small, terms that are rapidly oscillating will average out, and only the ones where the oscillation from the bath and the system nearly cancel each other will survive. If those surviving terms extract excitations from the system, then this will lead to cooling. This will be clearer in the momentum representation in the weakly interacting limit (see~\cref{sec:multifreq_longtime}).

\subsection{Total Hamiltonian}
\label{subsec:total_hamiltonian}

Once we have defined $H_S$, $H_B$, and $V_{SB}$, the total Hamiltonian is given by~\cref{eq:HSB} and the cooling map is given by \cref{eq:cooling_map0}. It is convenient to write the total Hamiltonian in terms of the Bogoliubov modes $\hat{a}_k$ [cf.~\cref{eq:Bogolibov}]. For that, we define the momentum mode operators for the bath:
\begin{equation}
    \hat b_k = \frac{1}{\sqrt{N}}\sum_{n=1}^N e^{i 2\pi k n/N} b_n.
\end{equation}
We then combine the modes with momentum $k$ and $-k$, so that we can write
\begin{equation}
    H_{SB} = \sum_{k=0}^{N/2} \alpha_k^\dagger H_{k} \alpha_k,
    \label{eq:HSB:Momentum}
\end{equation}
where for $k=1, \ldots, N/2-1$, the operator vector is $\alpha_k=(\hat{a}_{k},\hat{a}_{-k}^\dagger,\hat b_{k},\hat b_{-k}^\dagger)$ and $H_{k}$ is a $4\times 4$ matrix given in~\cref{app:Hk_derivation}.
For the specific case of local coupling ($nn=0$) with $\lambda_0=\mu_0=1$,
\begin{equation}
    {H}_{k} =
    \begin{pmatrix}
        \epsilon_k         &0                   &ge^{i\varphi_k}   &ige^{i\varphi_k}\\
        0                  &-\epsilon_k         &ig e^{i\varphi_k} &-ge^{i\varphi_k}\\
        ge^{-i\varphi_k}   &-i ge^{-i\varphi_k} &\Delta            &0\\
        -ige^{-i\varphi_k} &-ge^{-i\varphi_k}   &0                 &-\Delta
    \end{pmatrix},
    \label{eq:H_bogoliubov}
\end{equation}
for $k=1, \ldots, N/2-1$, while for $k=0,N/2$ one has to divide this expression by 2.

\subsection{Noise}

We will also consider the realistic case in which the system and the bath are in contact with an uncontrollable, external environment, and study how this affects the cooling and state preparation process. We will consider two different models to describe this situation: (i) a Lindblad master equation, which allows for strong simplifications; (ii) a more complex setup where the environment is itself described by a set of fermionic modes that are coupled to the system and the bath, but whose coupling cannot be engineered. The latter model will, however, be relegated to \cref{app:finite_noise} and we will focus on the former.

\subsubsection{Noise described by Master Equation}
\label{sec:adding_decoherence}

This model considers that all of the modes in the system and bath undergo some extra dynamics due to the environment, so that during each cycle, both are described by a master equation of the form
\begin{equation}
    \dot \rho_{SB} = \mathcal{L}_C(\rho_{SB}) + \mathcal{L}_E(\rho_{SB}),
\end{equation}
where $\mathcal{L}_C(\rho)=-i [H_{SB},\rho]$ is the cooling Lindbladian, and $\mathcal{L}_E(\rho_{SB})$ represents the environmental noise, given by
\begin{equation}
    \mathcal{L}_E = \kappa\sum_{n=1}^N \left[\mathcal{L}_{a_n} + \mathcal{L}_{a_n^\dagger} + \mathcal{L}_{b_n} + \mathcal{L}_{b_n^\dagger}\right].
    \label{eq:lindbladian}
\end{equation}
Here, $\kappa$ is the noise strength, and the superoperator $\mathcal{L}_{O}$ is defined as
\begin{equation}
    \mathcal{L}_O(\rho) = O\rho O^\dagger - \frac{1}{2}\{O^\dagger O,\rho\}.
    \label{eq:lindblad_superoperator}
\end{equation}
Physically, the noise term $\mathcal{L}_E$ describes processes where fermions are randomly introduced into or removed from the system and bath at a rate $\kappa$. We assume that the system and bath couple to the environment in the same way, and that the rates of particle gain and loss are equal.

Interestingly, $\mathcal{L}_C$ commutes with $\mathcal{L}_E$. This can be easily seen by moving to the interaction picture with respect to $H_{SB}$, which induces a canonical transformation on the operators $a_n,b_n$ since $H_{SB}$ is quadratic in those operators. But it is also easy to see that the noise Lindbladian $\mathcal{L}_E$ is invariant under such canonical transformations, which automatically implies that these Lindbladians commute with each other (see \cref{app:LcCommuteLe} for details). As a consequence, we can integrate the master equation over one cooling cycle as
\begin{equation}
    \mathcal{N}(\rho_S)= \tr_B\left[e^{-iH_{SB}t}\left(\tilde{\rho}_S \otimes \tilde{\rho}_B\right)e^{i H_{SB}t}\right],
    \label{eq:mapwithnoise}
\end{equation}
where $\tilde{\rho}_S = e^{\mathcal{L}_E t}(\rho_S)$ and
\begin{equation}
    \tilde{\rho}_B = e^{\mathcal{L}_E t}(\ket{\Omega}\bra{\Omega}) = \frac{1+e^{-2\kappa t}}{2}\ket{\Omega}\bra{\Omega}+\frac{1-e^{-2\kappa t}}{2}\ket{1}\bra{1}.
\end{equation}

A detailed derivation is given in \cref{app:secIV_noise}.
This noisy map $\mathcal{N}$ is similar to the noiseless cooling map $\mathcal{E}$ [\cref{eq:cooling_map0}], but with two modifications: the bath is initialized in a mixed state $\tilde{\rho}_B$, and the system state $\rho_S$ is first evolved under the noise channel $\mathcal{L}_E$ to $\tilde{\rho}_S$ before the cooling cycle.

\section{Cooling times and stationary states}
\label{sec:cooling_times_and_ss}

In this section, we use the simplicity of the model we are studying in order to make general statements about the cooling time and the final fidelities and relative energies. We will show that the scaling of some of these quantities in terms of the system size, $N$, can be easily estimated, and we will also provide simple expressions that can be used to obtain them.
In the following sections, we will evaluate these expressions for different cases. We will consider the weak-coupling limit and the case where the bath couples locally to the system (that is, $nn=0$).

We have deliberately chosen a model for the system, bath, and environment coupling that is translationally invariant. Consequently, all cooling maps (with and without noise) inherit this translational invariance. We can therefore decompose the maps as tensor products over momentum modes:
\begin{equation}
    \mathcal{E}=\bigotimes_{k=0}^{N/2} \mathcal{E}_k, \quad \mathcal{N}=\bigotimes_{k=0}^{N/2} \mathcal{N}_k,
    \label{eq:CPMk}
\end{equation}
where, as in \cref{subsec:total_hamiltonian}, we group modes $k$ and $-k$ together, indexed by $k\ge 0$. Since each map conserves parity, we can use a tensor product structure.
Similarly, the system Hamiltonian can be written as [cf.~\cref{eq:Hs}]
\begin{equation}
    H_{S} = \sum_{k=0}^{N/2} \epsilon_k h_k,
\end{equation}
where the mode Hamiltonians $h_k$ can be read off from~\cref{eq:Hs}: $h_k=\hat{a}^\dagger_k \hat{a}_k-\hat{a}_{-k}\hat{a}_{-k}^\dagger$ for $k=1,\ldots,N/2-1$ and $h_k=\hat{a}_k^\dagger \hat{a}_k-\hat{a}_k \hat{a}_k^\dagger$ for $k=0,N/2$.

This decomposition has important consequences: first, it allows us to efficiently compute the state at any given time numerically. Furthermore, we can already make some valuable statements about the cooling time and the steady state. Those statements apply to both the noiseless and the noisy case. For the following discussion, we will assume that each mode map $\mathcal{E}_k$ (or $\mathcal{N}_k$) has a unique fixed point, $\sigma_k$, which will be generically the case.

\subsection{Cooling time}

For simplicity, we assume that we start out with the vacuum state of the system (corresponding to the annihilation operators $a_n$). We consider here the noiseless map $\mathcal{E}$, but the same argument extends to the noisy map $\mathcal{N}$. Since for each $\mathcal{E}_k$ we have a unique fixed point, we have that for a large number of cycles $n\gg 1$,\footnote{We will not include constant factors in this discussion, since they do not change the scaling.}
\begin{equation}
    \|\mathcal{E}_k^n(\rho_k)-\sigma_k\|_1 \le e^{-\alpha_k n},
\end{equation}
where $\alpha_k=-\log(|\lambda_k|)>0$ is the cooling rate for mode $k$ and $\lambda_k$ is the second-largest eigenvalue (in absolute value) of the map $\mathcal{E}_k$ (see \cref{subsec:zeroth_order_mixing_time}). For simplicity, we omit constant prefactors as they do not affect the scaling behavior. Thus, for each value of $k=0, \ldots, N/2$, we define the number of cycles required to reach the steady state for each mode $k$ as
\begin{equation}
    n^c_k = \alpha_k^{-1}.
    \label{eq:n0k}
\end{equation}
We will use this quantity to characterize the cooling time for each of the modes, i.e., $n_k^c t$.

Let us now analyze global properties, such as the relative energy [\cref{eq:relative_energy}] or the fidelity [\cref{eq:fidelity}]. Since the fidelity trivially provides upper and lower bounds on the one-norm distance~\cite{Nielsen2011Quantum}, we will concentrate on the latter. Using the Kaleidoscope inequality, we obtain (see~\cref{subsec:global_convergence})
\begin{align}
    \|\mathcal{E}^n(\rho)-\bigotimes_k \sigma_k\|_1 &
    \leq \sum_{k=0}^{N/2} \|\mathcal{E}_k^n(\rho_k)-\sigma_k\|_1\nonumber\\
                                                    &\leq \sum_{k=0}^{N/2} e^{-\alpha_k n} < N e^{-\alpha n},
\end{align}
where $\alpha=\inf_N \min_k \alpha_k$ is the minimum cooling rate across all modes, and we neglect $\mathcal{O}(1)$ factors.

In the case $\alpha>0$, which can be expected outside the phase transition ($\theta\ne \pi/4$), we have that, in order to get $\epsilon$-close to the steady state, we just have to take $n^c= \alpha^{-1} \log(N/\epsilon)$ cycles. Thus, in that case the map is rapidly mixing and the cooling time only grows logarithmically with the system size $N$. Note that this argument also applies in the presence of noise as the noisy map $\mathcal{N}$ also decomposes as a tensor product [see~\cref{eq:CPMk}]. If $\alpha=0$, which can occur at the phase transition, the cooling time scaling depends on how the minimum $\alpha_k$ approaches zero with increasing system size $N$.

Let us now consider how fast the relative energy reaches its steady state value. Since the relative energy is related to the energy density (an intensive quantity), we analyze the convergence of the energy density. Using, again, translational invariance, we have that after $n$ cycles the error in the energy density will be bounded by (see~\cref{subsec:energy_convergence})
\begin{equation}
    \frac{1}{N}\sum_{k=0}^{N/2} \left|\tr\left(\epsilon_k h_k \left[\mathcal{E}_k^n(\rho_k)- \sigma_k\right]\right)\right|\le 2 \frac{\sum_{k=0}^{N/2} \epsilon_k e^{-\alpha_k n}}{N} \le e^{- \alpha n},
\end{equation}
where in the last inequality we have neglected terms of $O(1/N)$ and an overall constant.
$C$ is a constant of order $\mathcal{O}(1)$. Thus, if $\alpha>0$ and we aim for an energy density error of $\epsilon$, we need $n^c=\alpha^{-1}\log(1/\epsilon)$ cycles, independent of $N$.

So far, we have not imposed that the steady state $\sigma$ is close to the ground state (in fidelity or relative energy), but only analyzed for a given map how long it takes to approach its steady state. As we will see in the next section, to approach the ground state in steady state may require adjusting the cooling map $\mathcal{E}$ as $N$ changes. In such cases, the cooling rate $\alpha$ might also depend on $N$. As long as $\alpha$ decreases at most polynomially with $N$, we will have an efficient cooling method.

\subsection{Steady state}
\label{sec:steady-state}

So far, we have analyzed the time required to reach the steady state. Now we analyze to what extent the cooling has been successful, i.e., what is the final fidelity with the ground state of $H_S$, or the relative energy. We can leverage the simple structure of the maps [see~\cref{eq:CPMk}] to compute the fidelity or the relative energy in the steady state.

The total fidelity of the steady state is given by the product of fidelities for each mode:
\begin{equation}
    \mathcal{F}=\prod_{k=0}^{N/2} F_k,
\end{equation}
where $F_k=\langle \Psi_k|\sigma_k|\Psi_k\rangle$, and we have  decomposed the ground state as $|\Psi_{GS}\rangle= \bigotimes_k |\Psi_k\rangle$. Thus, in order to obtain $\mathcal{F}=O(1)$ independent of the system size, it is sufficient if each $F_k=1-O(1/N)$.

The total energy of the steady state is the sum of mode energies:
\begin{equation}
    E =\sum_{k=0}^{N/2} E_k = \sum_{k=0}^{N/2} \epsilon_k \tr( h_k \sigma_k).
\end{equation}
For modes $(k,-k)$ (with $k=1,\ldots,N/2-1$), the ground state energy contribution is $-\epsilon_k$. We thus define the relative energy as $e_k=(E_k+\epsilon_k)/\epsilon_k$, while for $k=0,N/2$ the expression is similar with $\epsilon_k$ replaced by $\epsilon_k/2$. The total relative energy is given by \cref{eq:relative_energy}, with $E_{GS}$ given in \cref{eq:EnergyGS}.

The mode steady states $\sigma_k$ can be found by solving the fixed-point equation $\mathcal{E}_k(\sigma_k)=\sigma_k$, a linear equation in a finite-dimensional space. For noiseless cooling, the Hilbert space for each pair $(k,-k)$ is at most four-dimensional, leading to a set of 16 linear equations. Parity symmetry can further simplify this. Furthermore, since the map is Gaussian, the steady state is also Gaussian, so that one can use the covariance matrix formalism to obtain exact expressions. In the noisy case (see~\cref{app:CM_derivation}), the same argument applies and one can also get analytical expressions. While these formulas for the relative energy and the fidelity are not always very transparent, we will illustrate them with various plots. We will also present efficient numerical computations and analytics for the case of a finite environment, shown in \cref{app:finite_noise}.

At this stage, one can already set some limits to the dissipative state preparation when the system Hamiltonian is switched off. In the simplest case when each ancillary mode only couples locally to one of the system modes (i.e., $nn=0$ in \cref{eq:system-bath-coupling_realspace}), the steady state is a product state in the original site representation. Thus, the total energy $E$ [cf. \cref{eq:system_hamiltonian_realspace}] will be lower bounded by $-N\sin(\theta)/2$\footnote{This can be easily seen as only the first term in \cref{eq:system_hamiltonian_realspace} will contribute to the energy of any parity-preserving product state, and this term is minimized by the vacuum state energy, $-N \sin(\theta)/2$.}, thereby giving a minimum achievable relative energy.

\subsubsection{Large \texorpdfstring{$N$}{N} limit}
\label{subsubsec:large_N_limit}

\begin{figure}[ht]
    \includegraphics[width=\linewidth]{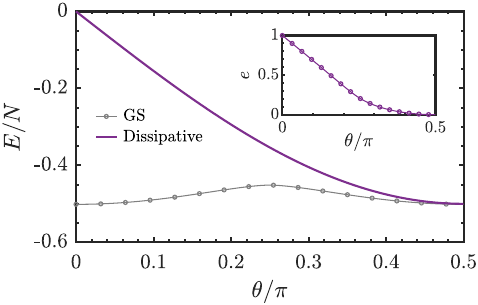}
    \caption{Minimal energy density obtainable by dissipative state preparation with $nn=0$ (purple solid line, using~\cref{eq:system_hamiltonian_realspace}). The corresponding relative energy (dashed line in the inset, using~\cref{eq:relative_energy}) and the energy of the ground state (gray symbols, labeled ``GS'', from~\cref{eq:EnergyGS}) are also shown. Dissipative state preparation performs best at $\theta=\pi/2$, where the ground state is already a product state, and worst around $\theta=0$.
    }
    \label{fig:Fig1}
\end{figure}

In the large $N$ limit, sums over momentum modes in \cref{eq:EnergyGS} can be approximated by an integral. Defining $x=2\pi k/N$ and $\Delta x=2\pi/N$, the ground state energy becomes
\begin{align}
    E_{GS} &= - \frac{N}{2\pi} \sum_{x=0}^{\pi} \sqrt{1+\sin(2\theta) \cos(x)} \Delta x\nonumber\\
           &\approx - N f(\theta),
\end{align}
where
\begin{equation}
    f(\theta)= \frac{1}{2\pi} \int_0^\pi \dd x \sqrt{1+\sin(2\theta) \cos(x)},
    \label{eq:energy_density_func}
\end{equation}
and we have ignored terms of $O(1)$ relative to $N$. This function $f(\theta)$ corresponds to the ground state energy density in the thermodynamic limit\footnote{It can be expressed by the complete elliptic integral of the second kind, $E(m) = \int_0^{\pi/2} \sqrt{1-m^2\sin^2\phi} \dd\phi$.}. \Cref{fig:Fig1} plots this energy density, along with the minimal energy achievable by DSP with local coupling ($nn=0$). As can be seen, dissipative state preparation performs best at $\theta=\pi/2$, where the ground state is a product state (vacuum), while it is worst at $\theta=0$. The exact ground state energy density is maximal at the transition point $\theta=\pi/4$.

\section{Weak interaction regime}
\label{sec:multifreq_longtime}

In this section, we consider the simplest case where each bath (ancillary) fermionic mode is coupled to one system mode (i.e., $nn=0$ in \cref{eq:system-bath-coupling_realspace}) and $\lambda_0=\mu_0=1$. As mentioned in \cref{sec:steady-state}, in this case DSP has very limited performance, and we will concentrate on the cooling scheme.

We will analyze several scenarios. In the first one, there is only one bath frequency $\Delta$ and one cycle time $t$ (\cref{subsec:single_time}). In the second one (\cref{subsubsec:randomized_times}), we have one $\Delta$ but different random cycle times $t_m$, with average $t$ which, as we will see, avoids resonant heating.
Finally, inspired by standard optical pumping and laser cooling techniques (see, e.g.,~\cite{Eschner2003Laser}), we will consider several cycles [see \cref{eq:cooling_map2}] where each of them has one out of $R$ values, $\Delta_r$ ($r=1,\ldots,R$), and a random cycle time with average $t$ (\cref{subsec:multi_freq}). We will operate in the limits
\begin{align}
    (g R t)^2      &\ll 1,\label{eq:weak_coupling_condition}\\
    (\Delta_r t)^2 &\gg 1.\label{eq:coolingcondition}
\end{align}
The first corresponds to the {\em weak coupling limit}, where the coupling constant of the system and the bath is sufficiently small. The second corresponds to the {\em cooling limit} where, as we will show, some of the modes cool down, i.e., their steady state will be close to the ground state.

\subsection{Single frequency}
\label{sec:single_frequency}

We consider two algorithms with a single bath frequency. In the first algorithm we repeatedly couple the bath for a fixed cycle time to the systems, whereas in the second algorithm the cycle time is chosen at random.

\subsubsection{Single cycle time}
\label{subsec:single_time}

We begin with the simplest case of a single cooling cycle with a bath characterized by a fixed frequency $\Delta$ (i.e., $R=1$). Using perturbation theory, we show in \cref{app:Single_freq steady state} that the cooling cycle map can be approximated by
\begin{equation}
    \mathcal{E}_k(\rho_k) =  e^{-i {h}_k t} e^{\mathcal{L}_k}(\rho_k) e^{i {h}_k t},
    \label{eq:LindbladE}
\end{equation}
where $\mathcal{L}_k=\mathcal{L}_{l_1}+\mathcal{L}_{l_2}$ is a Lindbladian [cf.~\cref{eq:lindblad_superoperator}] with jump operators
\begin{align}
    l_1 &= x_k \hat{a}_{k} - i y_k \hat{a}_{-k}^\dag, \label{eq:L1_def}\\
    l_2 &= x_k \hat{a}_{-k} - i y_k \hat{a}_k^\dag, \label{eq:L2_def}
\end{align}
and coefficients
\begin{align}
    x_k &= g\frac{1 - e^{i (\Delta-\epsilon_k) t}}{i (\Delta-\epsilon_k)}, \label{eq:xT_def}\\
    y_k &= g\frac{1 - e^{i (\Delta+\epsilon_k) t}}{i (\Delta+\epsilon_k)}. \label{eq:yT_def}
\end{align}
These terms arise from the time integration of oscillating exponentials due to the energy differences between the system and the bath~\footnote{Derivation details are presented in \cref{app:Single_freq steady state}.}.
The map $\mathcal{E}_k$ can be easily interpreted. Besides the system evolution, given by $e^{-i h_k t}$, the bath induces dissipation in each momentum mode. When $|x_k|\gg |y_k|$, modes $k$ and $-k$ are driven towards a vacuum state annihilated by $\hat{a}_{\pm k}$, which is the ground state for these modes, and thus they will be cooled. This condition is met when $|(\epsilon_k-\Delta)t|\ll 1$, in which case $|x_k|\approx gt$ and grows linearly with $t$, while $|y_k|\le g/\Delta$. Hence, imposing $|x_k|^2\gg |y_k|^2$ for those modes yields the cooling condition [\cref{eq:coolingcondition}].

\begin{figure}[ht]
    \begin{overpic}[width=\linewidth]{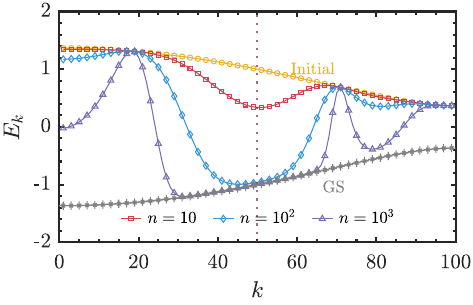}
        \put(1, 60){(a)}
    \end{overpic}
    \begin{overpic}[width=\linewidth]{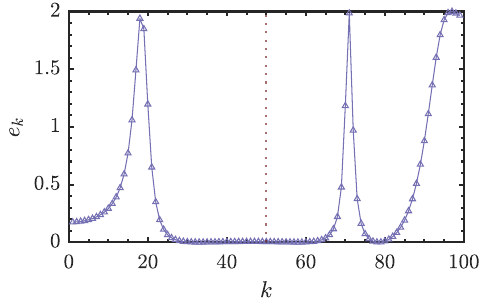}
        \put(1, 60){(b)}
    \end{overpic}
    \caption{(a) Snapshots of the mode energies $E_k$ at different stages of cooling with a single bath frequency $\Delta=\epsilon_{N/4}$. The initial state, which we chose here to be the most excited state, $\ket{\Psi_1}$, i.e. $\hat{a}_k^\dagger \ket{\Psi_1}=0$ for all $k$, is shown with circles, and after $n=10$, $100$, and $1000$ cycles. The ground-state (GS) energy $-\epsilon_k$ is plotted in gray. Other parameters are $\theta=\pi/3$, $N=200$, $g=0.01$, and $t=10$. The results for $n=1000$ almost coincide with those of the steady state. (b) Relative energy $e_k$ of each mode in steady state. The vertical dashed line (brown) marks $k=N/4$, for which $\Delta=\epsilon_k$. Modes with $k\approx \pm N/4$ are best cooled, in agreement with the resonance condition as predicted by \cref{eq:xT_def,eq:yT_def}. Some other modes are heated due to resonances when $(\epsilon_k-\Delta)t\sim 2\pi r$ with $r\in\mathbb{Z}\setminus\{0\}$, visible as peaks in the relative energy at modes near $k \approx 20$ and $k \approx 70$. We show in \cref{subsubsec:randomized_times} how these unwanted resonances can be avoided.}
    \label{fig:Fig2}
\end{figure}

In \cref{fig:Fig2} we plot the evolution of mode energies $E_k$, starting with the system in the state with the highest energy, $\ket{\Psi_1}$, for which $\hat{a}^\dagger_k\ket{\Psi_1}=0$, using a single chosen bath frequency $\Delta=\epsilon_{N/4}$.
The mode energies evolve from the initial state through intermediate stages after $n=10$, $n=100$, and $n=1000$ cycles, progressively approaching the ground state energy for modes near $k=N/4$.
This is also illustrated by the relative energy $e_k$ (see \cref{fig:Fig2}(b)) for $n=10^3$ with values approaching zero for modes where $\epsilon_k\approx\Delta$.

For some values of $k$, however, distinct peaks appear where the corresponding modes are not cooled at all but are instead heated.
For example, modes near $k \approx 20$ and $k \approx 80$ show relative energies reaching their maximal values $e_k \approx 2.0$.
These unwanted peaks occur at specific values of $k$ where $(\epsilon_k-\Delta)t\approx 2\pi r$ with $r\in\mathbb{Z}\setminus\{0\}$, which causes $x_k\approx 0$ in \cref{eq:xT_def}, so that instead of cooling there is heating. Given that these resonances depend on the cycle time, it is straightforward to circumvent them, as we will show in the next subsection.

\subsubsection{Randomized times}
\label{subsubsec:randomized_times}

We now consider a composed algorithm where each global cycle consists of $L$ subcycles. In each subcycle $m=1,\dots,L$, the system evolves for a random time $t_m$ chosen from a uniform distribution on the interval $[0,2t]$ with mean value $t$. We assume
\begin{equation}
    L \gg 1,
    \label{equationL}
\end{equation}
so that we can replace
\begin{align}
    \bar t    &= \frac{1}{L} \sum_m t_m \approx t,\\
    \bar{t^2} &= \frac{1}{L} \sum_m t^2_m \approx \frac{4t^2}{3}.
\end{align}

As long as the condition in \cref{eq:weak_coupling_condition} holds (with $R=1$), we can apply perturbation theory to the global cycle. Note that this requires, in principle, that $(g L\bar t)^2 \ll 1$. This is a very stringent condition. However, as we will see below, numerical results suggest that cooling persists even when this condition is violated, as long as the condition for a single elementary cycle, $(g \bar t)^2\ll 1$, holds, together with the cooling condition [\cref{eq:coolingcondition}].
This is consistent with the expectation that the overall cooling efficiency should primarily depend on the properties of individual cycles rather than $L$, the total number of times we pick a random cycle time, and whose main role is to average out unwanted resonances.

In the perturbative limit we obtain \cref{eq:LindbladE}, but now the Lindbladian $\mathcal{L}$ is the sum over $m$ Lindbladians of the form given in \cref{eq:L1_def}, with $t$ replaced by $t_m$ in $x_k$ and $y_k$ (see~\cref{app:long_time_averages}). In the limit given by~\cref{equationL}, we can replace the sum by an average with respect to the distribution of $t_m$. Furthermore, in the cooling limit [\cref{eq:coolingcondition}] this average washes out the time dependence of the Lindblad operators, yielding the following effective Lindbladian per cycle for $k$ (see~\cref{app:long_time_averages} for the derivation)
\begin{equation}
    \frac{1}{L}\mathcal{L}_k\rho_k= \gamma^{\rm c}_k (\mathcal{L}_{\hat{a}_k}  + \mathcal{L}_{\hat{a}_{-k}}) + \gamma^{\rm h}_k (\mathcal{L}_{\hat{a}_k^\dagger}  + \mathcal{L}_{\hat{a}_{-k}^\dagger}),
    \label{eq:mastercoolingave}
\end{equation}
where we can approximate the rates by
\begin{align}
    \gamma^{\rm c}_k &= \frac{2g^2}{(\Delta-\epsilon_k)^2+ \gamma_0^2}, \label{eq:gammac_def}\\
    \gamma^{\rm h}_k &= \frac{2g^2}{(\Delta+\epsilon_k)^2}, \label{eq:gammah_def}
\end{align}
and
\begin{equation}
    \gamma_0^{-2}=\frac{\bar t^2}{2}\approx \frac{2t^2}{3}.
    \label{eq:tau0a}
\end{equation}
The first terms in \cref{eq:mastercoolingave} cool the modes $k$ and $-k$, whereas the remaining terms heat them up. With this expression we can easily compute the cooling rate per elementary cycle [cf.~\cref{eq:n0k}] and obtain
\begin{equation}
    \alpha_k = (n^c_k)^{-1} = \gamma^{\rm c}_k + \gamma^{\rm h}_k,
    \label{eq:coolingrate}
\end{equation}
so that the total time is $n^c_k t$~\footnote{Derivation details are shown at the end of~\cref{app:long_time_averages}.}.
We emphasize that for a global cycle of $L$ steps, the rate would be $L\alpha_k$ but it is more clear if we just compute the rate per elementary cycle, since then it can be directly compared with the results of other sections.

The joint steady-state energy and relative energy for the modes $k$ and $-k$ are:\footnote{Note that here we take $\epsilon_k$ to be positive, and in the computation of the relative energy one is subtracting the ground state energy $-\epsilon_k$, and thus gets the plus sign here.}
\begin{align}
    E_k &= \epsilon_k \frac{-\gamma^{\rm c}_k+\gamma^{\rm h}_k}{\gamma^{\rm c}_k+\gamma^{\rm h}_k}, \label{eq:avg_longtime_ss_Ek}\\
    e_k &= \frac{E_k+\epsilon_k}{\epsilon_k} = \frac{2\gamma^{\rm h}_k}{\gamma^{\rm c}_k+\gamma^{\rm h}_k}.
    \label{eq:avg_longtime_ss1}
\end{align}

\begin{figure}[ht]
    \begin{overpic}[width=\linewidth]{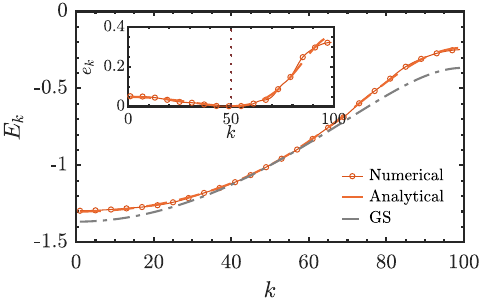}
        \put(1, 60){(a)}
    \end{overpic}
    \begin{overpic}[width=\linewidth]{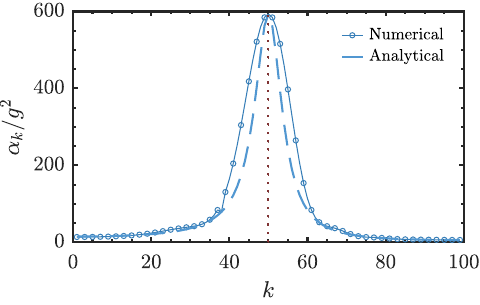}
        \put(1, 60){(b)}
    \end{overpic}
    \vspace*{-5mm}
    \caption{(a) Steady-state energy per mode $E_k$ versus $k$ in the randomized-time scenario for $\theta=\pi/3$, $N=200$, $g=10^{-4}$, $t=20$, $\Delta=\epsilon_{N/4}$, and $L=100$. The dashed dot line is the exact numerical solution; the solid line is given by our perturbative formula from \cref{eq:avg_longtime_ss_Ek}. The ground-state energy is shown in gray (labeled GS). The inset shows the relative energy $e_k$ from \cref{eq:avg_longtime_ss1}. (b) Cooling rate $\alpha_k / g^2$ as a function of $k$. The exact, numerical results (dots) agree well with the analytical prediction from \cref{eq:gammac_def,eq:gammah_def,eq:coolingrate} (blue line). The vertical dashed line marks $k=N/4$, where $\Delta=\epsilon_k$. Hence, we see that randomizing the cycle times washes out resonances away from the primary resonance point, allowing all modes to cool more efficiently.}
    \label{fig:Fig3}
\end{figure}

In \cref{fig:Fig3} we have computed exactly the steady-state energies $E_k$ and the cooling rates $\alpha_k$ for a map with $L=100$ random-time cycles. In addition to the numerical results, we have also presented the analytical approximations given by~\cref{eq:avg_longtime_ss_Ek,eq:avg_longtime_ss1}.
As can be seen from~\cref{fig:Fig3}(a) and its inset, the oscillations present in the single-frequency case (see~\cref{fig:Fig2}) have disappeared, and the cooling performance across the spectrum is now much more uniform compared to the single-frequency case (\cref{fig:Fig2}). The analytical result estimates well the value of both the energy [\cref{eq:avg_longtime_ss_Ek}] and the cooling rate [\cref{eq:coolingrate}].

As before, the modes for which $\Delta=\epsilon_k$ (marked by the vertical dashed line at $k=N/4$) are cooled most effectively, with a cooling rate $\alpha_{N/4}/g^2 \approx 590$ that is nearly two orders of magnitude higher than the rate away from resonance near the edge.
The relative energy still reaches its minimum of $e_k \approx 0.01$ near the resonant mode $k=\pm N/4$ and gradually increases to approximately 0.05 (0.3) at the left (right) edges of the spectrum ($k=0$ and $k=100$).

\begin{figure}[ht]
    \begin{overpic}[width=\linewidth]{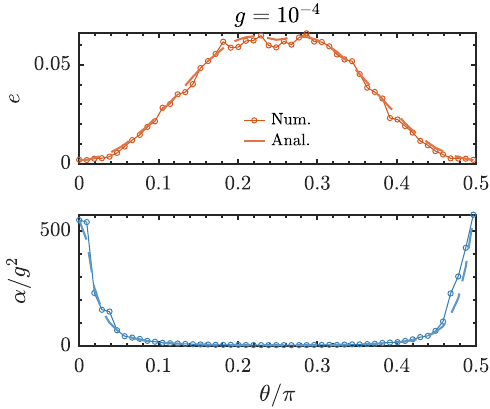}
        \put(2, 74){(a)}\label{fig:Fig4old:a}
        \put(2, 37){(b)}\label{fig:Fig4old:b}
    \end{overpic}
    \caption{Relative energy $e$ and cooling rate $\alpha/g^2$ as a function of the model parameter $\theta$ for $N=200$, $g=10^{-4}$, $t=20$, $\Delta=\epsilon_{N/4}$, and $L=100$. Cooling is slower around the quantum critical point $\theta=\pi/4$, where the relative energy peaks at $e\approx 0.065$ and the cooling rate reaches its minimum of $\alpha/g^2\approx 1$. Away from criticality, the system reaches much higher cooling rates and lower relative energies.
    The analytical approximation (blue) [\cref{eq:coolingrate}]
    and the exact numerical results (dot symbols) agree very well across the parameter range.
    }
    \label{fig:Fig4old}
\end{figure}

In \cref{fig:Fig4old} we plot the total relative energy [\cref{eq:relative_energy}] as a function of $\theta$ for the same parameters as in \cref{fig:Fig3}. The energies are relatively low, indicating sizable cooling for all modes. Cooling is better for $\theta \sim 0, \pi/2$, and worse around the phase transition at $\theta=\pi/4$, as expected. Also, the cooling rate slows down near that point. In fact, for $\theta=\pi/4$, as mentioned earlier, we have $\epsilon_{N/2}=0$, so that this particular mode (and those close to it) will not be cooled since $\gamma^{\rm c}_{N/2} = \gamma^{\rm h}_{N/2}$.

\begin{figure*}[ht]
    \begin{overpic}[width=\linewidth]{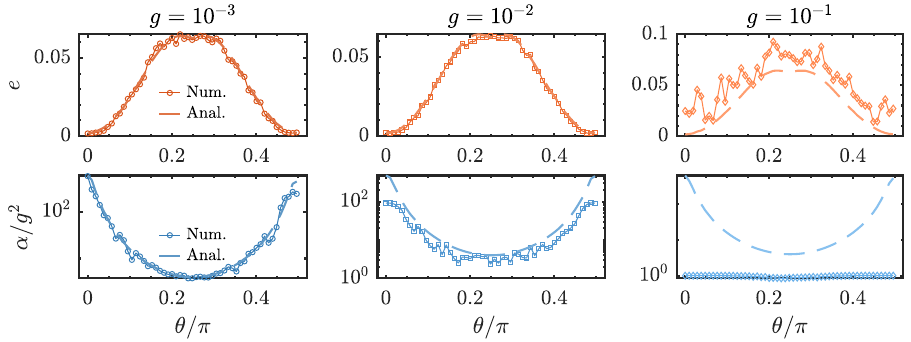}
        \put(5, 35){(a)}\label{fig:Fig4a}
        \put(38, 35){(b)}\label{fig:Fig4b}
        \put(68, 35){(c)}\label{fig:Fig4c}
        \put(5, 18){(d)}\label{fig:Fig4d}
        \put(38, 18){(e)}\label{fig:Fig4e}
        \put(70, 18){(f)}\label{fig:Fig4f}
    \end{overpic}
    \vspace*{-5mm}
    \caption{Comparison of cooling performance for different coupling strengths $g$. From left to right columns: $g=10^{-3}$, $g=10^{-2}$, and $g=10^{-1}$. (a)-(c): total relative energy $e$ as a function of $\theta$. (d-f): cooling rate $\alpha/g^2$ as a function of $\theta$. Other parameters are the same as in \cref{fig:Fig3}. Dots show numerical results while lines show analytical predictions from \cref{eq:avg_longtime_ss1,eq:coolingrate}.
    The cooling is slower around the quantum critical point $\theta=\pi/4$.
    One can observe that the agreement between the analytical approximations (blue) and the exact numerical results (purple) get better when $g$ decreases.
    For $g=10^{-3}$ (a,d), the relative energy reaches a minimum of $e \approx 0.002$ away from the critical point, with perfect agreement between numerics and analytics. At $g=10^{-2}$ (b,e), the analytical approximation of the relative energy remains highly accurate but the analytical approximation to the cooling rate starts to downgrade. For $g=10^{-1}$ (c,f), despite exceeding the weak coupling limit with $gt=2 > 1$, cooling still occurs with a minimum relative energy of $e \approx 0.1$, although the analytical approximation now shows significant deviations from the numerical results.
    }
    \label{fig:Fig4}
\end{figure*}

In \cref{fig:Fig3,fig:Fig4old} we have taken $g=10^{-4}, t=20$ and $L=100$, for which the condition $(g L t)^2\ll 1$ is satisfied and thus the perturbative argument is valid. This condition is very stringent since it implies that $g\propto 1/L$, and thus the cooling rate will get smaller by a factor of $L^2$, thus leading to extremely long cooling times. This would be particularly disadvantageous in a noisy environment. However, one would expect that cooling should persist as long as the conditions [\cref{eq:weak_coupling_condition,eq:coolingcondition}] are satisfied, independent of $L$. The reason for this is that the role of $L$ is to obtain averaged results. In \cref{fig:Fig4} we investigate to what extent cooling persists when we increase $g$. There, we plot the total relative energy [\cref{eq:relative_energy}] as a function of $\theta$ for the same parameters as in \cref{fig:Fig4old} and different values of $g$. For $g=10^{-2}$ (where $gLt = 20 > 1$ but still $gt = 0.2 < 1$), we obtain similar results as for $g=10^{-4}$ in the steady state energy, and which are consistent with the analytical formulae. The cooling rate, however, is somewhat modified, although one sees the benefit of increasing $g$. For $g=10^{-1}$ (for which $gt = 2 \agt 1$) cooling still persists, although the analytical formulae are not accurate anymore. Finally, for $g=1$ (not shown), basically no cooling takes place.

From the expressions of $\gamma^{\rm c,h}_k$ it is clear that a mode corresponding to momentum $k$ with $\epsilon_{k_0}\approx \Delta$ will be cooled. In order to get a better analytical understanding, let us consider two different sets of modes: (i) $|\epsilon_{k_0}- \Delta|\alt \gamma_0$; (ii) $\gamma_0\alt |\epsilon_{k_1}- \Delta|$. The relative energies in steady state and the cooling rates per cycle will be\footnote{Via an expansion of the Lorentzians in \cref{eq:gammac_def,eq:gammah_def} in the corresponding limits.}
\begin{align}
    \label{eq:ek0_ek1}
    e_{k_0}      &\approx \frac{3}{4(\Delta t)^2}, \quad  e_{k_1} \approx \frac{(\Delta-\epsilon_{k_1})^2}{\Delta^2+\epsilon_{k_1}^2},\\
    \alpha_{k_0} &\approx \frac{4}{3} g^2 t^2, \quad \alpha_{k_1} \approx \frac{2g^2}{(\Delta-\epsilon_{k_1})^2}.
    \label{eq:alpha01}
\end{align}
The first modes ($k_0$) get very cold for long cycle times $t$, although the set of modes for which that happens also decreases with time. However, note that not only these first modes are cooled; the modes ($k_1$) that fulfill $|\epsilon_{k_1}- \Delta|\ll \Delta$ are also cooled down according to \cref{eq:ek0_ek1}, although the rate and the final energy do not depend on the cycling time $t$, and thus they cannot be made arbitrarily large or small, respectively. The relative energy is maximal for the modes with $k=0,N/2$. All this explains more quantitatively the results of \cref{fig:Fig3}.

\subsection{Multi-frequencies}
\label{subsec:multi_freq}

We consider now the scenario where we have several cycles, each of them characterized by a value of $\Delta_r$ where $r=1,\ldots,R$. Thus, we have several frequencies in the bath, resonant with those of different system modes, which can thus be cooled simultaneously~\cite{Eschner2003Laser} (see also~\cite{Marti2025Efficient}). However, for a given value of the coupling strength $g$, the weak coupling condition \cref{eq:weak_coupling_condition} will eventually be violated, as if more frequencies are used, the total duration of the global cycle, $Rt$, will also increase. Thus, in order to cool further it may be necessary to decrease $g$ as we add more frequencies. As before, we will assume that the procedure is repeated $L$ times in order to obtain averaged results and avoid unwanted resonances. The times for each cycle, $t_{r,m}$ with $m=1,\ldots,L$, are randomly chosen from an interval $[0,2t]$.

The extension of the previous results to this case is straightforward. As before, we can use perturbation theory, average the resulting equations and obtain the master equation [\cref{eq:mastercoolingave}], but with the replacement $L\to RL$ and
\begin{equation}
    \gamma^{\rm c,h}_k = \frac{1}{R} \sum_{r=1}^R \gamma^{\rm c,h}_{r,k},
    \label{eq:gammach}
\end{equation}
with $\gamma^{\rm c,h}_{r,k}$ given by~\cref{eq:gammac_def,eq:gammah_def} and where, for each $r$, $\gamma_0$ in these equations is replaced by the same expression but where we sum the times $t_{r,m}$ over $m$ [cf.~\cref{eq:tau0a}].

\begin{figure}[ht]
    \begin{overpic}[width=\linewidth]{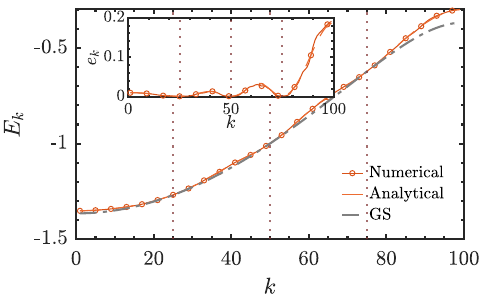}
        \put(0, 59){(a)}
    \end{overpic}
    \begin{overpic}[width=\linewidth]{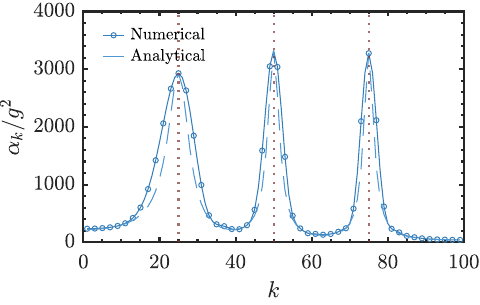}
        \put(0, 59){(b)}
    \end{overpic}
    \vspace*{-5mm}
    \caption{(a) Steady-state energy per mode $E_k$ versus $k$ using three distinct bath frequencies, $\Delta=\epsilon_{N/8},\epsilon_{N/4},\epsilon_{3N/8}$. We set $L=100$ cycles for each frequency, randomly distributed times $t_{r,m}$ with mean $t=50$, $g=10^{-4}$, $\theta=\pi/3$, and $N=200$. The main plot shows $E_k$ from~\cref{eq:avg_longtime_ss_Ek}, while the inset shows the relative energies $e_k$ from~\cref{eq:avg_longtime_ss1}. (b) Cooling rates $\alpha_k/g^2$ with the analytical approximation from~\cref{eq:coolingrate}. As before, dots correspond to the exact numerical calculation, lines correspond to the perturbative analytical results, and the ground-state (GS) is indicated by the gray line. Brown vertical dashed lines show the modes resonant with the three chosen frequencies, where cooling is strongest.
    }
    \label{fig:Fig5}
\end{figure}

In \cref{fig:Fig5}, we have computed the energy in steady state and the cooling rate per elementary cycle for $R=3$ different frequencies, and $L=100$ cycles for each of them. The frequencies are chosen as $\Delta_r=\epsilon_{k_r}$ with $k_r = r N/8$ for $r=1,2,3$ (marked by vertical dashed lines). All those frequencies are alternating in the exact calculation, i.e., the $(3m+r)$-th cycle uses $\Delta=\Delta_r$ ($r=1,2,3$, and $m=1, \ldots, L$).
In the relative energy plot (inset), one can clearly see that the modes resonant with the bath frequencies, i.e., $\epsilon_k \approx \Delta_r$, are cooled most effectively, reaching relative energies as low as $e_k < 0.01$ at these resonant points. The cooling rate also shows three distinct peaks at the corresponding resonant frequencies. As seen in \cref{fig:Fig5}, the analytical calculation predicts well the numerical results.

\begin{figure}[ht]
    \begin{overpic}[width=\linewidth]{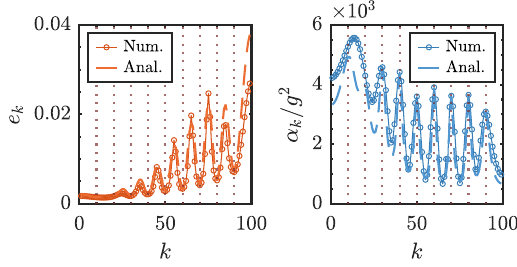}
        \put(0, 46){(a)}
        \put(53, 46){(b)}
    \end{overpic}
    \begin{overpic}[width=\linewidth]{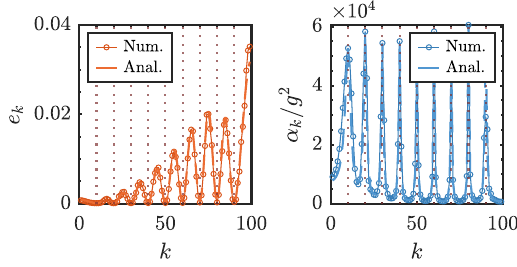}
        \put(0, 46){(c)}
        \put(53, 46){(d)}
    \end{overpic}
    \vspace*{-5mm}
    \caption{(a) Relative energy $e_k$ (left panel) and (b) cooling rate $\alpha_k/g^2$ versus $k$ for the same scenario as \cref{fig:Fig5} but with nine distinct frequencies, $\Delta=\{\epsilon_{x}, x=0.1,0.2,\ldots,0.9\}$. Increasing the number of resonances broadens the cooling range at the cost of slightly degraded analytic-numerical agreement. The numerical results (dots) agree well with the analytical results from~\cref{eq:avg_longtime_ss1,eq:coolingrate} (solid line). (c-d) The same result with $t=200$. The longer average pulse time shrinks $\gamma_0$ in \cref{eq:tau0a} and thus sharper resonances are visible.}
    \label{fig:Fig6}
\end{figure}

In \cref{fig:Fig6}(a-b), we performed the same computation but with nine different frequencies $\Delta=\{\epsilon_{x}, x=0.1,0.2,\ldots,0.9\}$, covering a broader range of the spectrum. On the one hand, the relative energies (left panel) get smaller, since more modes are cooled, with $e_k < 0.02$ achieved across most of the spectrum. On the other hand, the analytical approximation for the cooling rates (right panel) becomes less accurate. The reason is that the first peaks overlap with each other, which is neglected in our approximations. This can be resolved by choosing longer evolution times, as demonstrated in \cref{fig:Fig6}(c-d), where we increase $t$ to 200. The longer average pulse time leads to sharper resonances and better agreement between analytics and numerics.

\begin{figure}[ht]
    \includegraphics[width=\linewidth]{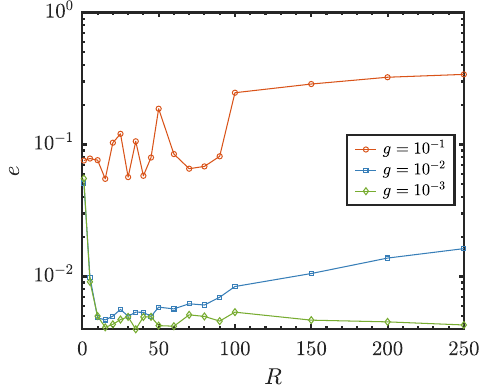}
    \vspace*{-5mm}
    \caption{Total relative energy $e$ versus the number of bath frequencies $R$ for $N=1000$, $\theta=\pi/3$, $t=50$ and three coupling strengths: $g=0.1$ (red), $g=0.01$ (magenta) and $g=0.001$ (blue). Each frequency is chosen to be equally spaced in momentum from $k=0$ up to $k=N/2$ (i.e., $\Delta_r=\epsilon_{k_r}$ with $k_r=N/2 \times r/(R+1)$). For $g=10^{-3}$, the relative energy decreases
    from $e \approx 0.025$ at $R=1$ to $e \approx 0.006$ at $R=250$, and remains stable despite $gRt=12.5 \gg 1$.
    For $g=10^{-2}$ the same trend is shown initially, although as the cooling condition is strongly violated, the relative energy increases at $R=250$. For $g=0.1$ the cooling is significantly worse with $e \approx 0.1-0.2$, as expected since $gt = 5 \agt 1$.
    }
    \label{fig:Fig8}
\end{figure}

The improved cooling across a wide range of modes, as shown in \cref{fig:Fig6}, translates to a lower total relative energy. To analyze this systematically, we now investigate how the total relative energy behaves as we increase the number of frequencies. In \cref{fig:Fig8}, we plot the total relative energy, $e$, as a function of the number of frequencies used, $R$, for three coupling strengths: $g=0.1$, $g=0.01$, and $g=0.001$ for $t=50$ and $N=1000$. For $g=10^{-1}$, the weak coupling condition [\cref{eq:weak_coupling_condition}] is violated even for a single frequency ($gt=5 \gg 1$), and thus the final relative energy is relatively high ($e \approx 0.1-0.2$) for all values of $R$. For $g=10^{-2}$ and $g=10^{-3}$, one clearly sees that as $R$ increases, the relative energy decreases. For $g=10^{-2}$, at around $R=10$ (where $gRt=5 > 1$), the relative energy starts increasing again, which can be attributed to the violation of the weak coupling condition. For $g=10^{-3}$, however, this does not seem to be the case and the relative energy seems to saturate at $e \approx 0.006$ as a function of $R$ even up to $R=250$ (where $gRt=12.5 > 1$).

\begin{figure}[ht]
    \begin{overpic}[width=\linewidth]{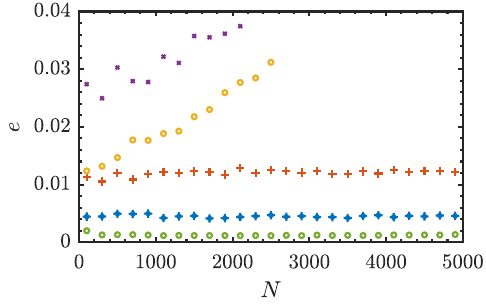}
        \put(2, 59){(a)}
    \end{overpic}
    \begin{overpic}[width=\linewidth]{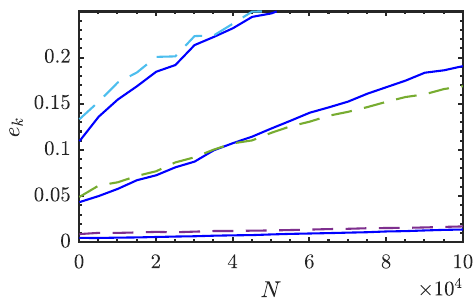}
        \put(2, 59){(b)}
    \end{overpic}
    \vspace*{-5mm}
    \caption{(a) Total relative energy $e$ versus the number of fermionic modes $N$ for $\theta=\pi/3$. For the two lower curves $g=10^{-2}$, and $t=20$ (`o') and $t=10$ (`x'). For the three upper curves, $g=10^{-3}$, and $t=20$ (`+'), $t=50$ (`*') and $t=200$ (`o'). (b) Relative energy $e_k$ for $k=N/8$ (solid lines) and $k=N/4$ (dashed lines) as a function of $N$. The curves from bottom to top correspond to $g=10^{-2}, 5\times 10^{-3}, 10^{-3}$ and $t=0.025/g$. In all plots, the number of frequencies scale with the system size as $R=N/5$. As before, the frequencies are equally spaced in momentum from $k=0$ up to $k=N/2$. As $N$ increases, all curves eventually show increasing relative energy $e_k$, though at different rates.}
    \label{fig:Fig8b}
\end{figure}

We have further investigated this behavior in \cref{fig:Fig8b}, where we have plotted the relative energy as a function of $N$, with $R=N/5$, for various values of $g$ and $t$. \cref{fig:Fig8b} (a) shows that, as expected, for $g=10^{-2}$ the relative energy is larger than for $g=10^{-3}$, and also that the energy becomes smaller for longer times $t$. As we increase the value of $N$ (and thus of $R$), the weak-coupling condition gets violated. In the case of $g=10^{-2}$, this is reflected in the increase of the relative energy, as in \cref{fig:Fig8}. However, for $g=10^{-3}$ the relative energy seems to saturate, even though $gtR\gg 1$ (for instance $gtR=0.2N$ for the curve with circles, and thus for $N=5000$ this value is 1000). Since it takes relatively long to compute the value of the energy as $N$ increases, in order to see if this trend continues for larger values of $N$ we have just computed the relative energy $e_k$ for some values of $k$. In \cref{fig:Fig8b}(b) we have plotted this quantity for two values of $k$ (solid and dashed lines), for $g=0.01,0.005,0.001$. One can see there that even for $g=10^{-3}$, the curve eventually bends and the relative energy increases for large values of $N$ ($N > 40,000$). However, this tendency is very smooth even though the weak-coupling condition is violated. Thus, in case we want to cool to very low energies, i.e., close to the ground state, we will have to increase the number of frequencies and reduce the coupling constant correspondingly, which will necessarily lead to longer cooling times.

In order to obtain simple analytical formulas, we define
\begin{align}
    \delta   &= (\epsilon_{\rm M} - \epsilon_{\rm m})/R ,\\
    \Delta_r &= \epsilon_{\rm m} + \delta (r - \frac{1}{2}),
    \label{eq:Deltar}
\end{align}
where $\epsilon_{\rm M, m} = \sqrt{1\pm \sin(2\theta)}$ are the maximum and minimum values of $\epsilon_k$. We will consider the regime $R^2 \gg 1$ and $\delta^2 t^2 \ll 1$. In this limit, the sums over $r$ can be replaced by integrals, yielding (see~\cref{app:multifrequency_cooling})
\begin{align}
    \gamma^{\rm h}_k &\approx \frac{2 g^2 }{(\epsilon_{\rm M}+\epsilon_k)(\epsilon_{\rm m} +\epsilon_k)},\\
    \gamma^{\rm c}_k &\approx \frac{2g^2 t \beta_k}{\epsilon_{\rm M}-\epsilon_{\rm m}},
\end{align}
where $\beta_k = \sqrt{2/3} [\tan^{-1}(z_M)-\tan^{-1}(z_m)]$, and $z_x=(\epsilon_x-\epsilon_k)t \sqrt{2/3} $. We then obtain
\begin{align}
    e_k      &= \frac{\epsilon_{\rm M}-\epsilon_{\rm m}}{\beta_k(\epsilon_{\rm m}+\epsilon_k)}\frac{1}{t},\\
    \alpha_k &= \frac{\beta_k g^2 t}{\epsilon_{\rm M} - \epsilon_{\rm m}}.
\end{align}
Again, the cooling rate $\alpha_k$ is defined per elementary cooling cycle, and the global rate including $L\times R$ elementary cycles is $\alpha_k LR$.

Here, we can distinguish two cases. In the first one,
\begin{equation}
    R\gg (\epsilon_{\rm M}-\epsilon_{\rm m})t \gg 1,
    \label{eq:Rggeps}
\end{equation}
we obtain that $\beta_k \in[\pi/2,\pi]$, and thus $e_k$ scales as $1/(\epsilon_{\rm M} t)$ while $\alpha_k$ scales as $(gt)^2/(\epsilon_{\rm M} - \epsilon_{\rm m})t$. In the second one,
\begin{equation}
    R\gg 1 \gg (\epsilon_{\rm M}-\epsilon_{\rm m})t,
    \label{eq:Rgg1}
\end{equation}
we find $\beta_k\approx (\epsilon_{\rm M}-\epsilon_{\rm m})t\sqrt{2/3}$, and hence $e_k$ scales as $1/(\epsilon_{\rm M} t)^2$ and $\alpha_k$ as $(gt)^2$. More details are provided in \cref{app:multifrequency_cooling}.

\subsection{Cooling to the ground state}
\label{subsec:cooling_to_gs}

An important question is whether it is possible to cool the system arbitrarily close to its ground state and, if so, what the necessary time scale is. We require that the fidelity $\mathcal{F}=O(1)$ be independent of the system size. As argued in \cref{sec:steady-state}, this amounts to demanding that for each mode $k$, the infidelity $\varepsilon_k = 1 - F_k = O(1/N)$. As seen in \cref{subsec:single_time}, this is not possible with a single bath frequency, since only those modes nearly resonant with that frequency are cooled effectively. We will show next that with multifrequency cooling, cooling to the ground state is possible, as suggested by the general theory of Davies~\cite{Davies1974Markovian}. However, in contrast to that theory--which typically requires a time growing exponentially with system size--we will show that for our specific model (when away from the phase transition) the cooling times only scale polynomially.

We simply have to take $R, \epsilon_{\rm M} t = O(N)$ so that in both regimes, given by~\cref{eq:Rggeps,eq:Rgg1} we have $\varepsilon_k=O(1/N)$, and thus $\mathcal{F}=O(1)$. To fulfill the weak-coupling condition [\cref{eq:weak_coupling_condition}], we must take $g=O(1/(R t))=O(1/N^2)$, so that the number of cycles is $n^c\sim \epsilon_{\rm M} t/(gt)^2 = O(N^3)$. This yields a total cooling time $T=n^c t= O(N^4)$, with the announced polynomial scaling with $N$.

\subsection{Impact of Noise}
\label{subsec:impact_noise}

Now we analyze how the results are modified in the presence of noise. We will use the model of~\cref{sec:adding_decoherence}, since given its simplicity it is possible to extend the analytical results derived above. As we explained there, we just have to replace the map in \cref{eq:cooling_map0} by~\cref{eq:mapwithnoise}, and use again the translational invariance to decompose the map into products for the different values of $k$ [\cref{eq:CPMk}]. This allows us to compute exactly the steady state and the cooling rates, as we did before.

\begin{figure}[ht]
    \begin{overpic}[width=\linewidth]{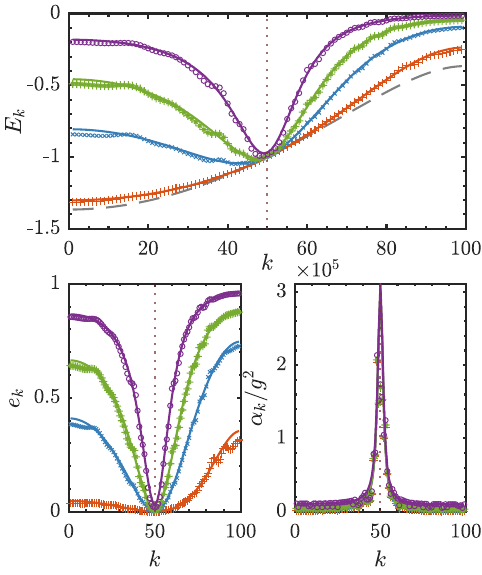}
        \put(4, 96){(a)}
        \put(4, 47){(b)}
        \put(45, 47){(c)}
    \end{overpic}
    \caption{(a) Energy $E_k$, (b) relative energy $e_k$, and (c) cooling rate $\alpha_k/g^2$ for the same parameters as in \cref{fig:Fig3} ($\theta=\pi/3$, $N=200$, $g=10^{-4}$, $t=20$, $\Delta=\epsilon_{N/4}$, and $L=100$), with varying noise strength $\kappa$. The ratios $\kappa/g^2$ are $[0,0.03,0.1,0.3,1]$ for the orange, blue, green, and purple lines. As noise increases, the energy spectrum progressively flattens and shifts away from the ground state, with the effect most pronounced for modes far from resonance. The solid lines correspond to the analytical approximations from \cref{eq:avg_longtime_ss_Ek,eq:avg_longtime_ss1,eq:coolingrate} with noisy rates $\gamma^{\rm c,h}_k + \kappa t$, which remain accurate across all noise strengths. The relative energies $e$ for the steady state of the full system are, from bottom to top: $0.052,0.292,0.479,0.673$.}
    \label{fig:Fig10}
\end{figure}

In the weak-interaction limit, which is now defined as [compare~\cref{eq:weak_coupling_condition}]
\begin{equation}
    (g R t)^2,\ \kappa t \ll 1,
\end{equation}
we again obtain that during an elementary cycle the master equation is given by the Lindbladian in \cref{eq:mastercoolingave}, with the replacement $L\to RL$ in the multifrequency case, and with $\gamma^{\rm c,h}_k$ replaced by $\gamma^{\rm c,h}_k + \kappa t$. The noise will not affect cooling as long as
\begin{equation}
    \gamma^{\rm c}_k \agt \kappa t.
    \label{eq:gammaagtkappa}
\end{equation}
Let us consider first the case of a single frequency, but with multiple cycles as in \cref{subsubsec:randomized_times}. In \cref{fig:Fig10} we have plotted the energy and the relative energy for the same parameters as~\cref{fig:Fig3} for $\kappa \le g^2$.
As the noise strength increases from $\kappa/g^2=0$ (purple) to $\kappa/g^2=1$ (green), the system energy progressively deviates from the noiseless case.
For small noise ($\kappa/g^2=0.03$), cooling remains effective near the resonant mode ($k=N/4$), with $e_k<0.05$, but shows degradation for distant modes. Remarkably, the resonant mode remains only weakly affected by noise even up to $\kappa/g^2\approx 1$, while the others, which were mildly cooled in its absence, are not cooled anymore. This is consistent with the cooling rate (bottom panel), which shows that the resonant mode still has the largest cooling rate and thus for that mode the condition [\cref{eq:gammaagtkappa}] is satisfied.

We can get some analytical understanding by coming back to formulas [\cref{eq:alpha01}] derived for modes $k_{0,1}$ fulfilling: (i) $|\epsilon_{k_0}- \Delta|\alt \gamma_0$ and  (ii) $\gamma_0\alt |\epsilon_{k_1}- \Delta|$, and requiring the low-noise condition [\cref{eq:gammaagtkappa}] (where $\gamma^{\rm c}\approx \alpha_{k_0}$). For modes $k_0$ (i), in the regime of weak coupling [\cref{eq:weak_coupling_condition}], cooling simply requires $g\gg \kappa$. For modes $k_1$ (ii), employing the cooling condition [\cref{eq:coolingcondition}], cooling requires $g^2\gg \kappa (\Delta-\epsilon_{k_1})^2/\Delta$, which is much more stringent since it involves $g^2$ instead of $g$.

\begin{figure*}[ht]
    \includegraphics[width=0.8\linewidth]{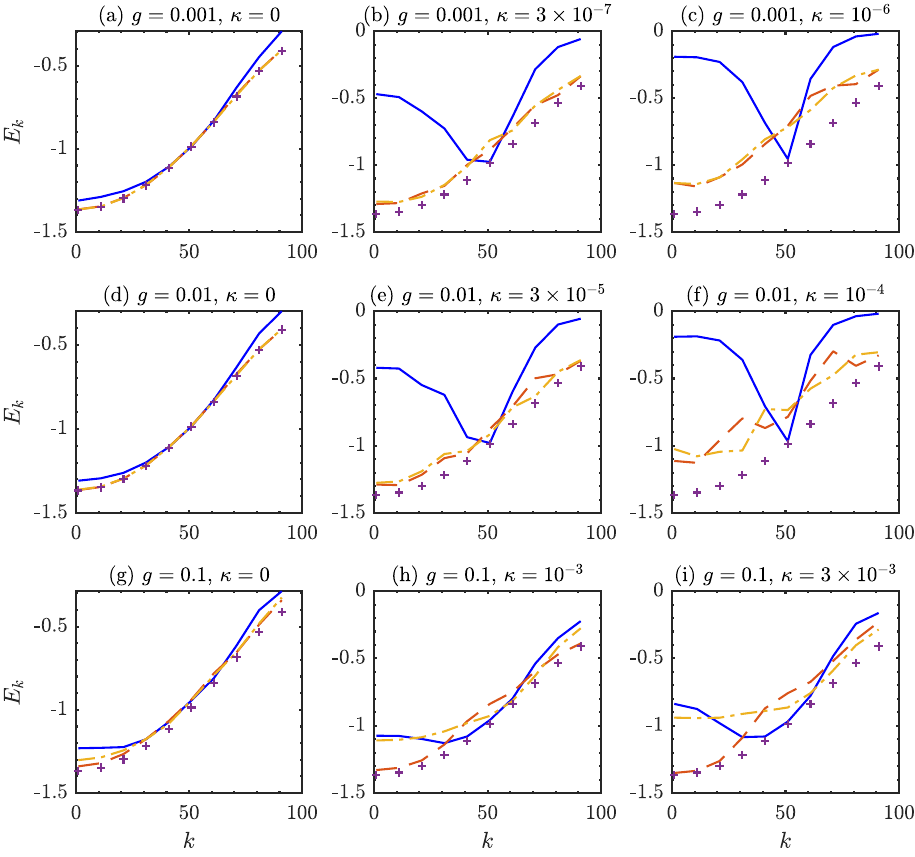}
    \caption{Energy $E_k$ as function of $k$ for $N=200$, $\theta=\pi/3$ and different values of $g$, $t$, $\kappa$ and $R$. Figures (a-c), (d-f) and (g-i) correspond to the coupling strengths $g=0.001$, $g=0.01$, and $g=0.1$, respectively, and cycle times $t=50$, $t=50$, and $t=10$, respectively. The number of bath frequencies are $R=1$ (solid lines), $R=11$ (dashed lines), and $R=21$ (dash-dotted lines). The crosses are the exact ground state energies ($-\epsilon_k$). For Figs.~(a,d,g), $\kappa=0$ (noiseless case), while for (b,c,e,f,h,i), the noise strengths are $\kappa=3\times 10^{-7},10^{-6},3\times 10^{-5}, 10^{-4}, 10^{-3}, 3\times 10^{-3}$, respectively. The energy spectrum shows progressive reheating with increasing noise strength, though this effect is partially mitigated by using multiple bath frequencies. Multi-frequency cooling ($R=11$ and $R=21$) consistently outperforms single-frequency cooling ($R=1$) across all parameter regimes.}
    \label{fig:FigNew3}
\end{figure*}

Let us now consider the multiple frequency scenario ($R>1$). In \cref{fig:FigNew3}, we investigate the interplay between noise, coupling strength, and the number of bath frequencies. The figure displays the steady-state energies $E_k$ for different parameter combinations. Each row corresponds to a fixed coupling strength $g$ ($0.001$, $0.01$, $0.1$), while each column corresponds to a fixed noise strength $\kappa$ (relative to $g^2$). Within each panel, different line styles represent different numbers of bath frequencies $R$ ($1$, $11$, $21$). First, let us look at the effect of noise for a fixed coupling strength $g$ and number of frequencies $R$ (i.e., moving horizontally across a row, looking at a specific line style). As expected, increasing the noise strength $\kappa$ leads to a degradation in cooling performance, with the steady-state energy moving further away from the ground state energy (purple crosses). This effect is visible across all coupling strengths. For instance, comparing panels (a), (b), and (c) for $g=0.001$, the dash-dotted line ($R=21$) closely follows the ground state in the noiseless case (a), but deviates significantly as noise increases (b, c).

Next, let us consider the effect of increasing the number of bath frequencies $R$ for fixed $g$ and $\kappa$ (i.e., comparing different line styles within a single panel). First, for the particular mode that was cooled down by a single frequency ($k=N/4$), adding more frequencies has a detrimental effect. This can be understood since the frequencies that are not resonant tend to heat that mode, which cannot be compensated by the resonant frequency. However, in general, using multiple frequencies ($R=11$ or $R=21$) leads to better overall cooling performance compared to a single frequency ($R=1$), as indicated by the energy being closer to the ground state. For example, in panel (c), the multi-frequency lines (dashed and dash-dotted) are significantly closer to the ground state across the entire spectrum compared to the single-frequency line (solid blue).

However, the advantage of using multiple frequencies decreases as the coupling strength $g$ or the noise strength $\kappa$ increases. For strong coupling ($g=0.1$, bottom row), the difference between $R=1$, $R=11$, and $R=21$ is less pronounced, especially in the presence of noise (panels (h), (i)). Despite these limitations, the results demonstrate a degree of robustness. Even for the relatively large coupling $g=0.1$ (bottom row), multi-frequency cooling still offers some improvement over single-frequency cooling, particularly in the noiseless case (panel (g)). This highlights the general utility of the multi-frequency approach combined with randomized times as a way to boost the cooling performance.

In order to get some analytical insight, we consider the situation analyzed in the previous section, with the frequencies $\Delta_r$ given in \cref{eq:Deltar}, so that all the modes can be simultaneously cooled down. For that, we assume the regime of~\cref{eq:Rggeps}. In such a case, the condition given by~\cref{eq:gammaagtkappa} results in
\begin{equation}
    g^2/\kappa \gg  (\epsilon_{\rm M}-\epsilon_{\rm m}).
    \label{eq:kappag}
\end{equation}
One has to recall that this is under the assumption that $R\gg 1$ and the weak coupling condition [\cref{eq:weak_coupling_condition}], so that as we increase $R$, one will have to reduce $g$ and the cooling condition [\cref{eq:kappag}] will be eventually violated. This will occur in case we want to cool down to the global ground state. If we are interested in the relative energy, then we can just fix the value of $R$, which will indicate for which values of $\kappa$ we can obtain it. The behavior of~\cref{fig:FigNew3} can be then understood by considering the interplay between the weak coupling condition [\cref{eq:weak_coupling_condition}] and the condition of the strength of the noise [\cref{eq:kappag}]. Increasing $R$ helps cool more modes effectively, but it also strengthens the requirement on $g$ to satisfy $(gRt)^2 \ll 1$. If $g$ is too large, the weak coupling approximation breaks down, limiting the benefit of multiple frequencies. Conversely, if $g$ is reduced to satisfy the weak coupling condition for large $R$, the system becomes more susceptible to noise according to~\cref{eq:kappag}, $\kappa (\epsilon_{\rm M} - \epsilon_{\rm m}) \ll g^2$.


\section{Generalization of Cooling and DSP}
\label{sec:nn_cooling}

In the previous sections, we aimed to understand the simple scenario of purely local couplings ($nn=0$), setting $\lambda_0=\mu_0=1$. We focused on the weak coupling limit $(gt)^2\ll1$ and considered the cooling limit $(\Delta t)^2\gg1$. We demonstrated that we can push the weak coupling limit quite far; however, the cooling limit requires long cycle times $t$. This makes noise effects, which are proportional to $\kappa t$, more relevant and creates experimental challenges.\\

In this section, we will focus on an improvement of the first strategy used in this paper; namely, using just one frequency $\Delta$ and one cycle time $t$. We will still consider the weak coupling limit $(gt)^2 \ll 1$ but focus now on the regime $t \approx 1$, which will permit both larger $g$ and, in turn, larger noise strength $\kappa$. The improvement will be due to the fact that we consider larger nearest-neighbor distances, i.e., $nn \ge 0$, and that we will optimize the coupling constants as well as the cycle time and the bath frequency. We will focus here on the energies of the steady state of the considered process.

The different nearest-neighbor distances are denoted with a full integer $0,1,2,\ldots$ when both bath sites $j=nn$ and $j=-nn$ are connected to the system, and with a half-integer $0.5,1.5,\ldots$ when only sites at $j=nn$ are coupled (with $j=-nn$ turned off). This half-integer notation reflects a configuration equivalent to a symmetric arrangement where the entire bath is shifted by half a site and the nearest-neighbor distance becomes a half-integer.

The energy of the steady state, derived in \cref{subapp:rho_ss_noiseless}, is now more involved than in \cref{subsec:single_time} as it takes into account multiple coupling terms $\lambda_j, \mu_j$ which will now play an important role. It is given by
\begin{align}
    E_\text{ss} &=\sum_k\epsilon_k\frac{-|A_k x_k|^2+|B_k y_k|^2}{|A_k x_k|^2+|B_k y_k|^2},\label{eq:ss_energy_nn}
\end{align}
where $x_k,y_k$ are defined in \cref{eq:xT_def,eq:yT_def}, and
\begin{align}
    A_k &=\sum_{j=-nn}^{nn}\left(\cos(\varphi_k)\lambda_j+i\sin(\varphi_k)\mu_j\right)e^{-i\frac{2\pi jk}{N}},\\
    B_k &=\sum_{j=-nn}^{nn}\left(-\sin(\varphi_k)\lambda_j+i\cos(\varphi_k)\mu_j\right)e^{-i\frac{2\pi jk}{N}}.
\end{align}
Here, $\varphi_k$ is defined in \cref{eq:varphik}.
We resort to numerical methods to find the optimal parameters for which the total relative energy $e$ is minimized. That is, we optimize over the set of nearest-neighbor couplings $\{\lambda_j,\mu_j\}$, the bath frequency $\Delta$, and the cycle time $t$ to determine
\begin{equation}
    \min_{\{\lambda_j, \mu_j, \Delta, t\}} e(\theta).
    \label{eq:cooling_optimization_procedure}
\end{equation}

First, we will focus on noiseless cooling and show how increasing $nn$ is helpful. We will treat two different optimizations. The first one, that we will call $\theta$-specific, is done for particular choices of $\theta$, i.e., individual instances of $H_S$, which corresponds to the minimization in \cref{eq:cooling_optimization_procedure}. The second optimization, which we call phase-averaged, is done for entire phases in the system. That is, we find a single choice of parameters for $\theta \in [0, \pi/4]$ and another for $\theta \in [\pi/4, \pi/2]$ which can be used for cooling given any Hamiltonian in that phase. This requires minimizing the integral of all relative energies in the phase, i.e.,
\begin{equation}
    \min_{\{\lambda_j, \mu_j, \Delta, t\}} \int_{\theta_1}^{\theta_2} e(\theta)\dd\theta.
    \label{eq:average_cooling_optimization_procedure}
\end{equation}
This latter optimization is motivated by the fact that sometimes, the specific instance of our model (in our case, the value of $\theta$) could be unknown or have some uncertainty, and we may only have information about the phase, which is much more general. We will compare both optimizations to each other in the case of $nn \ge 0$, and to the results of the previous strategies in the case $nn=0$.

Next, we will compare DSP and cooling, both in the case of phase-averaged and $\theta$-specific optimizations. We will show that adding more neighbors greatly helps both DSP and cooling, but that the efficiency of DSP remains below cooling. We perform a comparison in relative energy for both methods. For example, we show (see~\cref{fig:cooling_specific_results}) that the systems reach relative energies of order $10^{-6}$ both with DSP and cooling just by increasing the nearest neighbor distance by one ($nn=1$), which is around three orders of magnitude better than the best results obtained in \cref{sec:multifreq_longtime}, even with multiple frequencies and randomized times. This comes at the cost of optimizing the cooling process for the specific instances of the Hamiltonian.

Finally, we will tackle noisy cooling and DSP. We will show how noise alters the efficiency of the algorithms, compare it to the results of the first subsection, and show that re-optimization of the parameters improves the cooling performance.

All optimizations are performed using the BFGS algorithm, a quasi-Newton method suitable for solving unconstrained nonlinear optimization problems, as implemented in the \texttt{Optim.jl} package for Julia.

\subsection{Noiseless Cooling}

We show here that our numerical optimization of the cooling parameters yields results that outperform the strategies from previous sections, both when optimizing for specific values of $\theta$ and for entire phases, even with purely local couplings ($nn=0$).

\begin{figure}[ht]
    \begin{overpic}[width=\linewidth]{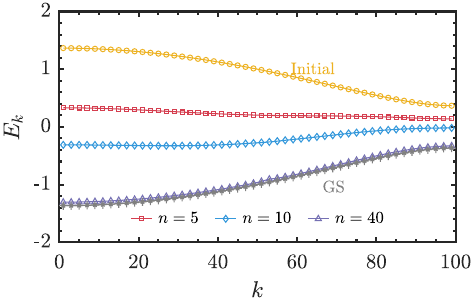}
        \put(1, 60){(a)}
    \end{overpic}
    \begin{overpic}[width=\linewidth]{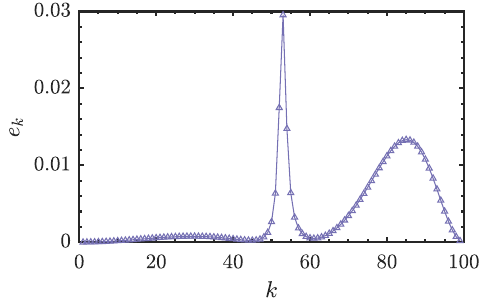}
        \put(1, 60){(b)}
    \end{overpic}
    \caption{(a) Snapshots of the mode energies $E_k$ at different stages of cooling for the $\theta$-specific optimal parameters at $\theta=\pi/3$ for $nn=0$. The initial state of the system, which we chose here to be the most excited state (as in \cref{fig:Fig2}), is shown with circles, as well as the state after $n=5$, $10$, and $40$ cycles. The ground-state (GS) energy $-\epsilon_k$ is plotted in gray. Other parameters are $N=200$ and $g=0.1$, similar to \cref{fig:Fig2} but with a larger coupling constant. The results for $n=40$ almost coincide with those of the steady state. (b) Relative energy $e_k$ of each mode in steady state.}
    \label{fig:Fig2_opt}
\end{figure}

A notable difference from the cooling strategies in \cref{sec:multifreq_longtime} is that here all modes are cooled equally and in less cycles. Furthermore, the energy peaks seen in \cref{fig:Fig2} do not appear (\cref{fig:Fig2_opt}). This can be explained as follows. Before, the value of $\Delta$ was chosen to cool one specific mode. Now, however, the notion of ``targeting'' specific parts of the spectrum with the bath frequency is lost. Additionally, the absence of accidental mode-specific reheating in \cref{fig:Fig2_opt} can be attributed to the fact that the optimal parameters avoid the resonance conditions $(\epsilon_k-\Delta)t=2\pi r$ that previously caused heating of specific modes.

To avoid such accidental resonances, choosing shorter times is ideal. For long cycle times, a small change in $\epsilon_k$ can imply a big change in $(\epsilon_k - \Delta)t$, which could bring it closer to some multiple of $2\pi$, as seen in \cref{fig:Fig2}. This happens, for example, when varying $\theta$ across a phase, so it is crucial to take it into account when considering phase-averaged optimizations. Small changes in $\epsilon_k$ also happen when increasing the system size $N$, since more energies appear between $\epsilon_{\rm m}$ and $\epsilon_{\rm M}$. A different way of avoiding the resonances could be to use time randomization, as in \cref{subsubsec:randomized_times}. However, this would have the effect of, yet again, increasing the total time, which in turn results in worse cooling in the presence of noise.

The optimal values of $\Delta$, therefore, do not correspond to any specific mode energy, nor do they maintain any simple relationship with the optimal cycle time $t$, since the effect of the couplings and the energy differences among the different modes is more relevant (as evidenced by~\cref{eq:ss_energy_nn}) than in the previous section, especially for $nn>0$. Another factor to take into account is the choice of the couplings $\lambda_j, \mu_j$, which also take on an important role in the steady state energy. The optimal couplings often show symmetry or antisymmetry, depending on the phase and number of neighbors allowed; this might be explained by the large amount of symmetries in our model. For example, symmetric couplings might be associated to a reflection symmetry in our model (changing site $1$ for site $N$, site $2$ for site $N-1$, etc.).

For the simplest case $nn=0$, the $\theta$-specific optimal couplings for $\theta\in[\pi/4,\pi/2]$ are always $\lambda_0=1,\ \mu_0=0$. For $\theta\in[0,\pi/4]$, however, the optimization for each $\theta$ yields a different result. When optimizing over entire phases (see \cref{tab:optimal_cool_avg}), the result for $\theta\in[\pi/4,\pi/2]$ is still $\lambda_0=1,\ \mu_0=0$, whereas for $\theta\in[0,\pi/4]$ it leans towards $\lambda_0=\mu_0=1$, the choice we made in the first sections.

\begin{figure}[ht]
    \includegraphics[width=\linewidth]{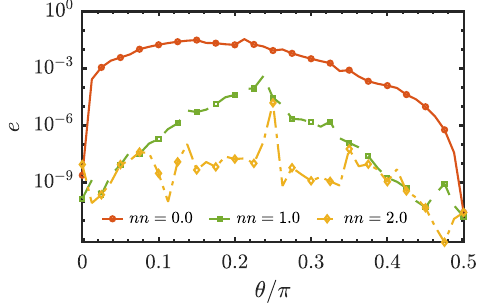}
    \caption{Optimization results for cooling with $\theta$-specific parameters with a system size $N=20$. The relative energy $e$ is plotted against $\theta/\pi$ for various nearest-neighbor coupling schemes ($nn$). For $nn \geq 1$, significant improvement in cooling efficiency is observed across $\theta$, with $nn=2$ achieving relative energies very close to the ground state (GS). Performance slightly decreases near the critical point ($\theta = \pi/4$) for all schemes, likely due to the closing energy gap. These results demonstrate the effectiveness of optimized $\theta$-specific cooling, particularly with $nn>0$.}
    \label{fig:cooling_specific_results}
\end{figure}

When allowing for $nn>0$, cooling achieves energies orders of magnitude better than with purely local couplings. The $\theta$-specific optimization (\cref{fig:cooling_specific_results}), generally works much better than phase-averaged optimization (\cref{fig:cooling_average_results}), which seems to find a limit on efficiency at $nn=1$.

\cref{tab:optimal_cool_avg} presents some more examples of optimal parameters for phase-averaged cooling with different nearest-neighbor coupling ranges, for a system size $N=20$. As shown in \cref{fig:cooling_average_results}, increasing the nearest neighbor distance by a half-integer ($nn=0.5$) provides considerable advantage in the $\theta \leq \pi/4$ phase. Additionally, the optimal parameters found for $N=20$ are shown to still perform well for increased system sizes in~\cref{app:scalability}.

\begin{table}[!ht]
    \centering
    \begin{tabular}{cccccccc}
        \toprule
        \multirow{2}{*}{$nn$} & \multirow{2}{*}{Phase} & \multirow{2}{*}{$\Delta$} & \multirow{2}{*}{$t$} & \multicolumn{4}{c}{Coupling parameters}                 \\
        \cmidrule(lr){5-8}
                              &                        &                           &                      & $\lambda_0$ & $\mu_0$ & $\lambda_{\pm1}$ & $\mu_{\pm1}$ \\
        \midrule
        \multirow{2}{*}{0}    & $\theta \leq \pi/4$    & 0.925                     & 3.05                 & 1.00        & 1.00    & ---              & ---          \\
                              & $\theta \geq \pi/4$    & 0.744                     & 3.33                 & 1.00        & 0.00    & ---              & ---          \\
        \midrule
        \multirow{2}{*}{0.5}  & $\theta \leq \pi/4$    & 0.688                     & 3.67                 & 1.00        & 0.53    & 1.00, ---        & -0.53, ---   \\
                              & $\theta \geq \pi/4$    & 0.793                     & 3.12                 & 1.00        & 0.03    & 0.34, ---        & 0.14, ---    \\
        \midrule
        \multirow{2}{*}{1}    & $\theta \leq \pi/4$    & 0.693                     & 3.70                 & 1.00        & 0.47    & 0.97, 0.01       & -0.51, 0.05  \\
                              & $\theta \geq \pi/4$    & 0.700                     & 3.71                 & 1.00        & 0.00    & 0.27, 0.27       & 0.15, -0.15  \\
        \bottomrule
    \end{tabular}
    \caption{Optimal parameters for phase-averaged cooling with different nearest-neighbor coupling ranges, for a system size $N=20$. The largest coupling is always normalized to 1. For $nn=0.5$ and $nn=1$, the values for $\lambda_{\pm1}$ and $\mu_{\pm1}$ are shown as pairs (positive index, negative index). Adding ``half a neighbor'' ($nn=0.5$) provides considerable advantage in the $\theta \leq \pi/4$ phase, while showing minimal improvement in the $\theta \geq \pi/4$ phase (see \cref{fig:cooling_average_results}).}
    \label{tab:optimal_cool_avg}
\end{table}

\begin{figure}[ht]
    \includegraphics[width=\linewidth]{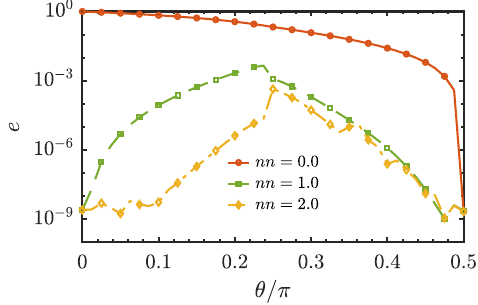}
    \caption{Same as~\cref{fig:cooling_specific_results} but for DSP. DSP performance is comparable to or better than cooling for $nn \geq 1$, but shows distinct behavior for $nn=0$, highlighting the importance of nonlocal couplings. These results illustrate the potential of optimized DSP for preparing low-energy states, especially with longer-range interactions.}
    \label{fig:stateprep_specific_results}
\end{figure}

\begin{figure}[ht]
    \includegraphics[width=\linewidth]{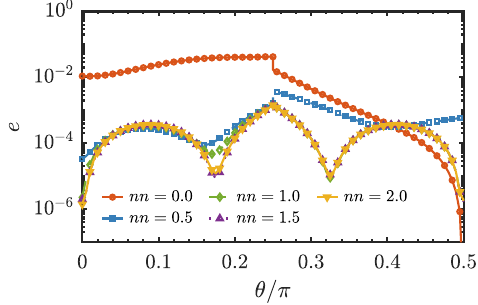}
    \caption{Optimization results for cooling with phase-averaged parameters for a system size of $N=20$. The relative energy $e$ is plotted against $\theta/\pi$. We showcase both integer and half-integer $nn$. The performance improves with increasing $nn$, and the energy density of the steady state depends less on $\theta$ than in the DSP case.}
    \label{fig:cooling_average_results}
\end{figure}

\begin{figure}
    \includegraphics[width=\linewidth]{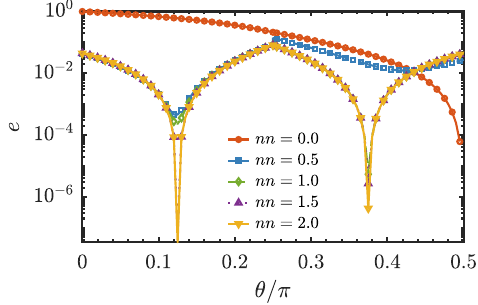}
    \caption{Same as~\cref{fig:cooling_average_results} but for DSP. In contrast to cooling, DSP exhibits distinct minima in the middle of each phase, with $nn=2$ achieving relative energies as low as $10^{-6}$, which seems to indicate that the state prepared has low energy all across the phase. This suggests that phase-averaged DSP may be particularly effective for certain system configurations, although the overall energy reduction is less pronounced than in cooling, especially near the critical point.}
    \label{fig:stateprep_average_results}
\end{figure}

\subsection{Comparison to DSP}

Comparing DSP with cooling, we find that both algorithms benefit substantially from longer-range couplings, but their relative performance depends on the optimization approach.

We first focus on phase-averaged optimization. The results of the optimization for nearest neighbor distances $nn=0,0.5,1,1.5 \text{ and }2$ are shown in \cref{fig:cooling_average_results,fig:stateprep_average_results} for cooling and DSP respectively. It can be seen that cooling outperforms DSP across the entire parameter range. With $nn=0.5$, cooling achieves a minimum relative energy of $10^{-4}$ throughout, while DSP reaches only $10^{-2}$ at best, with $e>10^{-1}$ near $\theta=0$ and $\theta=\pi/2$. It can also be seen that the effect of $nn=0.5$ is different for both phases. At $nn=1$, both DSP and cooling seem to reach a limit in efficiency in both phases for the range of $nn$ studied (see \cref{fig:cooling_average_results} and \cref{fig:stateprep_average_results}).

For $\theta$-specific optimization, shown for $nn=0,1\text{ and }2$ in \cref{fig:cooling_specific_results,fig:stateprep_specific_results} for cooling and DSP respectively, the difference between cooling and DSP gets smaller. For $nn=0$, DSP performs poorly except near $\theta=\pi/2$, since it is only able to produce product states (see \cref{subsubsec:large_N_limit}). However, for $nn=1$, DSP improves substantially to $e\approx10^{-6}$ at optimal $\theta$ values, and for $nn=2$, it reaches $e\approx10^{-9}$ at certain points, comparable to the best cooling results in \cref{fig:cooling_specific_results}.

\subsection{Noisy cooling and DSP}
\label{sec:noisy-regime}

This section presents numerical results that incorporate environmental noise with the depolarizing model explained in \cref{sec:adding_decoherence}. Here, we compare the results of cooling strategies for re-optimized parameters with the ones obtained in \cref{subsec:impact_noise}. For the case of a finite environment, we direct the reader to~\cref{subapp:finite_noise_numerics}, where we show that even though DSP and cooling may look comparable without noise (and non-local couplings), they behave very differently in case one adds noise.

Albeit we consider here a more general coupling between bath and system, the depolarizing channel still acts in the same way as described in the previous sections (see~\cref{subapp:noisy_cooling_ss} for the full derivation), adding a constant $2\kappa t$ term in the denominator for the relative energy $e_k$ as shown in \cref{eq:steady_state_energy_mode_noisy}. In contrast to \cref{sec:multifreq_longtime}, we consider here a smaller cycle time, while still using a single frequency; this allows us to increase $g$ in order to keep the effect of cooling larger than the effect of noise, while at the same time staying in the weak coupling regime. Additionally, we can re-optimize the parameters starting from the noiseless case, which results in a much lower energy than we had before (see~\cref{fig:Fig10}). This re-optimization is done by solving the modified $\theta$-specific optimization problem:
\begin{equation}
    \min_{\{\lambda_j, \mu_j, \Delta, t\}} e_{\text{noisy}}(\theta,\kappa),
    \label{eq:dep_noise_optimization}
\end{equation}
where $e_{\text{noisy}}(\theta,\kappa)$ is the relative energy in the presence of noise strength $\kappa$ (see~\cref{eq:steady_state_energy_mode_noisy}).

\begin{figure}
    \centering
    \includegraphics[width=\linewidth]{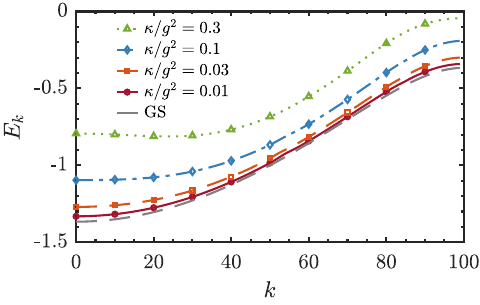}
    \caption{Energy $E_k$ for the same parameters as in \cref{fig:Fig3} ($\theta=\pi/3$, $N=200$), but $g=0.1$, with varying noise strength $\kappa$ chosen to match \cref{fig:Fig10}, and re-optimizing for $nn=0$ and each value of $\kappa$. The relative energies $e$ for the steady state of the full system are, from bottom to top: $0.025,0.066,0.187,0.413$.}
    \label{fig:spec}
\end{figure}

\Cref{fig:spec} shows the energy spectrum $E_k$ versus mode number $k$ for different noise strengths ranging from $\kappa/g^2=0.01$ to $\kappa/g^2=0.3$. We show for $nn=0$ how re-optimizing parameters can improve cooling efficiency in the presence of noise. For the lowest noise level ($\kappa/g^2=0.01$), the energy spectrum closely follows the ground state energy with only minor deviations.
As noise increases to $\kappa/g^2=0.03$, all modes remain well-cooled but uniformly shift upward from the ground state energies. At $\kappa/g^2=0.1$, the deviation becomes bigger, particularly for modes with $k<40$, where the energy is approximately 15-20\% higher than the ground state value. At $\kappa/g^2=0.3$, cooling is substantially compromised, with modes $k>80$ approaching zero energy and low-$k$ modes showing energy levels nearly 40\% above the ground state.

This uniform reheating across modes contrasts with the results in \cref{fig:Fig10}, where the non-optimized approach showed greater noise sensitivity for modes away from resonance. The effect of noise here is much more even across modes, reflecting the balanced cooling achieved through parameter optimization. The total relative energy in \cref{fig:spec} is also lower than in \cref{fig:Fig10}, showing the power of re-optimization. Additionally, we are able to push the value of $g$ up to $0.1$ thanks to the use of lower cycle times, which in turn allows for higher values of $\kappa$ while still retaining the weak coupling condition.

Regarding DSP, this model of noise is not sufficient. Since DSP can still work at times close to zero, taking the limit $t\rightarrow 0$ would allow us to return to the noiseless regime. In an experiment, however, this will not be feasible due to external constraints. Additionally, one would need to consider the increasing impact of noise as the cycle number $n$ grows.

\section{Conclusion and Outlook}
\label{sec:conclusion}

In this work, we studied how a many-body system can be cooled by repeatedly coupling it to a bath that is reset after some cycle time. We deliberately considered a simple Hamiltonian governing the evolution of the system, the bath, and their interaction to allow for an analytical derivation of the solutions. This approach enabled us to gain a deep understanding of the underlying processes that lead to cooling, as well as those that prevent it. Furthermore, it allowed us to determine how reheating processes can be circumvented by randomly choosing the cycle times and using multiple frequencies (see also Refs.~\cite{Marti2025Efficient, Mi2024Stable, Lloyd2025Quasiparticle}). Thus, the cooling protocol solely involves a simple coupling between the system and the bath, along with the choice of random cycle times and bath frequencies.

Additionally, we examined the impact of noise on the cooling process and showed that cooling with the simplest nearest-neighbor coupling between system and bath will be challenging. However, we also demonstrated how to combat the effect of noise with optimizations over various parameters (including the coupling strategy) of the cooling process. We explored several improvements to the original cooling algorithm. More precisely, we demonstrated that the cooling efficiency can be boosted by optimizing the cycle times and the coupling strategy. A local, but more sophisticated coupling strategy leads to lower energies than the simplest nearest-neighbor coupling. We also made comparisons with another state preparation scheme, namely dissipative state preparation (DSP)~\cite{Poyatos1996Quantum,Kraus2008Preparation,Diehl2008Quantum,Verstraete2009Quantuma}. As mentioned before, in contrast to DSP, the cooling algorithm cools the system to the {\it unknown} ground state of a continuously present Hamiltonian~\cite{Davies1974Markovian,Terhal2000Problem}. We demonstrated that both the optimized cooling and the DSP algorithms achieve low energies and that our cooling algorithm outperforms the DSP in various situations.

Overall, our findings provide valuable insights into the mechanisms of cooling many-body systems and offer practical strategies to enhance their efficiency through randomization and optimized coupling schemes.
For general interacting models, whose ground state can encode QMA-complete problems, polynomial time cooling to the ground state is clearly not guaranteed. However, the physical principles we identify---resonant energy transfer and its mitigation via randomization---are not limited to integrable systems. For systems with certain properties, such as being in a weakly interacting regime, our methods could potentially still provide an efficient pathway to low-energy states, a promising direction for future numerical and analytical studies
(see e.g.,~\cite{Zhan2025Rapid}).

Future research could explore combining our randomized, multi-frequency approach with time-dependent couplings, as presented in Lloyd et al.~\cite{Lloyd2025Quasiparticle} for the Google quantum processor experiment~\cite{Mi2024Stable} or by Matthies \textit{et al.}~\cite{Matthies2024Programmable}. Additionally, an intriguing avenue of study would be how error mitigation techniques, such as Zero Noise Extrapolation~\cite{Temme2017Error}, can enhance the robustness of these cooling algorithms on noisy quantum hardware. We are currently exploring the use of machine learning techniques to identify optimal coupling parameters for more complex Hamiltonians as well as couplings to the bath. While this work focuses on analog quantum computation, similar concepts can be extended to digital computations, where the robustness to algorithmic errors from digitization (e.g., Trotterization) in addition to gate-level hardware noise would need to be addressed.
Further investigating the cooling and DSP algorithms for Floquet computation~\cite{Lloyd2025Quasiparticle} and adapting these protocols to more complex Hamiltonians such as these from quantum chemistry~\cite{Li2024Dissipative} and classical optimization problems~\cite{Feng2024Escaping}--along with their implementation on experimental quantum platforms~\cite{Barreiro2011Opensystem,Schindler2013Quantum,Ma2019Dissipatively}--remain important directions for future research.

\begin{acknowledgments}
    We thank M. Hartmann for discussions.
    This research is part of the Munich Quantum Valley, which is supported by the Bavarian state government with funds from the Hightech Agenda Bayern Plus.
    The work is partially supported by the Deutsche Forschungsgemeinschaft (DFG, German Research Foundation) under Germany's Excellence Strategy - EXC-2111 - 390814868.
    D.M. and B.K. acknowledge funding from the BMW endowment fund and the Horizon Europe programmes HORIZON-CL4-2022-QUANTUM-02-SGA via the project 101113690 (PASQuanS2.1) and HORIZON-CL4-2021-DIGITAL-EMERGING-02-10 under grant agreement No.~101080085 (QCFD).
    S.L and J.I.C~acknowledge funding from the project FermiQP of the Bildungsministerium f\"ur Bildung und Forschung (BMBF).
\end{acknowledgments}

\textit{Note added.}---While finalizing this manuscript, we became aware of a related work by Zhan et al.~\cite{Zhan2025Rapid} on ground state preparation via dissipative dynamics.

\FloatBarrier

\clearpage
\appendix

\section{Derivation of the Mode-Dependent Hamiltonian}
\label{app:Hk_derivation}

In this appendix, we present the details about the block-diagonalization of the Hamiltonian considered in this work. We will first diagonalize the system Hamiltonian and then block-diagonalize the total Hamiltonian. Throughout this and other appendices, we will use dimensionless units for all energy scales and coupling constants.

\subsection{System Hamiltonian}

The system will be composed of an even number, $N$, of fermionic modes. The system Hamiltonian is given by
\begin{align}
    H_S &= \frac{1}{2} \sin\theta \sum_{n=1}^N (a_n^\dagger a_n - a_n a_n^\dagger)\nonumber\\
        &\quad + \frac{1}{2} \cos\theta \sum_{n=1}^N [a_n^\dagger (a_{n+1} + i a_{n+1}^\dagger) + \text{h.c.}],
    \label{eq:app:system-hamiltonian}
\end{align}
where $a_n$ ($a_n^\dagger$) denotes the fermionic annihilation (creation) operator at site $n$, with the periodic boundary condition $a_{N+1} = a_{1}$.

As mentioned in \cref{subsec:sys_model} of the main text, the system Hamiltonian is traceless and possesses various symmetries, which allow us to restrict, without loss of generality, $\theta$ to the interval $[0,\pi/2]$. This will be further proved in \cref{app:theta_mapping}.

We now derive the momentum-space formulation of \cref{eq:app:system-hamiltonian}. We begin by introducing the Fourier transform for an even number $N$ of fermionic modes:
\begin{align}
    \tilde{a}_k         &= \frac{1}{\sqrt{N}} \sum_{n=1}^N e^{i \frac{2\pi k n}{N}} a_n,\\
    \tilde{a}_k^\dagger &= \frac{1}{\sqrt{N}} \sum_{n=1}^N e^{-i \frac{2\pi k n}{N}} a_n^\dagger.
\end{align}
The operators $\tilde{a}_k$ and $\tilde{a}_k^\dagger$ are defined to be the annihilation and creation operators in momentum space. These definitions can be inverted using the orthogonality relation $\sum_{n=1}^N e^{i \frac{2\pi (k-q) n}{N}} = N \delta_{k,q}$ to obtain
\begin{equation}
    \label{eq:fourier_def_ann}
    a_n = \frac{1}{\sqrt{N}} \sum_{k=-N/2+1}^{N/2} e^{-i \frac{2\pi k n}{N}}  \tilde{a}_k,
\end{equation}
and similarly for $a_n^\dagger$. Here, $k$ runs over integer values from $-N/2+1$ to $N/2$.

Expressing $H_S$ in terms of the new operators leads to:
\begin{equation}
    H_S=\sum_{k=1}^{N/2-1} \tilde{\alpha}_k^\dagger \tilde{H}_k \tilde{\alpha}_k+ H_{k=0}+H_{k=N/2}.
\end{equation}
Here, we defined for any $k$ such that $1 \le k \le N/2-1$ the vector
\begin{align}
    \tilde{\alpha}_k =
    \begin{pmatrix}
        \tilde{a}_k\\
        \tilde{a}_{-k}^\dagger
    \end{pmatrix}
\end{align}
and the $2 \times 2$ matrix $\tilde{H}_k$ is given by
\begin{align}
    \tilde{H}_k
     &=
    \begin{pmatrix}
        w_k   &r_k\\
        r_k^* &-w_k
    \end{pmatrix}\nonumber\\
     &\equiv
    \begin{pmatrix}
        \sin\theta + \cos\theta \cos(\frac{2\pi k}{N})
         &
        \cos\theta \sin(\frac{2\pi k}{N})
        \\
        \cos\theta \sin(\frac{2\pi k}{N})
         &
        -\left[ \sin\theta + \cos\theta \cos(\frac{2\pi k}{N})\right]
    \end{pmatrix}.
    \label{eq:Hk_2x2_matrix}
\end{align}
Moreover, since terms such as $\tilde{a}_0^\dagger \tilde{a}_{-0}^\dagger$ vanish due to the fermionic anticommutation relation, we obtain only diagonal terms for both $H_{k=0,N/2}$. In particular, we have
\begin{align}
    H_{k=0}
     &=
    \frac{1}{2} \sin\theta
    \Bigl(\tilde{a}_0^\dagger \tilde{a}_0
    -
    \tilde{a}_0 \tilde{a}_0^\dagger
    \Bigr) +
    \frac{1}{2} \cos\theta
    \Bigl[
        \tilde{a}_0^\dagger \tilde{a}_0
        +
        \tilde{a}_0 \tilde{a}_0^\dagger
        \Bigr]
\end{align}
and
\begin{align}
    H_{k=N/2}
     &=
    \frac{1}{2} \sin\theta
    \Bigl(\tilde{a}_{N/2}^\dagger \tilde{a}_{N/2}
    -
    \tilde{a}_{N/2} \tilde{a}_{N/2}^\dagger
    \Bigr)\nonumber\\
     &\quad
    - \frac{1}{2} \cos\theta
    \Bigl(\tilde{a}_{N/2}^\dagger \tilde{a}_{N/2} +
    \tilde{a}_{N/2} \tilde{a}_{N/2}^\dagger
    \Bigr),
\end{align}
which is diagonal in the mode $\tilde{a}_{N/2}$. Using this way of grouping the terms in the Hamiltonian, we see that, apart from the cases $k=0,N/2$, we only need to sum over the positive values $k=1,2,\ldots,N/2-1$, as the negative $-k$ are automatically included.

\subsection{Bogoliubov transformation and diagonalization}

We now diagonalize each block Hamiltonian via a Bogoliubov transformation. We focus on the generic modes with $1 \leq k \leq N/2-1$, which appear in $2\times 2$ block form. The special cases at $k=0$ and $k=N/2$ are already diagonal as shown above.

\subsubsection{Diagonalizing the \texorpdfstring{$2\times 2$}{2 by 2} block for generic \texorpdfstring{$k$}{k}}

Because each $2\times 2$ matrix $\tilde{H}_k$ (\cref{eq:Hk_2x2_matrix}) is real and symmetric, it is diagonalizable through a real orthogonal transformation. The eigenvalues are $\pm \epsilon_k$ with

\begin{equation}
    \epsilon_k = \sqrt{w_k^2 + r_k^2}=\sqrt{1 + \sin2\theta \cos\frac{2\pi k}{N}},
    \label{eq:app:epsilonk}
\end{equation}

To determine the eigenvector corresponding to $+\epsilon_k$, we parameterize the real, normalized eigenvector as
\begin{align}
    \begin{pmatrix}
        x\\
        y
    \end{pmatrix} = \begin{pmatrix}
                        \cos(\varphi_k)\\
                        \sin(\varphi_k)
                    \end{pmatrix}.
    \label{eq:bog_positive_eigenvector_ansatz}
\end{align}
Substituting \cref{eq:bog_positive_eigenvector_ansatz} in the eigenvalue equation and solving for the ratio of components, we get
\begin{align}
    \frac{\sin(\varphi_k)}{\cos(\varphi_k)} = \tan(\varphi_k) = \frac{\epsilon_k - w_k}{r_k}.
    \label{eq:bog_ratio_sin_cos}
\end{align}
As the eigenvector to eigenvalue $-\epsilon_k$ is necessarily the real vector orthogonal to the one given in \cref{eq:bog_positive_eigenvector_ansatz}, we have that the real orthogonal matrix $U_k$ that diagonalizes $\tilde{H}_k$ is
\begin{align}
    U_k &=
    \begin{pmatrix}
        \cos(\varphi_k) &- \sin(\varphi_k)\\
        \sin(\varphi_k) &\quad\cos(\varphi_k)
    \end{pmatrix},
    \label{eq:bog_Uk_matrix}
\end{align}
Explicitly, we have $U_k^\mathsf{T} \tilde{H}_k U_k= \mbox{diag}(\epsilon_k, -\epsilon_k)$.
Moreover, we can obtain a more explicit final result for $\tan(2\varphi_k)$ by applying trigonometric identities and substituting the expressions for $\epsilon_k$, $w_k$, and $r_k$ into \cref{eq:bog_ratio_sin_cos}:
\begin{align}
    \tan2\varphi_k
     &=
    \frac{2 \tan(\varphi_k)}{1 - \tan^2(\varphi_k)}\nonumber\\
     &=
    \frac{\sin(\frac{2\pi k}{N})}{\tan\theta +\cos(\frac{2\pi k}{N})} = \frac{r_k}{w_k}.
    \label{eq:bog_tan_phi_k_final}
\end{align}

\subsubsection{Mapping to Bogoliubov quasiparticles}

This diagonalization at the matrix level defines a canonical transformation on the fermionic operators. Recall that
\begin{align}
    \tilde{\alpha}_k = \begin{pmatrix}
                           \tilde{a}_k\\
                           \tilde{a}_{-k}^\dagger
                       \end{pmatrix}.
\end{align}
The Bogoliubov transformation defined by
\begin{align}
    \hat{\alpha}_k = U_k^\mathsf{T} \tilde{\alpha}_k =
    \begin{pmatrix}
        \cos(\varphi_k) \tilde{a}_k
        +
        \sin(\varphi_k) \tilde{a}_{-k}^\dagger
        \\
        -\sin(\varphi_k) \tilde{a}_k
        +
        \cos(\varphi_k) \tilde{a}_{-k}^\dagger
    \end{pmatrix},
\end{align}
diagonalizes the Hamiltonian block $\tilde{H}_k$, i.e.,
\begin{align}
    \hat{\alpha}_k^\dagger
    \begin{pmatrix}
        + \epsilon_k &0\\
        0            &- \epsilon_k
    \end{pmatrix}
    \hat{\alpha}_k
    = + \epsilon_k \hat{a}_k^\dagger \hat{a}_k
    -
    \epsilon_k \hat{a}_{-k} \hat{a}_{-k}^\dagger,
\end{align}
for each $k\neq 0,N/2$.

\subsubsection{Diagonalized system Hamiltonian and ground-state energy}

Putting together all modes, including the special cases at $k=0$ and $k=N/2$, the Hamiltonian becomes
\begin{align}
    H_S
     &= \sum_{k=0}^{N/2} \epsilon_k h_k,
    \label{eq:bogoliubov_HS_diag}
\end{align}
where $\epsilon_k=\epsilon_{-k}$ and for the special cases $k=0$ or $k=N/2$ (those modes appear purely diagonally) we have $\hat{a}_k=\hat{a}_{-k}^\dagger$. The operators $h_k$ are given by  $h_k=\hat{a}^\dagger_k \hat{a}_k-\hat{a}_{-k}\hat{a}_{-k}^\dagger$ for $k=1,\ldots,N/2-1$ and $h_k=\hat{a}_k^\dagger \hat{a}_k-\hat{a}_k \hat{a}_k^\dagger$ for $k=0,N/2$.

Consequently, the ground state of $H_S$ is the vacuum state annihilated by all $\hat{a}_k$, i.e.,
\begin{align}
    \hat{a}_k\ket{\Psi_{GS}}=0
    \quad
    \text{for all }
    k,
\end{align}
and the ground-state energy is
\begin{align}
    E_{GS} = - \frac{1}{2}
    \sum_{k=-N/2+1}^{N/2}
    \epsilon_k,
\end{align}
with $\epsilon_k$ given in \cref{eq:app:epsilonk}.

\subsection{Bath Hamiltonian}

The system is coupled to a bath consisting of $N$ independent fermionic sites with Hamiltonian:
\begin{equation}
    H_B = \frac{\Delta}{2}\sum_{n=1}^N \left(b_n^\dagger b_n - b_n b_n^\dagger\right),
    \label{eq:bath-hamiltonian}
\end{equation}
where $b_n$ ($b_n^\dagger$) denotes the bath fermionic annihilation (creation) operator at site $n$, and $\Delta$ is the bath energy splitting. The bath is initialized in its ground state $\rho_B = \ket{\Omega}\bra{\Omega}$ at the beginning of each cooling cycle, and this ground state is annihilated by each $b_n$, i.e., $b_n\ket{\Omega}=0$.

We now transform \cref{eq:bath-hamiltonian} to momentum space, following the same steps as in the system Hamiltonian derivation. We use the same (inverse) Fourier transformation as before and obtain
\begin{align}
    H_B = \frac{\Delta}{2}
    \sum_{k=-N/2+1}^{N/2}
    (\tilde{b}_k^\dagger \tilde{b}_k -\tilde{b}_k \tilde{b}_k^\dagger).
    \label{eq:bath-hamiltonian-momentum-derived}
\end{align}
We now organize the sum in \cref{eq:bath-hamiltonian-momentum-derived} to pair positive and negative modes, yielding $2\times 2$ blocks for generic $k$ and diagonal terms for $k=0, N/2$:
\begin{align}
    H_B
     &=
    \sum_{k=1}^{N/2-1}
    \tilde{\beta}_k^\dagger \hat{H}_k \tilde{\beta}_k+
    \frac{\Delta}{2}
    (\tilde{b}_0^\dagger \tilde{b}_0 - \tilde{b}_0 \tilde{b}_0^\dagger)\nonumber\\
     &\quad  +
    \frac{\Delta}{2}
    (\tilde{b}_{N/2}^\dagger \tilde{b}_{N/2} - \tilde{b}_{N/2} \tilde{b}_{N/2}^\dagger),
\end{align}
where
\begin{align}
    \tilde{\beta}_k = \begin{pmatrix}
                          \tilde{b}_k\\
                          \tilde{b}_{-k}^\dagger
                      \end{pmatrix}, \quad
    \hat{H}_k = \begin{pmatrix}
                    \Delta &0\\
                    0      &-\Delta
                \end{pmatrix}.
\end{align}

\subsection{Coupling to the bath}

The system-bath coupling is assumed to be translation-invariant and extends up to a range of $nn$ nearest neighbors. It is given by
\begin{equation}
    {V}_{SB} = g \sum_{n=1}^N \sum_{j=-nn}^{nn} \left[\left(\lambda_j a_n^\dag {b}_{n+j} + i \mu_j a_n {b}_{n+j}\right) + \text{h.c.}\right],
    \label{eq:system-bath-coupling}
\end{equation}
where $g$ is the coupling strength, and $\lambda_j,\mu_j\le 1$ are dimensionless coupling constants to modes for different neighbor distances $j$.

To analyze this coupling in momentum space, we substitute the Fourier transforms of the system and bath operators (\cref{eq:fourier_def_ann}) into \cref{eq:system-bath-coupling}. After simplifying, this yields:
\begin{align}
    V_{SB} &= g \sum_{k=-N/2+1}^{N/2} \bigg[\left(\sum_{j=-nn}^{nn} \lambda_j e^{-i \frac{2\pi kj}{N}}\right) \tilde{a}_k^\dagger \tilde{b}_k\\
           &\quad + i\left(\sum_{j=-nn}^{nn} \mu_j e^{-i \frac{2\pi kj}{N}}\right) \tilde{a}_{-k} \tilde{b}_k + \text{h.c.}\bigg]
\end{align}
Organizing $V_{SB}$ into blocks pairing $k$ with $-k$ modes, we can write:
\begin{equation}
    V_{SB} = \sum_{k=0}^{N/2} \tilde{\alpha}_k^\dagger v_{SB,k} \tilde{\alpha}_k,
    \label{eq:app:HSBMomentum}
\end{equation}
where for each positive $k$ (except $k=0,N/2$), we define the four-component vector:
\begin{equation}
    \tilde{\alpha}_k = \begin{pmatrix}
        \tilde{a}_k\\
        \tilde{a}_{-k}^\dagger\\
        \tilde{b}_k\\
        \tilde{b}_{-k}^\dagger
    \end{pmatrix}.
\end{equation}

In \cref{sec:multifreq_longtime} of the main text, we consider only local couplings ($nn=0$) with $\lambda_0=\mu_0=1$, for which we have
\begin{equation}
    \label{eq:local_coupling_nn0}
    V_{SB} = g\sum_{k=-N/2+1}^{N/2} [\tilde{a}_k^\dagger \tilde{b}_k + i\tilde{a}_k \tilde{b}_{-k} + \tilde{b}_k^\dagger \tilde{a}_k - i\tilde{b}_{-k}^\dagger \tilde{a}_k^\dagger].
\end{equation}
That is, in this case we have for $1 \leq k \leq N/2-1$ the matrices $v_{SB,k}$ are given by
\begin{equation}
    v_{SB,k} = g\begin{pmatrix}
        0  &0  &1 &i\\
        0  &0  &i &-1\\
        1  &-i &0 &0\\
        -i &-1 &0 &0
    \end{pmatrix}.\label{eq:vSB_k}
\end{equation}
and for $k=0$ and $k=N/2$ we have
\begin{equation}
    v_{SB,0} = v_{SB,N/2} = v_{SB,k}/2
    \label{eq:vSB_0}
\end{equation}
with the corresponding vectors
\begin{equation}
    \tilde{\alpha}_0 = \begin{pmatrix}
        \tilde{a}_0\\
        \tilde{a}_0^\dagger\\
        \tilde{b}_0\\
        \tilde{b}_0^\dagger
    \end{pmatrix},\quad \tilde{\alpha}_{N/2} = \begin{pmatrix}
        \tilde{a}_{N/2}\\
        \tilde{a}_{N/2}^\dagger\\
        \tilde{b}_{N/2}\\
        \tilde{b}_{N/2}^\dagger
    \end{pmatrix}.
\end{equation}

\subsection{The total Hamiltonian}

The total Hamiltonian can now be written as the following sum over Hamiltonians of pairs of modes, $(k,-k)$:
\begin{equation}
    H_{SB} = H_S + H_B + V_{SB} = \sum_{k=0}^{N/2} \hat{\alpha}_k^\dagger h_{SB,k} \hat{\alpha}_k.\\
    \label{eq:app:HSB}
\end{equation}
for both local and non-local coupling. Let us first consider the case of local coupling ($nn=0$), as discussed in \cref{sec:multifreq_longtime}, where for generic $k$ ($1 \leq k \leq N/2-1$) we have
\begin{align}
    h_{SB,k} &=
    \begin{pmatrix}
        w_k &r_k  &g      &gi\\
        r_k &-w_k &gi     &-g\\
        g   &-gi  &\Delta &0\\
        -gi &-g   &0      &-\Delta
    \end{pmatrix},
\end{align}
with
\begin{align}
    w_k &= \sin\theta + \cos\theta\cos(2\pi k/N),\\
    r_k &= \cos\theta\sin(2\pi k/N).
\end{align}
For the special cases $k=0$ and $k=N/2$, we multiply the matrices $h_{S,k}$, $h_{B,k}$, and $v_{SB,k}$ by $1/2$ as shown in \cref{eq:vSB_0}, and the entries $w_k$ and $r_k$ are evaluated at the corresponding values of $k=0$ and $k=N/2$.
For the case where coupling is not local (i.e., $nn>0$), the derivation is similar. However, the entries of the matrices depend on the values of various coupling constants $\lambda_j$ and $\mu_j$. The final result for this case is presented below.

\subsection{Block diagonalization of the total Hamiltonian}

We apply the Bogoliubov transformation derived earlier for the system modes to the full Hamiltonian. The transformation matrix $U_k$ for the system part is extended to act on the full four-component space:
\begin{equation}
    \mathcal{U}_k =
    \begin{pmatrix}
        U_k &0\\
        0   &I_2
    \end{pmatrix}
\end{equation}
where $I_2$ is the $2\times 2$ identity matrix and $U_k$ is given in \cref{eq:bog_Uk_matrix}.
The transformed matrices $\hat{h}_{SB,k} = \mathcal{U}_k^\mathsf{T} h_{SB,k} \mathcal{U}_k$ for generic $k$ are given by
\begin{align}
    \hat{h}_{SB,k} &=
    \begin{pmatrix}
        \epsilon_k          &0                   &g e^{i\varphi_k}   &ig e^{i\varphi_k}\\
        0                   &-\epsilon_k         &ig e^{-i\varphi_k} &-g e^{i\varphi_k}\\
        g e^{-i\varphi_k}   &-ig e^{-i\varphi_k} &\Delta             &0\\
        -ig e^{-i\varphi_k} &-g e^{-i\varphi_k}  &0                  &-\Delta
    \end{pmatrix}.
\end{align}
For the special cases $k=0$ and $k=N/2$, we follow the same procedure but note the extra factor of 1/2. The final transformed Hamiltonian takes the block-diagonal form:
\begin{equation}
    H_{SB} = \sum_{k=0}^{N/2} \hat{\alpha}_k^\dagger \hat{h}_{SB,k} \hat{\alpha}_k,
\end{equation}
where $\hat{\alpha}_k = \mathcal{U}_k^\mathsf{T} \tilde{\alpha}_k$ are the transformed operators, and $\hat{h}_{SB,k}$ is the block Hamiltonian derived in \cref{subsec:total_hamiltonian} of the main text.

If the nearest neighbor distance extends beyond local couplings ($nn>0$), the procedure is the same and the Hamiltonian per each block $(k,-k)$ is given by
\begin{align}
    \hat{h}_{SB,k} &=
    \begin{pmatrix}
        \epsilon_k &0           &gA_k   &gB_k\\
        0          &-\epsilon_k &gB_k   &-g A_k\\
        g A_k^*    &g B_k^*     &\Delta &0\\
        gB_k^*     &-g A_k^*    &0      &-\Delta
    \end{pmatrix},\label{eq:hsb_momentum_space_nn>0}
\end{align}
where the coupling coefficients are:
\begin{align}
    A_k &=\sum_{j=-nn}^{nn}\left(\cos(\varphi_k)\lambda_j+i\sin(\varphi_k)\mu_j\right)e^{-i\frac{2\pi jk}{N}}
    \label{eq:Ak in hsb_k}\\
    B_k &=\sum_{j=-nn}^{nn}\left(-\sin(\varphi_k)\lambda_j+i\cos(\varphi_k)\mu_j\right)e^{-i\frac{2\pi jk}{N}}.
    \label{eq:Bk in hsb_k}
\end{align}

For the case $nn=0$, $\lambda_0=\mu_0=1$, we have $A_k = e^{i\varphi_k}$ and $B_k = ie^{-i\varphi_k}$.

\subsection{Restricting to $\theta\in[0,\pi/2]$}
\label{app:theta_mapping}

As explained in \cref{subsec:sys_model} of the main text, in case of local coupling ($nn=0$) and $\lambda_0=\mu_0=1$, we can restrict $\theta$ to the region $[0,\pi/2]$. Here, we show that this also holds for the general case of non-local couplings. To this end, we show that systems with $\theta$ outside this region can be mapped to equivalent systems within the region by appropriate transformations of the coupling parameters.

Let us first analyze the mapping $\theta \rightarrow -\theta$. Under this transformation, the system Hamiltonian $H_S$ [\cref{eq:system_hamiltonian_realspace}] undergoes sign changes in its terms. This transformation can be implemented through a mapping of fermionic operators where mode $k$ is mapped to $k\pm N/2 \mod N$. Specifically, $k=0$ is mapped to $k'=N/2$, mode $k=1$ to $k'=-N/2+1$, and vice versa, mode $k=-N/2+1$ to $k'=1$, etc.
For the system-bath coupling [\cref{eq:system-bath-coupling_realspace}], this transformation requires the coupling constants to transform as:
\begin{align}
    \lambda_j' &= \lambda_j(-1)^j,\\
    \mu_j'     &= \mu_j(-1)^j.
\end{align}

In the special case where $nn=0$, this mapping reduces to the symmetry explained in \cref{subsec:sys_model} of the main text since the only relevant coupling constants are $\lambda_0$ and $\mu_0$, which remain unchanged.

The other symmetry appears when considering the transformation $\theta'=\pi+\theta$. In this case, the two modes being mapped to one another are the same, $k'=k$, since this change in $\theta$ keeps the energy of each mode invariant. Specifically, we have:
\begin{align}
    \epsilon_k(\theta+\pi) &= \epsilon_k(\theta),\\
    \varphi_k(\theta+\pi)  &= \varphi_k(\theta),
\end{align}
where the second equation follows from $\tan(\theta+\pi)=\tan(\theta)$.
Since these quantities remain invariant, there is no need to modify the coupling constants $\lambda_j$ and $\mu_j$ under this transformation.

By combining these two symmetries, we can map any value of $\theta$ from the region $[-\pi/2,3\pi/2]$ to a value within the range $[0,\pi/2]$, with appropriate adjustments to the coupling constants. This extends the result mentioned in \cref{subsec:sys_model} of the main text to the general case of non-local coupling ($nn > 0$).

\section{Noise}
\label{app:secIV_noise}

In this appendix, we present the details about the cooling map in the presence of depolarizing noise as described in \cref{sec:adding_decoherence}. We start by showing that the cooling and noise Lindbladians commute and then derive the cooling map and the modified states on which the map acts.

\subsection{Proof that \texorpdfstring{$\mathcal{L}_C$}{LC} commutes with \texorpdfstring{$\mathcal{L}_E$}{LE}}
\label{app:LcCommuteLe}

The total evolution during a cooling cycle is governed by the master equation:
\begin{equation}
    \dot \rho_{SB} = \mathcal{L}_C(\rho_{SB}) + \mathcal{L}_E(\rho_{SB}),
\end{equation}
where $\mathcal{L}_C(\rho) = -i [H_{SB},\rho]$ generates the unitary evolution under the system-bath Hamiltonian $H_{SB}$, and $\mathcal{L}_E$ represents the environmental noise:
\begin{equation}
    \mathcal{L}_E = \kappa\sum_{n=1}^N \left[\mathcal{L}_{a_n} + \mathcal{L}_{a_n^\dagger} + \mathcal{L}_{b_n} + \mathcal{L}_{b_n^\dagger}\right].
    \label{eq:app:LE_def}
\end{equation}
Here, $\mathcal{L}_O(\rho) = O\rho O^\dagger - \frac{1}{2}\{O^\dagger O,\rho\}$.
Using the fermionic anticommutation relations, we can simplify the noise Lindbladian to:
\begin{equation}
    \mathcal{L}_E(\rho) = \kappa\sum_{n=1}^N \left[ (a_n\rho a_n^\dagger + a_n^\dagger\rho a_n - \rho) + (b_n\rho b_n^\dagger + b_n^\dagger\rho b_n - \rho) \right].
    \label{eq:LE_explicit_app}
\end{equation}

We will prove that $[\mathcal{L}_C, \mathcal{L}_E] = 0$.
The commutation relation is equivalent to showing that $\mathcal{L}_E$ is invariant under the unitary evolution generated by $H_{SB}$, i.e., that the following equality holds for $U(t) = e^{-i H_{SB} t}$:
\begin{equation}
    \mathcal{L}_E^I(\rho) \equiv U(t)^\dagger \mathcal{L}_E(U(t) \rho U(t)^\dagger) U(t) = \mathcal{L}_E(\rho),\label{eq:LE_invariance_condition}
\end{equation}
where $\mathcal{L}_E^I$ denotes the noise Lindbladian in the interaction picture with respect to $H_{SB}$. This can be easily seen as follows. Let us denote by $R=(a_1,\ldots, a_n, b_1,\ldots, b_n, a_1^\dagger,\ldots, a_n^\dagger, b_1^\dagger,\ldots, b_n^\dagger)$. Then, the fermionic anticommutation relations read $\{R_i,R_j^\dagger\}=\delta_{i,j}$. As the Hamiltonian $H_{SB}$ is quadratic in the fermionic operators, its action transforms $R$ into a vector $S=A R$, for some $4n\times 4n$ dimensional matrix $A$. Furthermore, $S$ has to obey the fermionic anticommutation relations, i.e., $\{S_i,S_j^\dagger\}=\delta_{i,j}$, which is fulfilled iff $A$ is unitary. Using this, it is straightforward to see that
\begin{equation}
    \mathcal{L}_E(\rho) =\sum_{i=1}^{4n}\left[ (R_i\rho R_i^\dagger  - 2 \rho) \right]=\sum_{i=1}^{4n}\left[ (S_i\rho S_i^\dagger  - 2 \rho) \right]= \mathcal{L}_E^I(\rho), \label{eq:LE_invariance_proof}
\end{equation}
which proves the statement. This commutation relation allows us to factorize the time evolution as follows
\begin{align}
    e^{(\mathcal{L}_C + \mathcal{L}_E)t} &= e^{\mathcal{L}_C t} e^{\mathcal{L}_E t} = e^{\mathcal{L}_E t} e^{\mathcal{L}_C t}.
\end{align}
Therefore, the noisy cooling map $\mathcal{N}$ for one cycle can be expressed as applying the noise evolution $e^{\mathcal{L}_E t}$ first, followed by the noiseless cooling map $\mathcal{E}$ (which corresponds to $e^{\mathcal{L}_C t}$ and tracing over the bath):
\begin{align}
    \mathcal{N}(\rho_S)
     &= \tr_B\left[e^{(\mathcal{L}_C + \mathcal{L}_E)t}(\rho_S \otimes \rho_B)\right] \nonumber\\
     &= \tr_B\left[e^{-iH_{SB}t} (\tilde{\rho}_S \otimes \tilde{\rho}_B) e^{i H_{SB}t}\right],
\end{align}
where $\tilde{\rho}_S = e^{\mathcal{L}_E t}(\rho_S)$ is the system density matrix evolved under noise, and $\tilde{\rho}_B = e^{\mathcal{L}_E t}(\rho_B)$ is the evolved bath state.

In order to determine $\tilde{\rho}_B$ we first note that the initial state of the bath is its ground state, i.e., $\rho_B = \ket{\Omega}\bra{\Omega} = \bigotimes_{n=1}^N \ket{0_n}\bra{0_n}$. Note further that $\mathcal{L}_E$ acts independently on each site $n$. Hence, we can determine $\tilde{\rho}_B = e^{\mathcal{L}_E t}(\rho_B)$ by focusing on bath modes individually.
We consider a single bath mode $b_n$ and solve the master equation $\dot{\rho}_{B,n} = \kappa (\mathcal{L}_{b_n} + \mathcal{L}_{b_n^\dagger}) \rho_{B,n}$ for the density matrix $\rho_{B,n}$ of site $n$, with initial condition $\rho_{B,n}(0) = \ket{0_n}\bra{0_n}$. Let $\rho_{B,n} = \begin{pmatrix} \rho_{00} & \rho_{01} \\ \rho_{10} & \rho_{11} \end{pmatrix}$. The equations for the matrix elements are:
\begin{equation}
    \dot{\rho}_{00} = \kappa (\rho_{11} - \rho_{00})
    = \kappa (1 - 2\rho_{00})
\end{equation}
which has the solution (recall that $\rho_{00}(0) = 1$),
\begin{align}
    \rho_{00}(t) = 1-\rho_{11}(t) = \frac{1+e^{-2\kappa t}}{2}.
\end{align}

Similarly, it is straightforward to see that the coherence terms $\rho_{01}$ and $\rho_{10}$ decay exponentially under $\mathcal{L}_E$. This can be easily seen by solving their respective ODEs, leading to
\begin{equation}
    \rho_{01}(t)=\rho_{01}(0)e^{-\kappa t}.
\end{equation}
with $\rho_{01}(0)=0$.
Thus, the bath state after the noise evolution for time $t$ is
\begin{align}
    \tilde{\rho}_B &= e^{\mathcal{L}_E t}(\ket{\Omega}\bra{\Omega})\nonumber\\
                   &= \bigotimes_{n=1}^N \left( \frac{1+e^{-2\kappa t}}{2}\ket{\Omega}\bra{\Omega}+\frac{1-e^{-2\kappa t}}{2}\ket{1}\bra{1} \right).
\end{align}
This mixed state is then used as the initial state of the bath for the cooling cycle.

Since the form of the noise is the same and independent for both the system and the bath, the time evolution equation for $\rho_S$ under noise alone is the same as that of the bath:
\begin{align}
    \frac{d\tilde{\rho}_S}{dt} &= \mathcal{L}_E(\tilde{\rho}_S) = \kappa\sum_{n=1}^N \left[\mathcal{L}_{a_n}(\tilde{\rho}_S) + \mathcal{L}_{a_n^\dagger}(\tilde{\rho}_S)\right].
\end{align}
The solution to this equation gives us $\tilde{\rho}_S = e^{\mathcal{L}_E t}(\rho_S(0))$ at time $t$, which is then used as the initial state for the cooling process in the numerical simulations reported in \cref{sec:multifreq_longtime,sec:nn_cooling} of the main text.

Contrary to the bath, we cannot separate the different $n$ modes and calculate the effect of the noise on each of them independently, so one would need to switch to momentum space and act on the 4 by 4 matrices for each $k,-k$ block of $\rho_S$. Additionally, we should take into account that the initial state of the system is different for every cycle. However, as will be shown in \cref{app:Single_freq steady state}, approximations can be made to simplify the noisy map $\mathcal{N}(\rho)$ so this calculation is not needed.

\subsection{Noisy map in momentum space}

For numerical simulations, it is convenient to express the master equation derived above in momentum space. The noise Lindbladian in momentum space is obtained by transforming the real-space Lindbladian $\mathcal{L}_E = \kappa\sum_{n=1}^N \left[\mathcal{L}_{a_n} + \mathcal{L}_{a_n^\dagger} + \mathcal{L}_{b_n} + \mathcal{L}_{b_n^\dagger}\right]$ using the Fourier transform. Since the Fourier transform is a unitary transformation (a specific canonical transformation), and $\mathcal{L}_E$ is invariant under canonical transformations as shown in \cref{app:LcCommuteLe}, the form remains the same but with momentum operators:
\begin{align}
    \mathcal{L}_E = \kappa\sum_{k=-N/2+1}^{N/2} \left[\mathcal{L}_{\tilde{a}_k} + \mathcal{L}_{\tilde{a}_k^\dagger} + \mathcal{L}_{\tilde{b}_k} + \mathcal{L}_{\tilde{b}_k^\dagger}\right].
\end{align}
Due to the translational invariance of both the system Hamiltonian and the noise, the density matrix in momentum space decomposes into a tensor product over momentum modes $(k,-k)$: $\tilde{\rho}_S = \bigotimes_{k=0}^{N/2} \tilde{\rho}_S^{(k)}$ and $\tilde{\rho}_B = \bigotimes_{k=0}^{N/2} \tilde{\rho}_B^{(k)}$.

Due to the block-diagonal structure of both $H_{SB}$ and $\mathcal{L}_E$ in momentum space, the noisy cooling map in momentum space can therefore be written as:
\begin{align}
    \mathcal{N}(\rho_S)         &= \bigotimes_{k=0}^{N/2} \mathcal{N}_k(\rho_S^{(k)}) \mbox{ with }\\
    \mathcal{N}_k(\rho_S^{(k)}) &= \tr_{B_k}\left[e^{-ih_{SB,k}t}(\tilde{\rho}_S^{(k)} \otimes \tilde{\rho}_B^{(k)})e^{i h_{SB,k}t}\right],
\end{align}
where $h_{SB,k}$ is the block Hamiltonian for momentum mode $k$ derived in \cref{subsec:total_hamiltonian} of the main text (and \cref{app:Hk_derivation}), and for each momentum mode pair $k$, the bath state evolves to:
\begin{align}
    \tilde{\rho}_B^{(k)} = &\left(\frac{1+e^{-2\kappa t}}{2}\ket{\Omega}\bra{\Omega}+\frac{1-e^{-2\kappa t}}{2}\ket{1}\bra{1} \right)^{\otimes 2}.
\end{align}
This formulation allows for efficient numerical implementation of the noisy cooling process, where each momentum block can be treated independently, as used in the simulations presented in \cref{sec:multifreq_longtime,sec:nn_cooling}.

\section{Derivation of convergence times}
\label{app:secIV_calculations}

In this appendix, we provide a detailed derivation of the scaling of the cooling time presented in \cref{sec:cooling_times_and_ss}.
We start out in \cref{subsec:zeroth_order_mixing_time} by showing how the distance between the state obtained after applying $n$ times a general completely positive map~(CPM), ${\cal E}$, to some input state and the unique stationary state of the CPM converges with $n$. In \cref{subsec:global_convergence} we then use the fact that the stationary state of our cooling map factorizes in momentum space to derive a bound on the cooling rate. In \cref{subsec:energy_convergence}, we determine the number of cycles required to achieve an energy density which is close to that of the stationary state.

\subsection{Mixing time and convergence to stationary state for generic CPM}
\label{subsec:zeroth_order_mixing_time}

In order to determine how fast the system converges to the stationary state, we use the relation between the transfer matrix and the CPM. Given a linear map, $T$, with $T(\cdot)=\sum_j L_j \cdot R_j$, the transfer matrix is defined as $\hat{T}=\sum_j L_j \otimes R_j^T$. The relation between the transfer matrix, $\hat{T}$, and the Choi-state, $E$ is $\hat{T}=d E^T$, where $E^T$ is defined via $\langle m,n|E^T | k,l\rangle=\langle m,k|E|n,l\rangle$, i.e., $\hat{T}=d E^{T_B} \text{SWAP}$, where $T_B$ denotes the partial transpose with respect to system $B$ and SWAP is defined via $\text{SWAP} |k,l\rangle=|l,k\rangle$ and $d$ denotes the dimension. It is straightforward to see that the concatenation of maps, $T_1 \circ T_2$ corresponds to the transfer matrix $\hat{T}_1 \cdot \hat{T}_2$. Note that in general $\hat{T}$ will not be Hermitian. Using the Jordan decomposition of $\hat{T}$, one can determine the action of $T^n$. Let us consider the dense set of non-defective matrices, which will be the generic case~\footnote{For non-diagonalizable (defective) matrices, the Jordan decomposition still allows us to analyze the convergence.}. For those we have that the algebraic multiplicity of each eigenvalue coincides with its geometric multiplicity.
They can be written as $\hat{T}=\sum_k \lambda_k |R_k\rangle \langle L_k|$, where $\langle L_k|R_l\rangle =\delta_{k,l}$, i.e., with biorthogonal right and left eigenvectors of $\hat{T}$. From this decomposition we have $\hat{T}^n = \sum_k \lambda_k^n |R_k\rangle \langle L_k|$. The corresponding map can be written as: $\mathcal{E}(\rho) = \sum_{m} \lambda_m \tilde{\mathcal{E}}_m(\rho)$, where $\tilde{\mathcal{E}}_m$ corresponds to the transfer matrix $ |R_m\rangle \langle L_m|$.

Therefore, we have
\begin{align}
    \mathcal{E}^n(\rho) = \sum_{m} \lambda_m^n \tilde{\mathcal{E}}_m(\rho).
    \label{eq:spectral_decomposition_map_n}
\end{align}
For a cooling process to be effective and for it to converge to a unique steady state, all eigenvalues other than $\lambda_1 = 1$ must satisfy $|\lambda_m| < 1$ for $m > 1$~\footnote{Recall that we consider here the case where we have a unique stationary state.} As the number of iterations $n$ increases, the terms $\lambda_m^n$ with $|\lambda_m| < 1$ will approach zero.

We can separate the term corresponding to $\lambda_1 = 1$ from the sum in \cref{eq:spectral_decomposition_map_n} to obtain
\begin{align}
    \mathcal{E}^n(\rho) &= \tilde{\mathcal{E}}_1(\rho) + \sum_{m=2} \lambda_m^n \tilde{\mathcal{E}}_m(\rho)\nonumber\\
                        &= \rho_\text{ss} + \sum_{m=2} \lambda_m^n \tilde{\mathcal{E}}_m(\rho).
\end{align}

To analyze the convergence to the steady state, we consider the difference between the evolved state $\mathcal{E}^n(\rho)$ and the steady state $\rho_\text{ss}$:
\begin{align}
    \mathcal{E}^n(\rho) - \rho_\text{ss} = \sum_{m=2} \lambda_m^n \tilde{\mathcal{E}}_m(\rho).
    \label{eq:difference_evolved_steadystate}
\end{align}
The rate of convergence is determined by how quickly this difference goes to zero as $n \to \infty$, which is governed by the eigenvalue $\lambda_2$, with second largest eigenvalue absolute value. Let us assume that there are more eigenvalues, $\lambda_2, \lambda_2', \lambda_2'', \ldots$ with the same magnitude $|\lambda_2|$. We denote the corresponding maps by $\tilde{\mathcal{E}}_2'$, etc. Then, for large $n$, we can approximate the difference as:
\begin{align}
    \mathcal{E}^n(\rho) - \rho_\text{ss} &\approx \lambda_2^n \tilde{\mathcal{E}}_2(\rho) + (\lambda_2')^n \tilde{\mathcal{E}}_2'(\rho) + (\lambda_2'')^n \tilde{\mathcal{E}}_2''(\rho) + \cdots \nonumber\\
                                         &= \lambda_2^n \left[ \tilde{\mathcal{E}}_2(\rho) + \tilde{\mathcal{E}}_2'(\rho) + \tilde{\mathcal{E}}_2''(\rho) + \cdots \right],
    \label{eq:difference_evolved_steadystate_approx}
\end{align}
where we have factored out $\lambda_2^n$ and redefined $\tilde{\mathcal{E}}_2'$, etc.~with a possible phase.
To quantify the convergence, we take the one-norm of the difference:
\begin{align}
    \|\mathcal{E}^n(\rho) - \rho_\text{ss}\|_1 &\approx \left\| \lambda_2^n \left[ \tilde{\mathcal{E}}_2(\rho) + \tilde{\mathcal{E}}_2'(\rho) + \tilde{\mathcal{E}}_2''(\rho) + \mathcal{O}\left(\frac{\lambda_3^n}{\lambda_2^n}\right)\right] \right\|_1 \nonumber\\
                                               &= |\lambda_2|^n \left\| \tilde{\mathcal{E}}_2(\rho) + \tilde{\mathcal{E}}_2'(\rho) + \tilde{\mathcal{E}}_2''(\rho) + \cdots \right\|_1\nonumber\\
                                               &= C |\lambda_2|^n = C e^{-n\alpha},
    \label{eq:norm_difference_evolved_steadystate}
\end{align}
with a constant $C$ and the cooling rate $\alpha$ as:
\begin{align}
    C      &= \left\| \tilde{\mathcal{E}}_2(\rho) + \tilde{\mathcal{E}}_2'(\rho) + \tilde{\mathcal{E}}_2''(\rho) + \cdots \right\|_1,\\
    \alpha &= -\log |\lambda_2| > 0.
    \label{eq:constant_C_definition}
\end{align}

Let us apply these bounds now to the map $\mathcal{E}_k$. More specifically, we consider a mode pair $(k,-k)$ and define a mode-specific cooling rate $\alpha_k = -\log |\lambda_k|$, where $\lambda_k$ is the second-largest eigenvalue of the map $\mathcal{E}_k$ in absolute value. This gives us the convergence:
\begin{equation}
    \|\mathcal{E}_k^n(\rho_k) - \sigma_k\|_1 \leq e^{-\alpha_k n},
    \label{eq:mode_convergence_rate}
\end{equation}
where $\sigma_k$ is the (unique) steady state of the map $\mathcal{E}_k$.
The number of cycles required to reach the steady state for mode $k$ is then:
\begin{equation}
    n^c_k = \alpha_k^{-1}.
    \label{eq:n0k_app}
\end{equation}

\subsection{Convergence of cooling map}
\label{subsec:global_convergence}

We now analyze the convergence of the global state $\mathcal{E}^n(\rho)$ to the global steady state $\rho_{S}^\text{ss} = \bigotimes_{k=0}^{N/2} \sigma_k$. Due to translational invariance, the cooling map $\mathcal{E}$ and its steady state can be decomposed into a tensor product of maps $\mathcal{E}_k$ and their corresponding steady states $\sigma_k$ for each momentum mode pair $k$. That is,
\begin{align}
    \mathcal{E}=\bigotimes_{k=0}^{N/2} \mathcal{E}_k, \quad \rho_{S}^\text{ss} = \bigotimes_{k=0}^{N/2} \sigma_k.
\end{align}
We assume the initial state is also a product state over the modes, $\rho_S = \bigotimes_{k=0}^{N/2} \rho_k$.

Our goal is to determine the global cooling time, i.e., how many cycles are needed for the global state $\mathcal{E}^n(\rho)$ to approach the global steady state $\rho_{S}^\text{ss} = \bigotimes_k \sigma_k$. We quantify this convergence using the one-norm distance
\begin{align}
    \|\mathcal{E}^n(\rho)-\bigotimes_k \sigma_k\|_1 = \|\bigotimes_{k=0}^{N/2}\mathcal{E}_k^n(\rho_k)-\bigotimes_{k=0}^{N/2}\sigma_k\|_1.
\end{align}
To bound this global distance, we will use the Kaleidoscope inequality. Let us denote by $\chi_i$ the tensor product state where the first $i$ modes are evolved and the remaining modes are in their steady state. That is,
\begin{equation}
    \chi_i = \left( \bigotimes_{j=0}^{i-1}\mathcal{E}_j^n(\rho_j) \right) \otimes \left( \bigotimes_{l=i}^{N/2} \sigma_l \right).
\end{equation}
We define $\chi_0 = \bigotimes_{k=0}^{N/2}\sigma_k$ and $\chi_{N/2+1} = \bigotimes_{k=0}^{N/2}\mathcal{E}_k^n(\rho_k)$. Then we can write the global distance as a telescoping sum:
\begin{align}
    \|\bigotimes_{k=0}^{N/2}\mathcal{E}_k^n(\rho_k)-\bigotimes_{k=0}^{N/2}\sigma_k\|_1
     &= \|\chi_{N/2+1} - \chi_0\|_1 \nonumber\\
     &= \left\| \sum_{i=1}^{N/2+1} (\chi_i - \chi_{i-1}) \right\|_1 \nonumber\\
     &\le \sum_{i=1}^{N/2+1} \|\chi_i - \chi_{i-1}\|_1,
\end{align}
where the inequality follows from the triangle inequality for the one-norm. Using that $\|\tau \otimes \rho - \tau \otimes \sigma\|_1 = \|\rho - \sigma\|_1$, we obtain
\begin{equation}
    \|\chi_i - \chi_{i-1}\|_1 = \|\mathcal{E}_{i-1}^n(\rho_{i-1}) - \sigma_{i-1}\|_1.
\end{equation}

Therefore, the global distance is bounded by the sum of the distances for each mode (also called the Kaleidoscope inequality):
\begin{align}
    \|\mathcal{E}^n(\rho)-\bigotimes_{k=0}^{N/2}\sigma_k\|_1 &\leq \sum_{k=0}^{N/2} \|\mathcal{E}_k^n(\rho_k)-\sigma_k\|_1.
\end{align}
Using \cref{eq:mode_convergence_rate}, we have:
\begin{align}
    \|\mathcal{E}^n(\rho) - \bigotimes_{k=0}^{N/2}\sigma_k\|_1 &\leq \sum_{k=0}^{N/2} e^{-\alpha_k n}.
\end{align}
To find a system-size independent cooling rate, we define the global cooling rate $ \alpha = \inf_N \min_k \alpha_k$. Assuming $\alpha > 0$, we can bound the sum by taking the minimum rate $\alpha$:
\begin{align}
    \sum_{k=0}^{N/2} e^{-\alpha_k n} \le \sum_{k=0}^{N/2} e^{-\alpha n} = (N/2 + 1) e^{-\alpha n} < N e^{-\alpha n},
\end{align}
where we have neglected $\mathcal{O}(1)$ factors for simplicity, focusing on the scaling with $N$. Thus, we obtain the global convergence rate:
\begin{equation}
    \|\mathcal{E}^n(\rho)-\bigotimes_{k=0}^{N/2}\sigma_k\|_1 \leq N e^{-\alpha n}.
\end{equation}
This shows that to reach a state that is $\epsilon$-close (in one-norm distance) to the steady state, i.e., $\|\mathcal{E}^n(\rho)-\rho_{S}^\text{ss}\|_1 \le \epsilon$, we need a number of cycles $n^c$ such that $N e^{-\alpha n^c} \approx \epsilon$, or equivalently,
\begin{equation}
    n^c \approx \frac{1}{\alpha} \log\left(\frac{N}{\epsilon}\right).
\end{equation}
Therefore, if $\alpha > 0$ (which is generally the case away from the phase transition at $\theta = \pi/4$), the number of cycles required to reach the steady state grows logarithmically with the system size $N$.

\subsection{Convergence of energy density}
\label{subsec:energy_convergence}

Next, we analyze the convergence of the relative energy by examining the convergence of the energy density. The error in the energy density after $n$ cycles can be bounded by
\begin{align}
    \left|\frac{1}{N}\sum_{k=0}^{N/2} \tr\left(\epsilon_k h_k \left[\mathcal{E}_k^n(\rho_k)- \sigma_k\right]\right)\right|\nonumber\\
    \le \frac{1}{N}\sum_{k=0}^{N/2} \left|\tr\left(\epsilon_k h_k \left[\mathcal{E}_k^n(\rho_k)- \sigma_k\right]\right)\right|,
\end{align}
by using the triangle inequality $|\tr(A+B)| \le |\tr(A)| + |\tr(B)|$.
Then, we use the inequality $|\tr(XY)| \le \|X\| \|Y\|_1$, where $\|X\|$ is the operator norm of $X$. We choose $X = \epsilon_k h_k$ and $Y = \mathcal{E}_k^n(\rho_k)- \sigma_k$ and will use the convergence of the state differences in the 1-norm as established in \cref{eq:mode_convergence_rate}.

The matrix $h_k = \hat{a}^\dagger_k \hat{a}_k-\hat{a}_{-k}\hat{a}_{-k}^\dagger$ has eigenvalues $\pm 1$ and thus operator norm $\|h_k\| = 1$.
Then, we have $\|\epsilon_k h_k\| = \epsilon_k$. Therefore, we bound:
\begin{align}
    \left|\tr\left(\epsilon_k h_k \left[\mathcal{E}_k^n(\rho_k)- \sigma_k\right]\right)\right|
     &\le \|\epsilon_k h_k\| \|\mathcal{E}_k^n(\rho_k)- \sigma_k\|_1 \nonumber\\
     &= \epsilon_k \|\mathcal{E}_k^n(\rho_k)- \sigma_k\|_1 \nonumber\\
     &\le \epsilon_k e^{-\alpha_k n},
\end{align}
by using the bound on mode convergence $\|\mathcal{E}_k^n(\rho_k)-\sigma_k\|_1 \le e^{-\alpha_k n}$ from \cref{eq:mode_convergence_rate}.

Substituting this back into the sum for the energy density error, we get the following bound
\begin{align}
    \frac{1}{N}\sum_{k=0}^{N/2} \left|\tr\left(\epsilon_k h_k \left[\mathcal{E}_k^n(\rho_k)- \sigma_k\right]\right)\right|
     &\le \frac{1}{N}\sum_{k=0}^{N/2} \epsilon_k e^{-\alpha_k n} \nonumber\\
     &\le \frac{1}{N}(\max_k \epsilon_k) \sum_{k=0}^{N/2} e^{-\alpha_k n} \nonumber\\
     &\le (\max_k \epsilon_k) e^{-\alpha n} \frac{N/2 + 1}{N} \nonumber\\
     &\le C e^{-\alpha n},
\end{align}
where in the last three steps we have taken the maximum value of $\epsilon_k$, used the minimum cooling rate $\alpha$, and $C = \max_k \epsilon_k$ is a constant of order $\mathcal{O}(1)$ since $\epsilon_k$ are mode energies and are bounded.
To achieve an energy density error of $\epsilon$, we need $C e^{-\alpha n^c} \approx \epsilon$, which gives
\begin{align}
    n^c &\approx \frac{1}{\alpha} \log\left(\frac{C}{\epsilon}\right)\nonumber\\
        &\approx \frac{1}{\alpha} \log\left(\frac{1}{\epsilon}\right),
\end{align}
neglecting the constant term $\frac{\log C}{\alpha}$. This shows that if $\alpha > 0$, the number of cycles required to reach a certain energy density error $\epsilon$ is independent of the system size $N$, scaling only logarithmically with the desired precision $\epsilon$.

If $\alpha = 0$, which can happen at a phase transition (e.g., at $\theta = \pi/4$ in our model), the cooling time scaling will depend on how the minimum cooling rate $\alpha_k$ approaches zero as the system size $N$ increases. If $\alpha$ decreases polynomially with $N$, the cooling time will scale polynomially with $N$.

\section{The cooling map and its steady state in the weak coupling regime}
\label{app:Single_freq steady state}

The aim of this appendix is to first derive via perturbation theory the cooling map in the weak coupling regime, where we consider a single frequency of the bath and a single cycle time. Then, we will determine its steady state and the corresponding energy. Finally, we will determine the energy of the steady state in the noisy case. We start out by describing the general perturbation theory, which we will then apply to the cooling map.

\subsection{The cooling map in the weak coupling regime}
\label{subapp:dyson_expansion_noiseless}

We begin by considering the general problem of quantum evolution under a Hamiltonian that can be split into an unperturbed part $H_0$ and a small perturbation $gV$:
\begin{equation}
    \dot{\rho}=-i[H,\rho],\quad H=H_0+gV,
\end{equation}
where the norm of $H_0$ and $V$ is assumed to be $\mathcal{O}(1)$, and the weak coupling condition $(gt)^2 \ll 1$ is satisfied. To perform the approximation, we first transform into the interaction picture with $U_I=e^{i H_0t}$:
\begin{equation}
    \dot{\rho_I}=-ig[V_I,\rho_I],\quad V_I=e^{i H_0t}Ve^{-iH_0t},\quad \rho_I=e^{i H_0t}\rho e^{-iH_0t}.
\end{equation}
Solving the differential equation for $\rho_I$ recursively and developing up to second order in $gt$ yields a Dyson expansion:
\begin{align}
    \rho_I(t) &=\rho_I(0)-ig\int_0^t[V_I(t'),\rho_I(t')]\dd t'\nonumber\\
              &=\rho_I(0)-ig\int_0^t[V_I(t'),\rho_I(0)]\dd t'+\mathcal{O}(g^2t^2).
\end{align}
By adding appropriate second-order terms, we can re-write the last expression as
\begin{equation}
    \rho_I(t)\simeq\left(\Id-ig\int_0^t V_I(t')\dd t'\right)\rho_I(0)\left(\Id - ig\int_0^t V_I(t')\dd t'\right)^\dag.
\end{equation}
Using that $\rho_I(0)=\rho(0)$ and returning back from the interaction picture we find
\begin{align}
    \rho(t) &= U(t)\rho(0)U(t)^\dag+\mathcal{O}(g^2t^2),\\
    U(t)    &= e^{-iH_0t}\left(\Id-ig\int_0^t V_I(t')\dd t'\right).
    \label{eq:Dyson_expansion}
\end{align}
Let us now apply this approximation to our map of interest, namely $\mathcal{E}_k$, i.e., the completely positive trace-preserving (CPTP) map for a single cooling cycle and a single pair of momenta, $(k,-k)$. This map is given by (see \cref{eq:hsb_momentum_space_nn>0}):
\begin{equation}
    \mathcal{E}_k(\rho_k) = \tr_B \left[ e^{-i h_{SB,k} t} \left( \rho_k \otimes \ket{\Omega}\bra{\Omega}_B \right) e^{i h_{SB,k} t} \right],
    \label{eq:CPTP_map}
\end{equation}
where $\rho_k$ is the system's density matrix, and $\ket{\Omega}_B=\ket{00}_B$ is the bath's ground state, formed by the vacuum of the two modes $\hat{b}_{\pm k}$ that appear in \cref{eq:hsb_momentum_space_nn>0}. The total Hamiltonian is $h_{SB,k} = h_{S,k} + h_{B,k} + gv_{SB,k}$, where we treat $gv_{SB,k}$ as a perturbation and $h_0 = h_{S,k} + h_{B,k}$ is the unperturbed Hamiltonian.

As we will see, using the approximation up to first order in \cref{eq:Dyson_expansion} will not be sufficient to derive a CPTP map. To this end, second-order terms are required. However, we will avoid computing them directly, by using a powerful result from Ref.~\cite{Wolf2008Dividing}. There, it has been shown that any divisible CPTP map can be written as an exponential of a Liouvillian. Since any map which can be infinitesimally generated is divisible, we will derive the cooling map in the weak coupling regime by first determining the first-order approximation of that map and then complete it by writing it in exponential form to ensure it is CPTP.

Expanding the evolution operator as shown in \cref{eq:Dyson_expansion}, we obtain:
\begin{equation}
    \mathcal{E}_k(\rho_k) \approx \tr_B \left[ e^{-i h_0 t} M_{k,t} \left( \rho_k \otimes \ket{\Omega}\bra{\Omega}_B \right) M_{k,t}^\dag e^{i h_0 t} \right],
    \label{eq:expanded_map}
\end{equation}
where the operator $M_{k,t}$ is defined as:
\begin{equation}
    M_{k,t} = \Id - i g\int_0^t e^{i h_0 \tau} v_{SB,k} e^{-i h_0 \tau}~ \dd\tau.
    \label{eq:MT_definition}
\end{equation}
Let us consider first the general case of $nn\geq 0$ and later simplify to the local coupling scenario. To calculate $M_{k,t}$ using the Hamiltonian given by \cref{eq:hsb_momentum_space_nn>0}, we use the following identity:
\begin{align}
    e^{i\epsilon_k \tau\hat{a}_k^\dag\hat{a}_k}\hat{a}_k^\dag e^{-i\epsilon_k \tau\hat{a}_k^\dag\hat{a}_k}=e^{i\epsilon_k \tau}\hat{a}_k^\dag,
\end{align}
which leads to:
\begin{align}
    \label{eq:Vk_interaction_picture}
     &e^{i h_0 \tau} v_{SB,k} e^{-i h_0 \tau}=\left(-A_k^*e^{i(\epsilon_k-\Delta)\tau}\hat{a}_k+B_k^*e^{i(\epsilon_k+\Delta)\tau}\hat{a}_{-k}^\dag\right)\hat{b}_{k}^\dag\nonumber\\
     &+\left(-A_k e^{i(\epsilon_k-\Delta)\tau}\hat{a}_{-k}+B_k e^{i(\epsilon_k+\Delta)\tau}\hat{a}_{k}^\dag\right)\hat{b}_{-k}^\dag+ \text{h.c.}.
\end{align}
Evaluating the integral, we find:
\begin{align}
    M_{k,t} &= \Id -i \left( l_1 \hat{b}_k^\dag + l_2 \hat{b}_{-k}^\dag + \text{h.c.} \right),
    \label{eq:MT_with_jump_operators}
\end{align}
where we have introduced the time-dependent operators $l_1$ and $l_2$, defined as:
\begin{align}
    l_1 &= A^*_k\, x_k\, \hat{a}_k + B^*_k\, y_k\, \hat{a}_{-k}^\dag, \label{eq:L1_def_app}\\
    l_2 &= A_k x_k\, \hat{a}_{-k} - B_k\, y_k \hat{a}_k^\dag, \label{eq:L2_def_app}
\end{align}
and the coefficients $x_k$ and $y_k$ are:
\begin{align}
    x_k &= g \int_0^t e^{i(\epsilon_k-\Delta)\tau} \dd\tau = g\frac{1 - e^{i (\Delta-\epsilon_k) t}}{i (\Delta-\epsilon_k)}, \label{eq:xT_def:app}\\
    y_k &= -g\int_0^t e^{i(\epsilon_k+\Delta)\tau} \dd\tau = g\frac{1 - e^{i (\epsilon_k + \Delta) t}}{i (\epsilon_k + \Delta)}. \label{eq:yT_def:app}
\end{align}
These functions arise from the time integration of oscillating exponentials in \cref{eq:Vk_interaction_picture} due to the energy differences between the system and bath.

Since the bath starts in the ground state $\ket{\Omega}_B$, any terms in $M_{k,t}$ containing $\hat{b}_k$ or $\hat{b}_{-k}$ (i.e., the Hermitian conjugate terms in \cref{eq:MT_with_jump_operators}) will annihilate the bath state and thus vanish upon acting on $\ket{\Omega}_B$. Moreover, after tracing out the bath, non-diagonal terms such as $\ket{01}\bra{10}$ or $\ket{10}\bra{01}$ do not contribute due to the orthogonality of the bath states. Therefore, the approximation of $\mathcal{E}_k(\rho_k)$ as given in \cref{eq:CPTP_map} simplifies to:
\begin{align}
    e^{-i h_k t} \left( \rho_k + l_1 \rho_k l_1^\dag + l_2 \rho_k l_2^\dag \right) e^{i h_k t}.
    \label{eq:nonCPTP_map}
\end{align}
Note that we omitted here the time dependence of the operators $l_1,l_2$ for brevity.

As mentioned before, the map in \cref{eq:nonCPTP_map} is not CPTP, because we have only kept terms up to first order in the Dyson expansion.
To ensure the map is CPTP, we need to include additional terms, which naturally arise from the second-order expansion of $M_{k,t}$ (as done in \cref{app:CM_derivation}). However, we can also correct it without computing the second order in the expansion by recognizing that the correct evolution (up to the unitary evolution with $h_k$) is governed by a (sum of) Lindbladians, as explained above (see Ref.~\cite{Wolf2008Dividing}).
\begin{align}
    \mathcal{E}_k(\rho_k) &= e^{-i h_k t} e^{\mathcal{L}_k}(\rho_k) e^{i h_k t},
\end{align}
where $\mathcal{L}_k = \mathcal{L}_1 + \mathcal{L}_2$, and the Lindblad superoperators $\mathcal{L}_i(\rho_k)$ are defined as:
\begin{align}
    \mathcal{L}_1(\rho_k) &= l_1 \rho_k l_1^\dag - \frac{1}{2} \{ l_1^\dag l_1, \rho_k \},\\
    \mathcal{L}_2(\rho_k) &= l_2 \rho_k l_2^\dag - \frac{1}{2} \{ l_2^\dag l_2, \rho_k \}.
\end{align}
This ensures that the map is CPTP while matching the first-order expansion in \cref{eq:nonCPTP_map}. The complete cooling map for a single cycle and a pair $\pm k$ is therefore given by:
\begin{align}
    \label{eq:approxiMap}
    \mathcal{E}_k(\rho_k) &= e^{-i h_k t} \left( \rho_k + \mathcal{L}_1(\rho_k) + \mathcal{L}_2(\rho_k) \right) e^{i h_k t} \nonumber\\
                          &\approx e^{-i h_k t} e^{\mathcal{L}_k}(\rho_k) e^{i h_k t}.
\end{align}
This map has a particularly simple interpretation: apart from the system evolution, given by $e^{-i h_k t}$, we have dissipative terms resulting from the coupling to the bath, which lead to cooling as explained in the main text.

Note that a phase can be added to these jump operators, since every term in the Lindbladian has both $l_i$ and $l_i^\dag$. For example, the results of \cref{sec:single_frequency} make use of an extra factor $e^{-i\varphi_k}$ to simplify the expressions of $l_1$ and $l_2$.

When looking at the full system, we can create the map as a tensor product of all of the maps for each $k$-mode, yielding the following approximation:
\begin{equation}
    \mathcal{E}(\rho)\approx e^{-i H_S t} e^{\mathcal{L}_C}(\rho) e^{i H_S t}.
    \label{eq:full_cooling_map_approx}
\end{equation}

In the special case of $nn=0$ with $\lambda_0=\mu_0=1$, the jump operators simplify to:
\begin{align}
    l_1 &= e^{-i\varphi_k}(x_k\hat{a}_k-iy_k\hat{a}_{-k}^\dag),\\
    l_2 &= e^{i\varphi_k}(x_k\hat{a}_{-k}-iy_k\hat{a}_{k}^\dag).
\end{align}
Clearly, these phases can be ignored here, as they cancel out in the Lindbladian.
\subsection{The steady state in the noiseless weak coupling regime}
\label{subapp:rho_ss_noiseless}
Next, we will determine the steady state of the map $\mathcal{E}_k(\rho_k)$, which satisfies the fixed point equation $\mathcal{E}_k(\rho_\text{ss}^{(k)})=\rho_\text{ss}^{(k)}$.
Clearly, this can be done in a brute force way, as it is given by a $4\times 4$ matrix. Using the invariant subspaces under the action of $\mathcal{E}_k$ leads to the following parametrization of the steady state density matrix
\begin{equation}
    \rho_\text{ss}^{(k)}=\begin{pmatrix}
        m       &0 &0 &\beta\\
        0       &n &0 &0\\
        0       &0 &p &0\\
        \beta^* &0 &0 &q
    \end{pmatrix}.
\end{equation}
The basis states for this matrix representation are $\ket{00}$,\ldots, $\ket{11}$, where the first and second digits represent the occupation of mode $k$ and $-k$, respectively.

The entries will be lengthy expressions in terms of the parameters of the map, but as long as we only want to determine the energy, there is no need to determine them explicitly. With the conditions obtained from the trace condition and from the fixed-point equation, we obtain the following result for energy (per two modes):
\begin{align}
    E_{ss}^k &=\tr[h_{k}\rho_\text{ss}^{(k)}]=\frac{\epsilon_k}{2}(-2\rho_\text{ss}^{0,0}+2\rho_\text{ss}^{1,1})\nonumber\\
             &= \epsilon_k(q-m)=\epsilon_k\frac{-|A_k x_k|^2+|B_k y_k|^2}{|A_k x_k|^2+|B_k y_k|^2},
    \label{eq:steady_state_energy_mode}
\end{align}
where $A_k, B_k$ are given in \cref{eq:Ak in hsb_k,eq:Bk in hsb_k} and $x_k, y_k$ are given in \cref{eq:xT_def:app,eq:yT_def:app}.

Note that in the case $nn=0,\ \lambda_0=\mu_0=1$, studied in \cref{sec:multifreq_longtime}, this equation reduces to
\begin{align}
    E_{ss}^k &=\epsilon_k\frac{-|x_k|^2+|y_k|^2}{|x_k|^2+| y_k|^2},
    \label{eq:steady_state_energy_mode_nn0}
\end{align}
and perfect cooling for a mode can be achieved (the energy becomes $ \approx -\epsilon_k$) if $|x_k|^2 \gg |y_k|^2$, which is the case if the energy of the mode and the bath frequency are on resonance, i.e., $(\epsilon_k-\Delta)t\ll 1$. In this case, $|y_k|\leq g/\Delta$ but $|x_k|\approx gt$, which leads to the cooling of that mode for long times, i.e., the \textit{cooling limit} $(\Delta t)^2\gg 1$.

There also exist accidental resonances $y_k=0$ that cool down specific modes when $(\Delta+\epsilon_k)t=2\pi r$. More importantly, there is accidental reheating if $x_k\approx 0$, which is the case for $(\Delta-\epsilon_k)t \approx 2\pi r$. As explained in the main text, the fact that these resonances depend on the cycle time allows us to suppress them by considering time randomization.

Another result worth mentioning here is that, when reducing \cref{eq:steady_state_energy_mode} to DSP, i.e., taking $h_{S,k}=0$ (equivalently, $\epsilon_k=0$), the dependence on $t$ and $\Delta$ also disappears, yielding the simpler result:
\begin{equation}
    E_\text{ss}^k(\text{DSP})=\epsilon_k\frac{-|A_k|^2+|B_k|^2}{|A_k|^2+|B_k|^2}.
    \label{eq:DSP_steadystate_energy}
\end{equation}

As can be seen, the only way that DSP can work is by tuning the couplings, and the result will greatly depend on their choice.

\subsection{The steady state energy in the noisy weak coupling regime}
\label{subapp:noisy_cooling_ss}

We now study how environmental noise affects the cooling process and modifies the steady state energy. We focus on the depolarizing noise model described in~\cref{sec:adding_decoherence} of the main text, and leave the case of finite noise for \cref{app:finite_noise}. The effect of an infinite bath is modeled by the following Lindbladian:
\begin{equation}
    \mathcal{L}_E=\kappa\sum_{n=1}^N \mathcal{L}_{a_n}+\mathcal{L}_{a_n^\dag}+\mathcal{L}_{b_n}+\mathcal{L}_{b_n^\dag},
\end{equation}
where the superoperator $\mathcal{L}_{O}$ is defined as
\begin{equation}
    \mathcal{L}_O(\rho) = O\rho O^\dagger - \frac{1}{2}\{O^\dagger O,\rho\},
\end{equation}
and the constant $\kappa$ will model the noise strength. We now use the fact that the cooling and noise maps commute (as proven in~\cref{app:secIV_noise}). This allows us to write the cooling map taking noise into account as
\begin{align}
    \mathcal{N}(\rho)=\tr_B\left[e^{-iH_{SB}t}\left(e^{\mathcal{L}_E^S}(\rho)\otimes e^{\mathcal{L}_E^B}(\rho_B)\right)e^{i H_{SB}t}\right].
\end{align}
Now we expand this expression keeping only the terms up to order $\mathcal{O}(\kappa t,g^2t^2)$ and find
\begin{align}
    \mathcal{N}(\rho) &\approx \tr_B[e^{-iH_{0}t}M_{k,t}(\rho+\mathcal{L}_E^S(\rho))\nonumber\\
                      &\quad\otimes (\rho_B+\mathcal{O}(\kappa t))M_{k,t}^\dag e^{i H_{0}t}],
\end{align}
where $M_{k,t}$ is the perturbative operator defined in~\cref{eq:MT_with_jump_operators}. Expanding the terms in this map, we find
\begin{align}
    \mathcal{N}(\rho)
     &\approx \tr_B\left[e^{-iH_{0}t}M_{k,t}\rho\otimes\rho_B M_{k,t}^\dag e^{i H_{0}t}\right]\nonumber\\
     &\quad + \tr_B\left[e^{-iH_{0}t}M_{k,t}\mathcal{L}_E^S(\rho)\otimes\rho_B M_{k,t}^\dag e^{i H_{0}t}\right]\nonumber\\
     &\quad + \tr_B\left[e^{-iH_{0}t}M_{k,t}\rho\otimes\mathcal{O}(\kappa t) M_{k,t}^\dag e^{i H_{0}t}\right].
\end{align}
The first term corresponds to the noiseless cooling map $\mathcal{E}(\rho)$. For the second term, to order $\mathcal{O}(g^0)$, $M_{k,t} \approx \Id$, so this term contributes $e^{-iH_{S}t}\mathcal{L}_E^S(\rho)e^{i H_{S}t}$. The third term contributes terms of order $\mathcal{O}(g^2\kappa t^3)$ which we can neglect. Therefore:
\begin{align}
    \mathcal{N}(\rho) &\approx \mathcal{E}(\rho) + e^{-iH_{S}t}\mathcal{L}_E^S(\rho)e^{i H_{S}t}\nonumber\\
                      &\approx e^{-iH_{S}t}e^{\mathcal{L}_C}(\rho)e^{i H_{S}t} + e^{-iH_{S}t}\mathcal{L}_E^S(\rho)e^{i H_{S}t}\nonumber\\
                      &\approx e^{-iH_{S}t}(e^{\mathcal{L}_C}(\rho) + \mathcal{L}_E^S(\rho))e^{i H_{S}t}\nonumber\\
                      &\approx e^{-iH_{S}t}e^{\mathcal{L}_C+\mathcal{L}_E^S}(\rho)e^{i H_{S}t},
\end{align}
where we have used that $\mathcal{E}(\rho) \approx e^{-iH_{S}t}e^{\mathcal{L}_C}(\rho)e^{i H_{S}t}$ from~\cref{eq:full_cooling_map_approx}, and in the last step we have used that $\mathcal{L}_C$ and $\mathcal{L}_E^S$ commute to combine them into a single exponential.

This shows that the cooling map including noise can be approximated as a Lindbladian with the original cooling jump operators plus additional jump operators from the noise channel. Solving for the steady state energy with this combined Lindbladian, which is done in a similar fashion to the noiseless case of \cref{subapp:rho_ss_noiseless}, modifies \cref{eq:steady_state_energy_mode} to yield:
\begin{equation}
    E_{ss}^k= \epsilon_k\frac{-|A_k x_k|^2+|B_k y_k|^2}{|A_k x_k|^2+|B_k y_k|^2+2\kappa t}.
    \label{eq:steady_state_energy_mode_noisy}
\end{equation}
which now shows an extra term in the denominator due to the action of the infinite bath, which modifies the cooling process and tries to drive the state to the maximally mixed state.

This result also extends to multiple frequencies and averaged times, and causes a modification in the values of the cooling and heating rates. In this case, the cooling and noise Lindbladians have the same terms but with different strengths (as seen in~\cref{eq:averaged_lindbladian}), so we can sum them to
\begin{align}
    (\mathcal{L}_C+\mathcal{L}_E)(\rho_k) &= \gamma_{k,\text{noisy}}^c \sum_{i=\pm k} \left( \hat{a}_i \rho_k \hat{a}_i^\dag - \frac{1}{2} \left\{ \hat{a}_i^\dag \hat{a}_i, \rho_k \right\} \right)        \nonumber\\
                                          &\quad + \gamma_{k,\text{noisy}}^h \sum_{i=\pm k} \left( \hat{a}_i^\dag \rho_k \hat{a}_i - \frac{1}{2} \left\{ \hat{a}_i \hat{a}_i^\dag, \rho_k \right\} \right),
\end{align}
where the new cooling and heating rates are given by
\begin{equation}
    \gamma_{k, \text{noisy}}^{\rm c,h}=\gamma_{k}^{\rm c,h}+\kappa t.
\end{equation}

\section{Working with the correlation matrix formalism}
\label{app:CM_derivation}

As we are throughout considering Gaussian states, all the previous derivations could have been performed using the correlation matrix (CM) formalism.
This is a well-suited method for systems described by quadratic Hamiltonians evolving initial Gaussian states, as such states remain Gaussian throughout the evolution~\cite{Weedbrook2012Gaussian}.
Gaussian states are fully characterized by their first and second moments, specifically the correlation matrix $\gamma$ or covariance matrix $\Gamma$ which we will both use for convenience in different contexts. In this appendix, we provide an alternative analysis of the cooling dynamics using this method.

The correlation matrix $\gamma$ for a set of $M$ fermionic operators $\hat{\alpha}_i$ ($i=1, \dots, M$) associated with a state $\rho$ is defined entry-wise as:
\begin{equation}
    \gamma_{ij} = \frac{1}{2}\tr(\rho\,[\hat{\alpha}_i,\hat{\alpha}_j^\dagger]),
    \label{eq:CM_definition}
\end{equation}
where $[\cdot,\cdot]$ denotes the commutator and $\hat{\alpha}_i$ are the components of the operator vector $\hat{\vec{\alpha}}$.

The CM formalism simplifies operations like partial trace and adding auxiliary systems. For example, combining independent systems corresponds to taking the direct sum of their respective CMs.

Due to the translational invariance discussed in \cref{app:Hk_derivation}, the total Hamiltonian $H_{SB}$ decomposes into independent blocks $h_{SB,k}$ (see \cref{eq:hsb_momentum_space_nn>0}) acting on momentum modes $(k, -k)$. We can analyze each block independently using the operator vector $\vec{\hat{\alpha}}_k = (\hat{a}_{k}, \hat{a}_{-k}^\dagger, \hat{b}_{k}, \hat{b}_{-k}^\dagger)^T$.

A cooling cycle involves evolving the initial state $\rho_S \otimes \rho_B^0$ (with CM $\gamma_{S,k}(0) \oplus \gamma_{B,k}^{0}$) for time $t=T/2$ under the joint Hamiltonian $h_{SB,k}$ and then tracing out the bath. The evolution of the CM $\gamma_{SB,k}$ under $h_{SB,k}$ for a time $t$ is given by~\cite{Bravyi2005Lagrangian}
\begin{equation}
    \gamma_{SB,k}(t) = e^{-2i h_{SB,k}t}\gamma_{SB,k}(0)e^{2i h_{SB,k}t},
    \label{eq:gamma_evolution}
\end{equation}
where an extra factor 2 appears compared to the evolution of density matrices due to the definition of CM in \cref{eq:CM_definition}.
For notational convenience, we absorb the factor 2 into the time by setting $T=2t$, where $t$ is the cycle time used in the main text. We then partition the exponential of the Hamiltonian into system and bath components:
\begin{equation}
    e^{-ih_{SB,k}T} =
    \begin{pmatrix}
        A_{S,k}  &A_{SB,k}\\
        A_{BS,k} &A_{B,k}
    \end{pmatrix}.
    \label{eq:evolution_blocks}
\end{equation}
Note that this partitioning is purely for organizational purposes—each submatrix depends on the full Hamiltonian $h_{SB,k}$. Only when the coupling $V_{SB}=0$ would these submatrices evolve independently according to their respective Hamiltonians.
Using this block structure, we can express the cooling map in the CM formalism. Starting with the definition of the map in \cref{eq:cooling_map0} from the main text, we obtain by taking the system block of the evolved total CM:
\begin{align}
    \mathcal{E}_k(\gamma_{S,k}) &= \tr_B\left[e^{-i h_{SB,k} T}(\gamma_{S,k} \oplus \gamma_{B,k}^{0})e^{i h_{SB,k} T}\right] \label{eq:gamma_m_plus_1}\\
                                &= A_{S,k} \gamma_{S,k} A_{S,k}^\dag + A_{SB,k} \gamma_{B,k}^{0} A_{SB,k}^\dag, \label{eq:cooling_map_cm}
\end{align}
where $\gamma_{B,k}^{0}$ is the CM of the bath ground state.

The steady state $\gamma_{S,k}^\text{ss}$ is the fixed point of this map, i.e., $\mathcal{E}_k(\gamma_{S,k}^\text{ss}) = \gamma_{S,k}^\text{ss}$, which gives us:
\begin{align}
    \gamma_{S,k}^\text{ss} &= A_{S,k} \gamma_{S,k}^\text{ss} A_{S,k}^\dag + A_{SB,k} \gamma_{B,k}^{0} A_{SB,k}^\dag. \label{eq:cooling-map-CM-notvectorized}
\end{align}
This is a linear equation for the matrix $\gamma_{S,k}^\text{ss}$ that can be solved by vectorization, a technique where the matrix equation is converted into a system of linear equations. To this end, we represent the CM as a vector $\ket{\gamma} = \vec{\gamma}$ and transform and solve the matrix equation of the form (assumed to be unique and invertible):
\begin{equation}
    \ket{\gamma_{S,k}^\text{ss}} = (A_{S,k} \otimes A_{S,k}^*)\ket{\gamma_{S,k}^\text{ss}} + (A_{SB,k} \otimes A_{SB,k}^*)\ket{\gamma_{B,k}^{0}},
\end{equation}
which leads to the solution
\begin{equation}
    \ket{\gamma_{S,k}^\text{ss}} = (\Id - A_{S,k} \otimes A_{S,k}^*)^{-1}(A_{SB,k} \otimes A_{SB,k}^*)\ket{\gamma_{B,k}^{0}}.
    \label{eq:cooling-map-CM}
\end{equation}

The formula in \cref{eq:cooling-map-CM} can be applied to each mode individually due to the block structure of the CM. The total CM of the system will then be the direct sum of CMs over all modes.

The mixing time of this map, which determines how quickly the system approaches the steady state, is determined by the largest eigenvalue of $A_{S,k} \otimes A_{S,k}^*$ (excluding the eigenvalue 1 associated with the steady state); the closer this eigenvalue is to 1, the slower the convergence.
The energy of a given mode for a state with correlation matrix $\gamma$ is computed as $E_k=\tr[h_{S,k}\gamma_{S,k}]$, where $h_{S,k}$ is the 2-by-2 top-left submatrix in $h_{SB,k}$.

For the Hamiltonian after the Bogoliubov transformation, with the bath in its ground state, the initial CM is:
\begin{equation}
    \gamma_{B,k}^{0} = \gamma_{S,k}^{GS} = \frac{1}{2}\begin{pmatrix}
        -1 &0\\
        0  &1
    \end{pmatrix}.
    \label{eq:gamma_b=gamma_gs}
\end{equation}
Thus, substituting into \cref{eq:cooling-map-CM}, the steady state can be determined for each mode individually due to the Gaussian nature of the evolution. The total CM of the system will then be the direct sum of the CMs over all modes.

To simplify our derivations, we introduce a phase transformation $a_n'= e^{i\pi/4}a_n$, $b_n'= e^{i\pi/4}b_n$, which preserves the canonical anticommutation relations of the fermionic operators. After applying this transformation, the Hamiltonian in the Bogoliubov basis [see \cref{eq:hsb_momentum_space_nn>0}] becomes:
\begin{align}
    H_{\text{Bog}}^k &= \vec{\alpha}'^\dag
    \begin{pmatrix}
        \epsilon_k &0           &gA_k   &-gC_k\\
        0          &-\epsilon_k &gC_k   &-gA_k\\
        gA^*_k     &gC^*_k      &\Delta &0\\
        -gC^*_k    &-gA^*_k     &0      &-\Delta
    \end{pmatrix}
    \vec{\alpha}',\\ \mbox{ with }
    \vec{\alpha}'    &=(\hat{a}_k'^{\dag},\hat{a}_{-k}',\hat{b}_k'^{\dag},\hat{b}_{-k}')^\dag,
    \label{eq:H_bogoliubov_appendix}
\end{align}
with the coupling coefficients
\begin{align}
    A_k &= \sum_{j=-nn}^{nn} e^{-i\phi_k j}(\lambda_j\cos\varphi_k+i\mu_j\sin\varphi_k),\\
    C_k &= \sum_{j=-nn}^{nn}e^{-i\phi_k j}(\mu_j\cos\varphi_k+i\lambda_j\sin\varphi_k).
\end{align}
Here, $\phi_k = 2\pi k/N$ and $\varphi_k$ is the Bogoliubov angle defined in \cref{eq:varphik} in the main text.

Next, we apply another unitary transformation within the system subspace and within the bath subspace, defined by the matrix $\Omega$ acting on pairs of operators:
\begin{equation}
    U_\Omega=\begin{pmatrix}
        \Omega &0\\
        0      &\Omega
    \end{pmatrix},\ \Omega = \frac{1}{\sqrt{2}} \begin{pmatrix}
        1 &1\\
        i &-i
    \end{pmatrix},
    \label{eq:omega_transform}
\end{equation}
leading to the Hamiltonian matrix we will work with in this Appendix:
\begin{equation}
    h_{SB,k} = i \begin{pmatrix}
        0          &-\epsilon_k &0      &gf_k\\
        \epsilon_k &0           &gp_k   &0\\
        0          &-gp_k^*     &0      &-\Delta\\
        -gf_k^*    &0           &\Delta &0
    \end{pmatrix},
    \label{eq:final_Hsb_pm_k}
\end{equation}
where the transformed coupling constants are:
\begin{align}
    f_k &= -e^{i\varphi_k} \sum_{j=-nn}^{nn} (\lambda_j + \mu_j) e^{-i\phi_k j},\\
    p_k &= e^{-i\varphi_k} \sum_{j=-nn}^{nn} (\lambda_j - \mu_j) e^{-i\phi_k j}.
\end{align}
This form of the Hamiltonian is going to be particularly convenient for perturbative analysis.
We can get rid of the minus sign in $f_k$ by setting $\tilde{\lambda}_j = -\mu_j$ and $\tilde{\mu}_j = -\lambda_j$, which leaves $p_k$ invariant. Additionally, the bath CM (transformed according to the extra unitary $\Omega$) will now be, in vectorized form:
\begin{equation}
    \ket{\gamma_{B,k}^{0}}=\begin{pmatrix}
        0\\
        i/2\\
        -i/2\\
        0
    \end{pmatrix}.
\end{equation}
We now assume weak coupling, that is, $(gt)^2\ll 1$ as discussed in \cref{sec:multifreq_longtime}, allowing us to treat the coupling terms as perturbations. The terms $f_k$ and $p_k$ are thus rescaled to be $\mathcal{O}(1)$. In the following, we will omit the $k$ subscript for brevity, since the derivation does not depend on the specific mode.

To perform the perturbative analysis, we express the total Hamiltonian as $h_{SB,k} = h_0 + g v_{SB,k}$, where $h_0$ contains the uncoupled system and bath terms, and $v_{SB,k}$ contains the coupling terms. We then apply a Dyson expansion of the time-evolution operator $e^{-i h_{SB,k} T}$ up to second order in $g$:
\begin{align}
    \label{eq:CM_dyson_expansion}
     &e^{-i h_{SB,k} T} \simeq e^{-i h_0 T} \left[ 1 - i g \int_0^T  \dd t' e^{i h_0 t'} v_{SB,k} e^{-i h_0 t'} \right. \nonumber\\
     &\quad \left. - g^2 \int_0^T  \dd t' \int_0^{t'}  \dd t'' e^{i h_0 t'} v_{SB,k} e^{-i h_0 t'} e^{i h_0 t''} v_{SB,k} e^{-i h_0 t''} \right].
\end{align}
We can write our exponential in block matrix form as in \cref{eq:evolution_blocks}. To second order in $g$, these submatrices can be expressed as:
\begin{align}
    A_{S,k}  &\simeq A_{S,k}^{(0)} + g^2 A_{S,k}^{(2)}, \label{eq:As_expansion}\\
    A_{SB,k} &\simeq g A_{SB,k}^{(1)}. \label{eq:Asb_expansion}
\end{align}
where the zeroth, first, and second order terms can be computed from the integrals in \cref{eq:CM_dyson_expansion}. The zeroth-order term represents the free evolution of the system:
\begin{equation}
    \label{eq:As0}
    A_{S,k}^{(0)} = \begin{pmatrix}
        \cos(\epsilon_k T)  &\sin(\epsilon_k T)\\
        -\sin(\epsilon_k T) &\cos(\epsilon_k T)
    \end{pmatrix},
\end{equation}
which corresponds to rotation in the phase space of the system.

The first-order term in the system-bath coupling is:
\begin{align}
    \label{eq:Asb1}
    A_{SB,k}^{(1)} &= \begin{pmatrix}
                          a_1 &a_2\\
                          a_3 &a_4
                      \end{pmatrix},\\
    a_1            &=x_1 \left[\cos(\Delta T) - \cos(\epsilon_k T)\right],\nonumber\\
    a_2            &=x_1 \sin(\Delta T) + i x_2 \sin(\epsilon_k T),\nonumber\\
    a_3            &=x_1 \sin(\epsilon_k T) + i x_2 \sin(\Delta T),\nonumber\\
    a_4            &=-i x_2 \left[\cos(\Delta T) - \cos(\epsilon_k T)\right],\nonumber
\end{align}
where $x_1$ and $x_2$ depend on the coupling parameters $f_k, p_k$ and the energy scales $\epsilon_k, \Delta$:
\begin{align}
    x_1 &= \frac{\epsilon_k p_k - \Delta f_k}{\epsilon_k^2 - \Delta^2}, \quad
    x_2 = -i \frac{\epsilon_k f_k - \Delta p_k}{\epsilon_k^2 - \Delta^2}. \label{eq:x1_x2_definition}
\end{align}
The second-order term in the system evolution is:
\begin{align}
    A_{S,k}^{(2)} &= \begin{pmatrix}
                         a_{11} &a_{12}\\
                         a_{21} &a_{22}
                     \end{pmatrix}, \label{eq:As2}
\end{align}
where the entries $a_{ij}$ are given by:
\begin{align}
    \label{eq:As2_entries}
    a_{11} &= |x_1|^2 \left[\cos(\Delta T) - \cos(\epsilon_k T)\right] \nonumber\\
           &\quad + \frac{T}{2} \sin(\epsilon_k T) \left(i f_k x^*_2 - p_k x^*_1\right), \nonumber\\
    a_{12} &= i x_1 x^*_2 \sin(\Delta T) - \frac{\sin(\epsilon_k T)}{2} \left(|x_1|^2 + |x_2|^2\right) \nonumber\\
           &\quad - i \frac{\Delta}{\epsilon_k} \sin(\epsilon_k T) \Im\left(i x_1 x^*_2\right) + \left(p_k x^*_1 - i f_k x^*_2\right) \frac{T}{2} \cos(\epsilon_k T), \nonumber\\
    a_{21} &= i x_2 x^*_1 \sin(\Delta T) + \frac{\sin(\epsilon_k T)}{2} \left(|x_1|^2 + |x_2|^2\right) \nonumber\\
           &\quad - i \frac{\Delta}{\epsilon_k} \sin(\epsilon_k T) \Im\left(i x_1 x^*_2\right) - \left(p_k x^*_1 - i f_k x^*_2\right) \frac{T}{2} \cos(\epsilon_k T), \nonumber\\
    a_{22} &= |x_2|^2 \left[\cos(\Delta T) - \cos(\epsilon_k T)\right] \nonumber\\
           &\quad + \frac{T}{2} \sin(\epsilon_k T) \left(i f_k x^*_2 - p_k x^*_1\right). \nonumber
\end{align}
To find the steady state, we solve the fixed-point equation given by \cref{eq:cooling-map-CM}. In the weak coupling limit, the matrix inverse $(\Id - A_{S,k} \otimes A_{S,k}^*)^{-1}$ can be approximated as:
\begin{equation}
    (\Id - A_{S,k} \otimes A_{S,k}^*)^{-1} \approx \frac{1}{g^2 Q} \begin{pmatrix}
        1 &0  &0  &1\\
        0 &1  &-1 &0\\
        0 &-1 &1  &0\\
        1 &0  &0  &1
    \end{pmatrix},
    \label{eq:inverse_matrix}
\end{equation}
with
\begin{align}
    Q = &\ 2(|x_1|^2 + |x_2|^2)(1 - \cos(\Delta T) \cos(\epsilon_k T))\nonumber\\
        &+4 \sin(\Delta T) \sin(\epsilon_k T) \Im(x_1 x^*_2).\label{eq:Q_definition}
\end{align}
Performing the matrix multiplication in \cref{eq:cooling-map-CM} cancels out the factors of $g$ and the denominators in $x_1$ and $x_2$, leading to the expression for the steady state energy that matches the results derived using the density matrix approach in \cref{app:Single_freq steady state}.

\subsection{Adding noise}

The CM formalism can be readily extended to incorporate the depolarizing noise model discussed in~\cref{sec:adding_decoherence}.

We use the formalism described by Bravyi and König~\cite{Bravyi2012Classical}, which is based on Majorana operators. Let the $4N$ Majorana operators be defined as $c_{2n-1}=a_n+a_n^\dag$, $c_{2n}=i(a_n-a_n^\dag)$ for the system operators ($n=1,\ldots,N$) and $c_{2N+2n-1}=b_n+b_n^\dag$, $c_{2N+2n}=i(b_n-b_n^\dag)$ for the bath operators ($n=1,\ldots,N$).

In the Majorana representation, the master equation takes the form:
\begin{align}
    \dot{\rho}=-i[H_{SB},\rho]+\sum_\mu \kappa_\mu (2L_\mu\rho L_\mu^\dag-\{L_\mu^\dag L_\mu,\rho\}),
\end{align}
where $H_{SB}=\frac{i}{4}\sum_{jk}\textbf{H}_{jk}c_j c_k$ and the Lindblad operators are $L_{\mu}=\sum_j l_{\mu j}c_j$.

Bearing in mind that we can still work in the block-diagonal Hamiltonian for modes $\pm k$, then $N=2$ and the index $j$ is just a reformulation of our notation.
In this picture, it has been shown~\cite{Bravyi2012Classical} that the evolution of the covariance matrix $\Gamma$ associated with $\rho$ follows:
\begin{align}
    \frac{\dd}{\dd t}\Gamma(t) &=X\Gamma(t)+\Gamma(t)X^T+Y\label{eq:CM_differential_eq},\\
    X                          &=-\textbf{H}-2(M+M^*),\\
    Y                          &=4i(M^*-M),\\
    M_{jk}                     &=\sum_{\mu}l_{\mu j}l_{\mu k}^*.
\end{align}
In our case, the Lindblad operators $L_\mu$ are of four different types, corresponding to the creation and annihilation operators for both system and bath modes:
\begin{align}
    L_{4j-3} &=\frac{\sqrt{\kappa}}{4}(c_{2j-1}-ic_{2j}),\\
    L_{4j-2} &=\frac{\sqrt{\kappa}}{4}(c_{2j-1}+ic_{2j}),\\
    L_{4j-1} &=\frac{\sqrt{\kappa}}{4}(c_{2N+2j-1}-ic_{2N+2j}),\\
    L_{4j}   &=\frac{\sqrt{\kappa}}{4}(c_{2N+2j-1}+ic_{2N+2j}).
\end{align}
Calculating $M$ is straightforward: only the diagonal elements $j=k$ contribute, since the off-diagonal elements have contributions $\pm i\frac{\kappa}{16}$ that cancel each other out. Therefore, $M=\kappa\Id$, which gives $Y=0$. This also confirms that our chosen Lindbladian commutes with any fermionic quadratic Hamiltonian, as previously shown in \cref{app:secIV_noise} using the density matrix approach.

The differential equation from \cref{eq:CM_differential_eq} has a steady state that we denote as $\Gamma_\text{ss}$. The solution to this equation gives the time evolution:
\begin{equation}
    \Gamma(t)=\Gamma_\text{ss}+e^{Xt}(\Gamma(0)-\Gamma_\text{ss})e^{X^T t},
\end{equation}
which, in our particular case, translates to:
\begin{equation}
    \Gamma(t)=\Gamma_\text{ss}+e^{-4\kappa t}e^{-\textbf{H}t}(\Gamma(0)-\Gamma_\text{ss})e^{\textbf{H}t}e^{-4\kappa t}.\label{eq:CM_k_time_evol}
\end{equation}
This shows that the effect of the environment is to add a damping factor $e^{-4\kappa t}$ to the ideal evolution.

As the steady state of the system and bath is what we want to determine, it remains to trace out the bath. Following from~\cref{eq:CM_differential_eq}, the fixed point can be chosen to be $\Gamma_\text{ss}=0$ (the maximally mixed state) since $Y=0$, and this in turn simplifies~\cref{eq:CM_k_time_evol}:
\begin{equation}
    \Gamma(t)=e^{-4\kappa t}e^{-\textbf{H}t}\Gamma(0)e^{\textbf{H}t}e^{-4\kappa t}.\label{eq:CM_simplified_time_evol}
\end{equation}
Now that we know the effect of the environment, we can return to the correlation matrix formalism and simply follow the same steps outlined for the ideal scenario in~\cref{app:CM_derivation}, where now $A_{S,k}$ and $A_{SB,k}$ are modified by the damping factor $e^{-4\kappa t}$. Taking into account the change $T=2t$, we obtain
\begin{equation}
    \ket{\gamma_S^\text{ss}} = (\Id -e^{-4\kappa T} A_{S,k} \otimes A^*_{S,k})^{-1} (e^{-4\kappa T}A_{SB,k} \otimes A^*_{SB,k}) \ket{\gamma_0^B}.
    \label{eq:steady_state_equation_vectorized}
\end{equation}

The noise considered here will always be smaller or equal to the bath coupling $g$, so we can once again use the weak coupling limit and keep only the highest orders in $gt,\kappa t$.

Firstly, we approximate the exponential by a Taylor series and only keep the first-order terms $\mathcal{O}(g^2t^2,\kappa t)$; this implies that $e^{-2\kappa T}A_{S,k} \simeq A_{S,k}^{(0)}(1-2\kappa T) + g^2 A_{S,k}^{(2)}$ and $e^{-2\kappa T}A_{SB,k} \simeq gA_{SB,k}^{(1)}$, which slightly modifies the inverse of $(\Id-A_{S,k}\otimes A^*_{S,k})$. This simplification of results, analogous to the one in \cref{app:Single_freq steady state}, is equivalent to considering that the change undergone by the bath as a result of the environment is negligible, since it will have a higher-order contribution on the system.

\section{Averaging Over Randomized Times}
\label{app:long_time_averages}
In this Appendix, we will continue working in the weak coupling regime. We now consider the scenario where, for each cooling cycle $j$, the cycle time $t_j$ is chosen from a random uniform distribution $[0, 2t]$ with mean value $t$. We are interested in understanding how averaging over the times affects the steady state of the system, particularly when $t$ is large.

To analyze this, we consider the concatenation of multiple such maps with different cycle times. For simplicity, we begin by examining two consecutive maps. Applying the CPTP map twice, we have
\begin{align}
     &\mathcal{E}_{t_2}( \mathcal{E}_{t_1}(\rho_k) ) \approx \mathcal{E}_{t_2} \left( e^{-i h_k t_1}\left( \rho_k + \mathcal{L}_{t_1}(\rho_k) \right)e^{i h_k t_1} \right) \nonumber\\
     &\approx e^{-i h_k (t_1 + t_2)} \left( \rho_k + \mathcal{L}_{t_1}(\rho_k) + \tilde{\mathcal{L}}_{t_1, t_2}(\rho_k) \right) e^{i h_k (t_1 + t_2)},
\end{align}
where the modified superoperator $\tilde{\mathcal{L}}_{t_1, t_2}$ accounts for the effect of the second cooling cycle on the system:
\begin{align}
    \tilde{\mathcal{L}}_{t_1, t_2}(\rho_k) &= e^{i h_k t_1} \mathcal{L}_{t_2} \left( e^{-ih_k t_1} \rho_k e^{i h_k t_1} \right) e^{-i h_k t_1} \nonumber\\
                                           &= \sum_{i=1}^2 \tilde{l}_{i2} \rho_k \tilde{l}_{i2}^\dag - \frac{1}{2} \left\{ \tilde{l}_{i2}^\dag \tilde{l}_{i2}, \rho_k \right\},
\end{align}
with the transformed jump operators given by
\begin{align}
    \tilde{l}_{i2} = e^{i h_k t_1} l_{i}(t_2) e^{-i h_k t_1}, \quad i = 1,2.
\end{align}
From here, it can be deduced that if $L\gg 1$ maps with random cycle times $t_j$ are concatenated, the total map will have a sum of Lindbladians $\tilde{\mathcal{L}}_{j}$, $j\in\{1,\ldots,L\}$ of the following form:
\begin{align}
    \mathcal{E}_{\text{total}}(\rho_k) &\approx e^{-i h_k (\sum_{j=1}^L t_j)} \left( \rho_k + \sum_{j=1}^L \tilde{\mathcal{L}}_{j}(\rho_k) \right) e^{i h_k (\sum_{j=1}^L t_j)}\nonumber\\
                                       &\approx e^{-i h_k L t} \left( \rho_k + \sum_{j=1}^L \tilde{\mathcal{L}}_j(\rho_k) \right) e^{i h_k L t},
\end{align}
where $\tilde{\mathcal{L}}_j(\rho_k)$ represents the contribution from the $j$-th cooling cycle:
\begin{align}
    \tilde{\mathcal{L}}_{j}(\rho_k) &= g^2\sum_{i=1}^2 \tilde{l}_{i j} \rho_k \tilde{l}_{i j}^\dag - \frac{1}{2} \left\{ \tilde{l}_{i j}^\dag \tilde{l}_{i j}, \rho_k \right\},
\end{align}
and the transformed jump operators for each cycle are:
\begin{align}
    \tilde{l}_{1j} &= A^*_k\, x_k(t_j)\, e^{-i \epsilon_k \tilde{T}_j} \hat{a}_k + B^*_k\, y_k(t_j)\, e^{i \epsilon_k \tilde{T}_j} \hat{a}_{-k}^\dag,\\
    \tilde{l}_{2j} &= A_k\, x_k(t_j)\, e^{-i \epsilon_k \tilde{T}_j} \hat{a}_{-k} - B_k\, y_k(t_j)\, e^{i \epsilon_k \tilde{T}_j} \hat{a}_k^\dag,
\end{align}
with $\tilde{T}_j = \sum_{m=1}^{j-1} t_m$ being the cumulative time up to the $(j-1)$-th cycle (for $j=1$ we set it to $0$). We work for now in the general case $nn\geq0$, but we will remark the final results for the $nn=0$ case as well.

For $j = 1$, we have $\tilde{\mathcal{L}}_1 = \mathcal{L}_{t_1}$. For $j > 1$, averaging over the different times $t_j$ causes the rapidly oscillating terms $e^{\pm 2i \epsilon_k \tilde{T}_j}$ to average out to zero.
This is because $\tilde{T}_j$ is a sum of random variables, and for large $j$, the phase $2\epsilon_k \tilde{T}_j$ becomes uniformly distributed over $[0, 2\pi)$, causing the average of the exponential to vanish.
This effectively eliminates the cross terms between modes $\hat{a}_k$ and $\hat{a}_{-k}$.

This averaging process leads to a diagonal Lindbladian for the averaged map, where only terms proportional to $\hat{a}_k^\dag \hat{a}_k$ and $\hat{a}_{-k}^\dag \hat{a}_{-k}$ survive. The averaged Lindbladian $\overline{\tilde{\mathcal{L}}_{j}(\rho_k)}$ becomes:
\begin{align}
    \overline{\tilde{\mathcal{L}}_{j}(\rho_k)}
     &= \frac{1}{(2t)^j} \int_{0}^{2t} \dd t_{j - 1} \cdots \int_{0}^{2t} \dd t_1 \tilde{\mathcal{L}}_j(\rho_k)\nonumber\\
     &= \gamma_k^{\rm c} \sum_{i=\pm k} \left( \hat{a}_i \rho_k \hat{a}_i^\dag - \frac{1}{2} \left\{ \hat{a}_i^\dag \hat{a}_i, \rho_k \right\} \right) \nonumber\\
     &\quad + \gamma_k^{\rm h} \sum_{i=\pm k} \left( \hat{a}_i^\dag \rho_k \hat{a}_i - \frac{1}{2} \left\{ \hat{a}_i \hat{a}_i^\dag, \rho_k \right\} \right),
    \label{eq:averaged_lindbladian}
\end{align}
where the rates $\gamma_k^{\rm c}$ and $\gamma_k^{\rm h}$ are given by averaging over the random times:
\begin{align}
    \gamma_k^{\rm c} &= \frac{1}{2t} \int_0^{2t} g^2 |A_k x_{k,t'}|^2 \dd t'\nonumber\\
                     &= g^2|A_k|^2 \left(\frac{2}{(\epsilon_k - \Delta)^2}-\frac{\sin\left[(\Delta-\epsilon_k)2t\right]}{(\Delta-\epsilon_k)^3t}\right),
\end{align}
and similarly for the heating rate $\gamma_k^{\rm h}$:
\begin{align}
    \gamma_k^{\rm h} &= \frac{1}{2t} \int_0^{2t} g^2 |B_k y_{k,t'}|^2 \dd t'\nonumber\\
                     &= g^2|B_k|^2 \left(\frac{2}{(\epsilon_k + \Delta)^2}-\frac{\sin\left[(\Delta+\epsilon_k)2t\right]}{(\Delta+\epsilon_k)^3t}\right).
\end{align}

While these are the exact results, they are cumbersome to work with. We can approximate them using another function that behaves in the same way in the resonant limit $|\Delta-\epsilon_k|t\ll 1$ and in the off-resonant limit $|\Delta-\epsilon_k|t\gg 1$. In this case, a Lorentzian can be shown numerically to work well (see~\cref{fig:Fig3}):
\begin{align}
    \gamma_k^{\rm c} &= \frac{2g^2|A_k|^2}{(\epsilon_k - \Delta)^2+\gamma_0^2},\\
    \gamma_k^{\rm h} &= \frac{2g^2|B_k|^2}{(\epsilon_k + \Delta)^2+\gamma_0^2},\\
    \gamma_0^2       &= \frac{3}{2t^2},
\end{align}
where in particular, for the case $nn=0$, $|A_k|^2=|B_k|^2=1$ as mentioned in the main text (see~\cref{eq:gammac_def,eq:gammah_def,eq:tau0a}). We assume that, since we aim to cool, $\Delta t\gg 1$ and therefore we can neglect $\gamma_0$ in $\gamma_k^{\rm h}$. However, in the cooling term, $\gamma_0$ becomes the leading term when we approach resonant conditions.
For $j=1$, the diagonal terms give the same result, but the cross terms do not vanish, since we only average over $l_{i1} l_{i1}^\dag$. These terms have weights $\gamma_\text{cross}$ and $\gamma^*_\text{cross}$, with:
\begin{equation}
    \gamma_\text{cross} = \frac{A_k B^*_k}{\epsilon_k^2 - \Delta^2}.
\end{equation}
However, in the limit $L \gg 1$, the single off-diagonal contribution from $j=1$ is negligible compared to the sum of the diagonal terms $L \gamma_k^{\rm c,h}$. Thus, we can safely neglect these off-diagonal terms. Moreover, the concatenation of the $L$ maps can be approximated by a single averaged map applied $L$ times. That is,
\begin{align}
    \mathcal{E}_{\text{total}}(\rho_k) &\approx \mathcal{E}_{\text{avg}}^L(\rho_k) = e^{-i h_{S,k} Lt} \left( \rho_k + L \mathcal{L}^\text{avg}(\rho_k) \right) e^{i h_{S,k} Lt},\\
    \mathcal{L}^\text{avg}(\rho_k)     &= \gamma_k^{\rm c} \sum_{i=\pm k} \left( \hat{a}_i \rho_k \hat{a}_i^\dag - \frac{1}{2} \left\{ \hat{a}_i^\dag \hat{a}_i, \rho_k \right\} \right)        \nonumber\\
                                       &\quad + \gamma_k^{\rm h} \sum_{i=\pm k} \left( \hat{a}_i^\dag \rho_k \hat{a}_i - \frac{1}{2} \left\{ \hat{a}_i \hat{a}_i^\dag, \rho_k \right\} \right).
    \label{eq:averaged_map}
\end{align}
This Lindbladian describes a process where the system relaxes towards a steady state determined by the rates $\gamma_k^{\rm c,h}$. Note that, crucially, the jump operators no longer couple the modes $\pm k$. Hence, they can be treated individually.

The resulting steady state is a thermal-like state with populations dependent on these rates, identical for both modes. Explicitly, the steady-state density matrix $\rho_{\text{ss}}^{(k)}$, energy $E_{\text{ss}}^k$ and fidelity for the $(k,-k)$ mode are:
\begin{align}
     &\rho_{\text{ss}}^{(k)} = \left( m_k \ket{\Omega}\bra{\Omega} + (1 - m_k) \ket{1}\bra{1} \right)^{\otimes 2},\\
     &m_k = \frac{\gamma_k^{\rm c}}{\gamma_k^{\rm c} + \gamma_k^{\rm h}}, \label{eq:ss_density_matrix}\\
     &E_{\text{ss}}^k = \epsilon_k \left( \frac{-\gamma_k^{\rm c} +\gamma_k^{\rm h}}{\gamma_k^{\rm c} + \gamma_k^{\rm h}} \right),\quad \mathcal{F}_k=m_k^2.
    \label{eq:avg_longtime_ss_app}
\end{align}

Furthermore, we can verify that dissipative state preparation (DSP) will not be effective in this regime. In DSP, setting $\epsilon_k = 0$ for the evolution eliminates the possibility of performing the rotating wave approximation (RWA), and thus obtaining a result close to the ground state would require fine-tuning of the couplings.

Furthermore, we can now calculate the convergence rate for the Lindbladian in \cref{eq:averaged_map} explicitly. We start by considering a general density matrix for some $k$-mode (since now we can treat $\pm k$ separately) and approximate the action of the average map after one cycle:
\begin{align}
    \rho_k^0(0)                         &=\begin{pmatrix}
                                              m_k^0(0)         &\beta_k^0(0)\\
                                              {\beta_k^0}^*(0) &1-m_k^0(0)
                                          \end{pmatrix},\\
    \mathcal{E}_\text{avg}(\rho_k^0(0)) &=\begin{pmatrix}
                                              m_k^0(t)          &\beta_k^0(t)\\
                                              {\beta_k^0}^* (t) &1-m_k^0(t)
                                          \end{pmatrix},\\
    m_k^0(t)                            &\approx\frac{\gamma_k^{\rm c}}{\gamma_k^{\rm c}+\gamma_k^{\rm h}}+e^{-(\gamma_k^{\rm c}+\gamma_k^{\rm h})}\left(m_k^0(0)-\frac{\gamma_k^{\rm c}}{\gamma_k^{\rm c}+\gamma_k^{\rm h}}\right),\\
    \beta_k^0(t)                        &\approx\beta_k^0(0)e^{-(\gamma_k^{\rm c}+\gamma_k^{\rm h})/2+i\epsilon_k t}.
\end{align}
The matrix $\mathcal{E}_\text{avg}(\rho_k^0(0))$ is the new starting point for the next cycle, i.e. $\rho_k^1(0)$. Repeating the map $n$ times yields the following result:
\begin{align}
    \mathcal{E}_\text{avg}^n(\rho_k^0(0)) &=\begin{pmatrix}
                                                m_k^n(t)         &\beta_k^n(t)\\
                                                {\beta_k^n}^*(t) &1-m_k^n(t)
                                            \end{pmatrix},\\
    m_k^n(t)                              &\approx\frac{\gamma_k^{\rm c}}{\gamma_k^{\rm c}+\gamma_k^{\rm h}}+e^{-(\gamma_k^{\rm c}+\gamma_k^{\rm h})n}\left(m_k^0(0)-\frac{\gamma_k^{\rm c}}{\gamma_k^{\rm c}+\gamma_k^{\rm h}}\right),\\
    \beta_k^n(t)                          &\approx\beta_k^0(0)e^{-(\gamma_k^{\rm c}+\gamma_k^{\rm h})n/2+i\epsilon_k t n}.
\end{align}
Thus, the convergence to $m_k\equiv m_k^\infty(t)$ is driven by an exponential in $-\alpha_k n=-(\gamma_k^{\rm c}+\gamma_k^{\rm h})n$. This also shows explicitly that, since in this particular case the action of $\exp(-ih_{S,k} t)$ commutes with the action of $\mathcal{L}^\text{avg}$, we fulfill the following identity:
\begin{equation}
    \left(e^{-ih_{S,k} t}e^{\mathcal{L}^\text{avg}}(\rho_k)e^{i h_{S,k} t}\right)^n=e^{-ih_{S,k} n t}e^{n\mathcal{L}^\text{avg}}(\rho_k)e^{i h_{S,k} n t}.
\end{equation}

\section{Multifrequency cooling}
\label{app:multifrequency_cooling}

In this Appendix we present some details of the calculations used to derive the results of \cref{subsec:multi_freq}, starting from the generalization of the previous Appendix to multiple frequencies.

Starting from \cref{eq:gammach} in the main text, the average cooling and heating rates are given by:
\begin{equation}
    \gamma^{\rm c,h}_k = \frac{1}{R} \sum_{r=1}^R \gamma^{\rm c,h}_{r,k},
\end{equation}
where $\gamma^{\rm c,h}_{r,k}$ are the rates for a single frequency $\Delta_r$, given by \cref{eq:gammac_def,eq:gammah_def} in the main text:
\begin{align}
    \gamma^{\rm c}_{r,k} &= \frac{2g^2}{(\Delta_r-\epsilon_k)^2+ \gamma_0^2},\\
    \gamma^{\rm h}_{r,k} &= \frac{2g^2}{(\Delta_r+\epsilon_k)^2}.
\end{align}
In the limit $R \gg 1$, we can approximate the sum over frequencies by an integral. We define the frequency spacing $\delta = (\epsilon_{\rm M} - \epsilon_{\rm m}) / R$, and the frequencies are chosen as $\Delta_r = \epsilon_{\rm m} + \delta (r - \frac{1}{2})$, where $\epsilon_{\rm M,\ m} = \sqrt{1\pm \sin(2\theta)}$ are the maximum and minimum values of $\epsilon_k$ as defined in the main text.

In the limit $R\gg 1$, the average cooling and heating rates can be expressed as an integral:
\begin{align}
    \gamma_k^{\rm h} &= \frac{1}{R} \sum_{r=1}^R \frac{2g^2}{(\Delta_r+\epsilon_k)^2} \nonumber\\
                     &\approx \frac{1}{(\epsilon_{\rm M}-\epsilon_{\rm m})} \int_{\epsilon_{\rm m}}^{\epsilon_{\rm M}} \frac{2g^2}{(\Delta+\epsilon_k)^2} \dd\Delta,\\
    \gamma_k^{\rm c} &= \frac{1}{R} \sum_{r=1}^R \frac{2g^2}{(\Delta_r-\epsilon_k)^2+ \gamma_0^2} \nonumber\\
                     &\approx \frac{1}{(\epsilon_{\rm M}-\epsilon_{\rm m})} \int_{\epsilon_{\rm m}}^{\epsilon_{\rm M}} \frac{2g^2}{(\Delta-\epsilon_k)^2+ \gamma_0^2} \dd\Delta.
\end{align}

Evaluating the integral for the heating rate $\gamma_k^{\rm h}$ yields
\begin{align}
    \gamma_k^{\rm h} &\approx \frac{1}{(\epsilon_{\rm M}-\epsilon_{\rm m})} \int_{\epsilon_{\rm m}}^{\epsilon_{\rm M}} \frac{2g^2}{(\Delta+\epsilon_k)^2} \dd\Delta\nonumber\\
                     &= \frac{2g^2}{(\epsilon_{\rm M}+\epsilon_k)(\epsilon_{\rm m}+\epsilon_k)}.
\end{align}
Evaluating the integral for the cooling rate $\gamma_k^{\rm c}$ with the substitution $u = (\Delta - \epsilon_k) / \gamma_0$ and $\dd\Delta = \gamma_0 \dd u$ yields
\begin{align}
    \gamma_k^{\rm c} &\approx \frac{1}{(\epsilon_{\rm M}-\epsilon_{\rm m})} \int_{\epsilon_{\rm m}}^{\epsilon_{\rm M}} \frac{2g^2}{(\Delta-\epsilon_k)^2+ \gamma_0^2} \dd\Delta\nonumber\\
                     &= \frac{2g^2 t \beta_k}{(\epsilon_{\rm M}-\epsilon_{\rm m})}.
\end{align}
where we defined
\begin{align}
    \beta_k &=\left[\tan^{-1}\left(\frac{\Delta-\epsilon_k}{\gamma_0}\right)\right]_{\epsilon_{\rm m}}^{\epsilon_{\rm M}}\sqrt{\frac{2}{3}}\nonumber\\
            &= \sqrt{\frac{2}{3}} \left[ \tan^{-1}(z_M) - \tan^{-1}(z_m) \right],
\end{align}
and $z_M = (\epsilon_{\rm M}-\epsilon_k)t \sqrt{\frac{2}{3}}$ and $z_m = (\epsilon_{\rm m}-\epsilon_k)t \sqrt{\frac{2}{3}}$.
This constant can be approximated depending on the regime we are in.

If $R\gg|\epsilon_{\rm M}-\epsilon_{\rm m}|t\gg1$, we have $(\epsilon_{\rm M}-\epsilon_{\rm m})t \gg \gamma_0^{-1}$, so $z_M$ and $z_m$ are large, and this implies that at least one of the energies $\epsilon_{M,m}$ fulfills $\epsilon_{M,m}t\gg1$. Thus, its associated $\tan^{-1}$ can be approximated as $\pm\pi/2$, and the maximum value attainable happens when both can be approximated as such, giving a value of $\pi$. Therefore, in this regime we have that $\sqrt{3/2}\ \beta_k\in[\pi/2,\pi]$.

In the second regime we considered here, namely $R\gg1\gg|\epsilon_{\rm M}-\epsilon_{\rm m}|t$, we have $(\epsilon_{\rm M}-\epsilon_{\rm m})t \ll \gamma_0^{-1}$, so $z_M$ and $z_m$ are small.
We can use the Taylor expansion $\arctan(z) \approx z$ for small $z$. Then,
\begin{align}
    \beta_k &\approx \sqrt{\frac{2}{3}} \left[ z_M - z_m \right]\nonumber\\
            &= \frac{2}{3} t (\epsilon_{\rm M}-\epsilon_{\rm m}).
\end{align}

Using these derivations we analyzed in the main text the performance of the multifrequency cooling algorithm.

\section{Effect of noise caused by a finite environment}
\label{app:finite_noise}

In \cref{sec:adding_decoherence} we defined a common way of modeling noise in system dynamics, via a master equation. However, as discussed in \cref{sec:nn_cooling}, this kind of noise can be easily overcome in DSP by setting $t\rightarrow 0$. Therefore, in this Appendix we introduce and study, both numerically and analytically, a different noise model, which showcases the advantage of cooling over DSP in the presence of an external environment.
\subsection{Model}
\label{subapp:finite_noise_model}

We describe the influence of an uncontrollable, external environment by connecting both the system and the bath to separate sets of $N$ independent fermionic sites, denoted $E_1$ and $E_2$. The coupling terms are local and designed to mimic spontaneous absorption and emission processes:
\begin{align}
    V_{SE_1} &= \kappa'\sum_{n=1}^N \left(a_n {c}_n^\dag + \text{h.c.}\right), \label{eq:V_SE}\\
    V_{BE_2} &= \kappa'\sum_{n=1}^N \left({b}_n {d}_n^\dag + \text{h.c.}\right), \label{eq:V_BE}
\end{align}
and the environment Hamiltonians are
\begin{align}
    H_{E_1} &= \frac{\Delta_E}{2} \sum_{n=1}^N \left({c}_n^\dag {c}_n - c_n {c}_n^\dag\right), \label{eq:H_E1}\\
    H_{E_2} &= \frac{\Delta_E}{2}\sum_{n=1}^N \left({d}_n^\dag {d}_n - d_n {d}_n^\dag \right), \label{eq:H_E2}
\end{align}
where the fermionic operators $c_n$ and $d_n$ act on the $n$-th fermionic site of environments $E_1$ and $E_2$, respectively. The environments are characterized by the energy splitting $\Delta_E$, the coupling strength $\kappa'$, and an initial state, $\rho_E$. We will assume that they are the same for both environments. Note that now $\kappa'$ has a different meaning than the $\kappa$ in the master equation of \cref{sec:adding_decoherence}, since before $\kappa$ was a rate and now $\kappa'$ is a coupling.

During a cooling cycle, the entire system (system $S$, bath $B$, and environments $E_1, E_2$) evolves under the total Hamiltonian:
\begin{equation}
    H_{\text{tot}} = H_{SB} + H_{E_1} + H_{E_2} + V_{SE_1} + V_{BE_2},
    \label{eq:H_tot}
\end{equation}
after which both environments and the bath are traced out and reset. In this description, the bath and the environments are treated similarly, except that the environment parameters and initial states are not controlled. Hence, the map corresponding to a cooling cycle is given by
\begin{equation}
    \mathcal{N}'(\rho_S) = \tr_{E_1 E_2 B}\left[e^{-iH_\text{tot}t}\left(\rho_S \otimes \rho_B\otimes \rho_{E_1}\otimes\rho_{E_2}\right)e^{i H_\text{tot}t}\right].
    \label{eq:cooling_mapdiss}
\end{equation}
This map $\mathcal{N}'$ is related to the master equation-based noisy map $\mathcal{N}$, but it has more free parameters.
Still, it is very different from the one considered in \cref{sec:adding_decoherence}. For instance, if $\rho_{E_{1,2}}$ is the vacuum of $c_n, d_n$ operators, in the perturbative limit the environment can only extract particles from the system, unlike the master equation, where the environment can also introduce them.
To be more specific, we will consider translationally-invariant initial states $\rho_{E_{1,2}}$ dependent on a parameter $p_E\in[-1,1]$ that can be expressed in the following form for each $(k,-k)$ block:
\begin{align}
    \rho_{E_{1,2}}   &=\bigotimes_k\rho_{E_{1,2}}^k,\\
    \rho_{E_{1,2}}^k &=\left[\frac{1+p_E}{2}\ket{\Omega}\bra{\Omega}+\frac{1-p_E}{2}\ket{1}\bra{1}\right]^{\otimes2}.
    \label{eq:rho_E_def}
\end{align}
This parameter $p_E$ encompasses a wide range of options, from the ground state ($p_E=1$) to the most excited state ($p_E=-1$) and the maximally mixed state ($p_E=0$).Additionally, the form in \cref{eq:rho_E_def}  still allows us to work in each $(k,-k)$ block independently, as we have been doing throughout the paper.

\subsection{Steady state of the system}
We will now follow a similar derivation as shown in \cref{subapp:dyson_expansion_noiseless}, with the difference that our approximation via the Dyson expansion now includes additional terms $\kappa' V_{SE_1},\ \kappa' V_{BE_2}$ coming from the action of the finite environments:
\begin{align}
    \mathcal{E}_k(\rho_k) &\approx \tr_{BE_1E_2} \left[ e^{-i h_0 t} M_{k,t}' \left( \rho_k \otimes \rho_{BE_1E_2} \right) M_{k,t}'^\dag e^{i h_0 t} \right],\label{eq:expanded_map_finite_noise}
\end{align}
where
\begin{align}    \rho_{BE_1E_2} &=\ket{\Omega}\bra{\Omega}_B\otimes\rho_{E_{1}}^k\otimes\rho_{E_{2}}^k,\\
                 M_{k,t}'       &=\Id-i\int_0^t e^{i h_0 \tau}(gv_{SB}^k+\kappa' (v_{SE_1}^k+v_{BE_2}^k))e^{-i h_0 \tau}\dd \tau.
\end{align}
This expansion, analogous to the noiseless case, would result in three contributions to the system evolution. However, the contribution that would come from the interaction between the bath $B$ and its environment $E_2$ disappears when applying the partial trace over $BE_1E_2$ and considering terms only up to $\mathcal{O}(g^2t^2,\kappa'^2t^2,g\kappa' t^2)$. Note that, since $\kappa'$ is a coupling strength, it is different from the rate $\kappa$ used in depolarizing noise, and terms must be considered up to the same order as terms involving $g$. Therefore, we are left with two Lindbladians acting on the system: the original one for the noiseless cooling $\mathcal{L}_C$ (generated by $v_{SB}^k$), and an additional one resulting from the interaction with $E_1$, which we call $\mathcal{L}_{E_1}$ (generated by $v_{SE_1}^k$).

The final result for the map $\mathcal{N}'_k$ is very similar to the one for depolarizing noise, namely
\begin{equation}
    \mathcal{N}'(\rho)\approx e^{-iH_{S}t}e^{\mathcal{L}_C+\mathcal{L}_{E_1}}(\rho) e^{i H_{S}t}.
\end{equation}
Extending the result of \cref{eq:steady_state_energy_mode} to accommodate the finite environment is straightforward, as the Lindbladian generated by $E_1$ has a similar shape as the one generated by $B$ but different $A_k^{(i)},B_k^{(i)}$ and $\Delta^{(i)}$. Additionally, one has to take into account that now the environment is initially not in the ground state, but in some initial state defined by $p_E$ [\cref{eq:rho_E_def}].
The corresponding Lindbladian is:
\begin{equation}
    \mathcal{L}_{E_1}=\frac{1+p_E}{2}(\mathcal{L}_{l_1^E}+\mathcal{L}_{l_2^E})+\frac{1-p_E}{2}(\mathcal{L}_{l_1^{E\dag}}+\mathcal{L}_{l_2^{E\dag}}),
\end{equation}
where the jump operators $l_{1,2}^E$ are defined analogously to the jump operators $l_{1,2}$ in \cref{eq:L1_def_app,eq:L2_def_app}, coming from the bath interaction. Now, however, we have extra jump operators $l_{1,2}^{E\dag}$ stemming from the fact that the environment is initially not in the ground state (note that choosing $p_E=1$ would get rid of these extra terms). Therefore, we can group the terms in both Lindbladians and again calculate the steady state, which translates into a sum in the numerator and denominator and the following result for the steady-state energy:
\begin{equation}
    E_{ss}^k= \epsilon_k\frac{\sum_{i=B,E_1}p_i(-|A^{(i)}_k x_k^{(i)}|^2+|B^{(i)}_k y_k^{(i)}|^2)}{\sum_{i=B,E_1}(|A^{(i)}_k x^{(i)}_k|^2+|B^{(i)}_k y_k^{(i)}|^2)}.\label{eq:steady_state_energy_multibath}
\end{equation}
where the $p_i$ characterize the initial state of both the bath ($p_B=1$) and the environment $E_1$ ($p_E\in[-1,1]$).

We can qualitatively see the effect that $p_i$ has on the energy by disconnecting the environment (i.e., setting $\kappa'=0$). We then focus on the case of a bath that is not in the ground state: for $p_B=-1$, $B$ is fully excited and the energy in the steady state flips sign compared to the $p_B=1$ case. Thus, we heat up instead of cooling. For the case of $p_B=0$, the bath state is maximally mixed and the energy of the steady state becomes zero, similarly to the case of the depolarizing noise.

Note that in this Appendix, we will always use $\lambda_0^E=1,\ \mu_0^E=0$ in the environment (as defined in \cref{eq:V_SE}), thus the corresponding coupling coefficients are $A_k^E=\cos\varphi_k$, $B_k^E=-\sin\varphi_k$.

As a final remark, if we want to work in the CM formalism (see \cref{app:CM_derivation}), the addition of a finite environment translates into an equation similar to the result of \cref{eq:cooling-map-CM}, but with a sum of contributions from the bath and the environment in the fixed-point equation:
\begin{align}
    \ket{\gamma_{S,k}^\text{ss}} &= \frac{A_{SB,k}\otimes A^*_{SB,k}\ket{\gamma_{0,k}^B}+A_{SE_1,k}\otimes A^*_{SE_1,k}\ket{\gamma_{0,k}^{E_1}}}{\Id-A_{S,k}\otimes A^*_{S,k}}\nonumber\\
                                 &= \frac{A_{SB,k}\otimes A^*_{SB,k}+p_E A_{SE_1,k}\otimes A^*_{SE_1,k}}{\Id-A_{S,k}\otimes A^*_{S,k}}\ket{\gamma_{0,k}^{B}}.
    \label{eq:noisy-cooling-map-CM}
\end{align}
Here, we use the simplification $\gamma_0^{E_1}=p_E\gamma_0^B$ since we assume the bath to be in the ground state and that we are in the eigenbasis for both bath and environment. Solving \cref{eq:noisy-cooling-map-CM} yields \cref{eq:steady_state_energy_multibath}.

\subsection{Numerical results}
\label{subapp:finite_noise_numerics}
\begin{figure}[ht]
    \centering
    \includegraphics[width=0.48\textwidth]{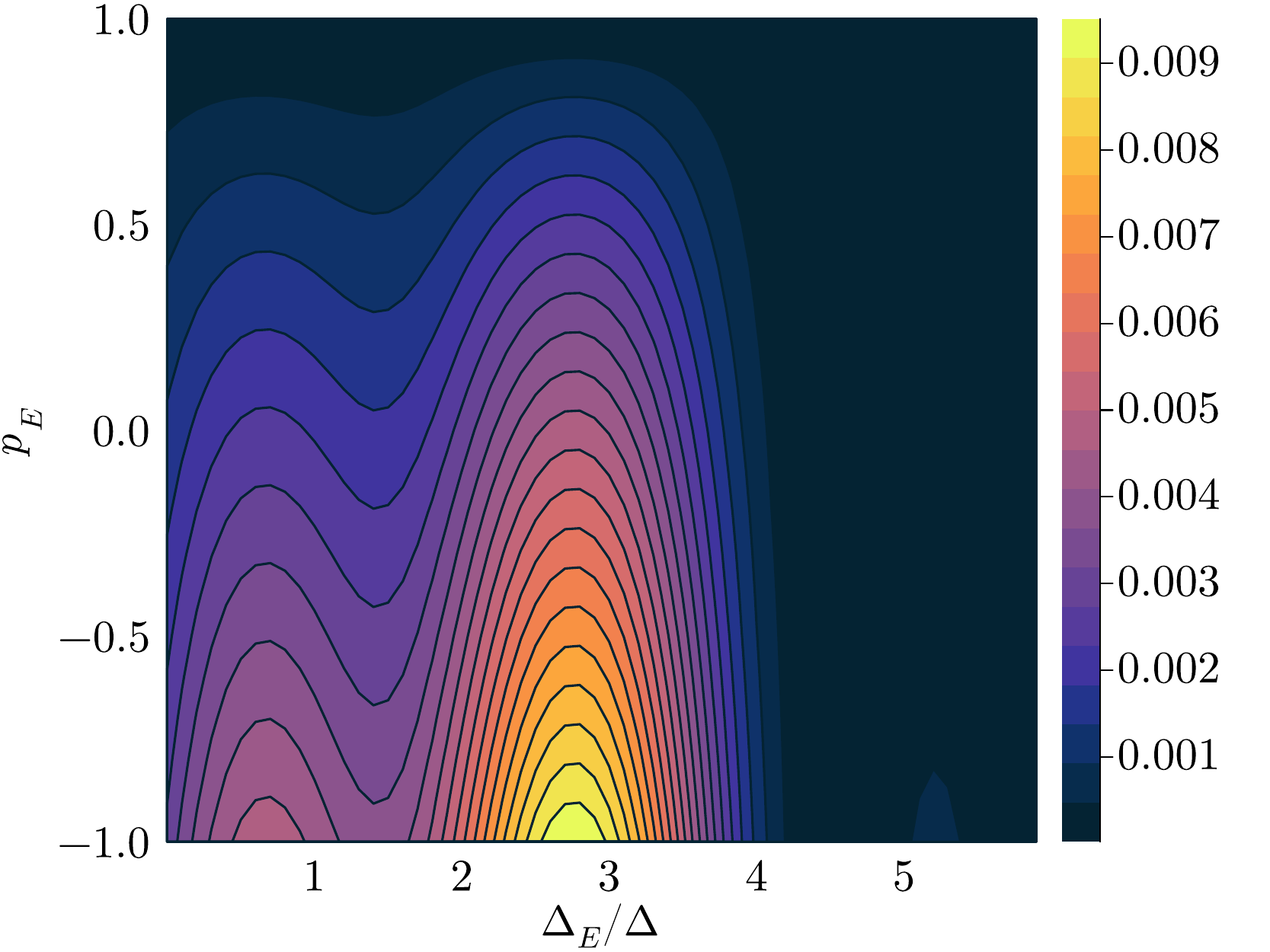}
    \caption{Contour plot of the relative energy $e$ (indicated by the colorbar) for the cooling protocol as a function of the relative environment energy splitting $\Delta_E/\Delta$ and the environment initial state parameter $p_E$. Parameters used are the noiseless-optimal ones for $N=20$, $nn=2$, $\theta=1.0$, with noise strength $\kappa'/g=0.04$. Darker blue regions indicate lower relative energy (better cooling), while lighter regions show higher energy due to environmental reheating.}
    \label{fig:DeltaEsweep_cooling}
\end{figure}

Given the finite noise model and the corresponding steady state from the previous subsection, we will now numerically analyze cooling and DSP under such a model, as well as covering re-optimization of parameters analogously to~\cref{sec:nn_cooling}. The analysis is different from the case of depolarizing noise, since in this case the steady state energy depends both on the parameters of the environment $(\Delta_E, \kappa')$ and on its initial state, characterized by $p_E$. The modified optimization becomes:

\begin{equation}
    \min_{\{\lambda_j, \mu_j, \Delta, t\}} e_{\text{noisy}}(\kappa', \Delta_E, p_E,\theta).
    \label{eq:finite_noise_optimization}
\end{equation}
The numerical results reveal several interesting features. Both the environment energy gap $\Delta_E$ and the initial state parameter $p_E$ of the environment affect the reheating rate. The environment in the most excited state ($p_E=-1$) has the worst effect both in cooling and DSP, as it introduces more reheating. For cooling, the gap of the bath also plays an important role: for $\Delta \ll \Delta_E$, there is minimal effect of the environment, whereas for $\Delta_E \approx \Delta$ there is a small increase in relative energy, as seen in \cref{fig:DeltaEsweep_cooling}. The effect of noise, however, does not drive the system too far from the ground state, showing an increase in energy density of approximately $10^{-2}$ in the worst-case scenario.

\begin{figure}[ht]
    \centering
    \includegraphics[width=0.48\textwidth]{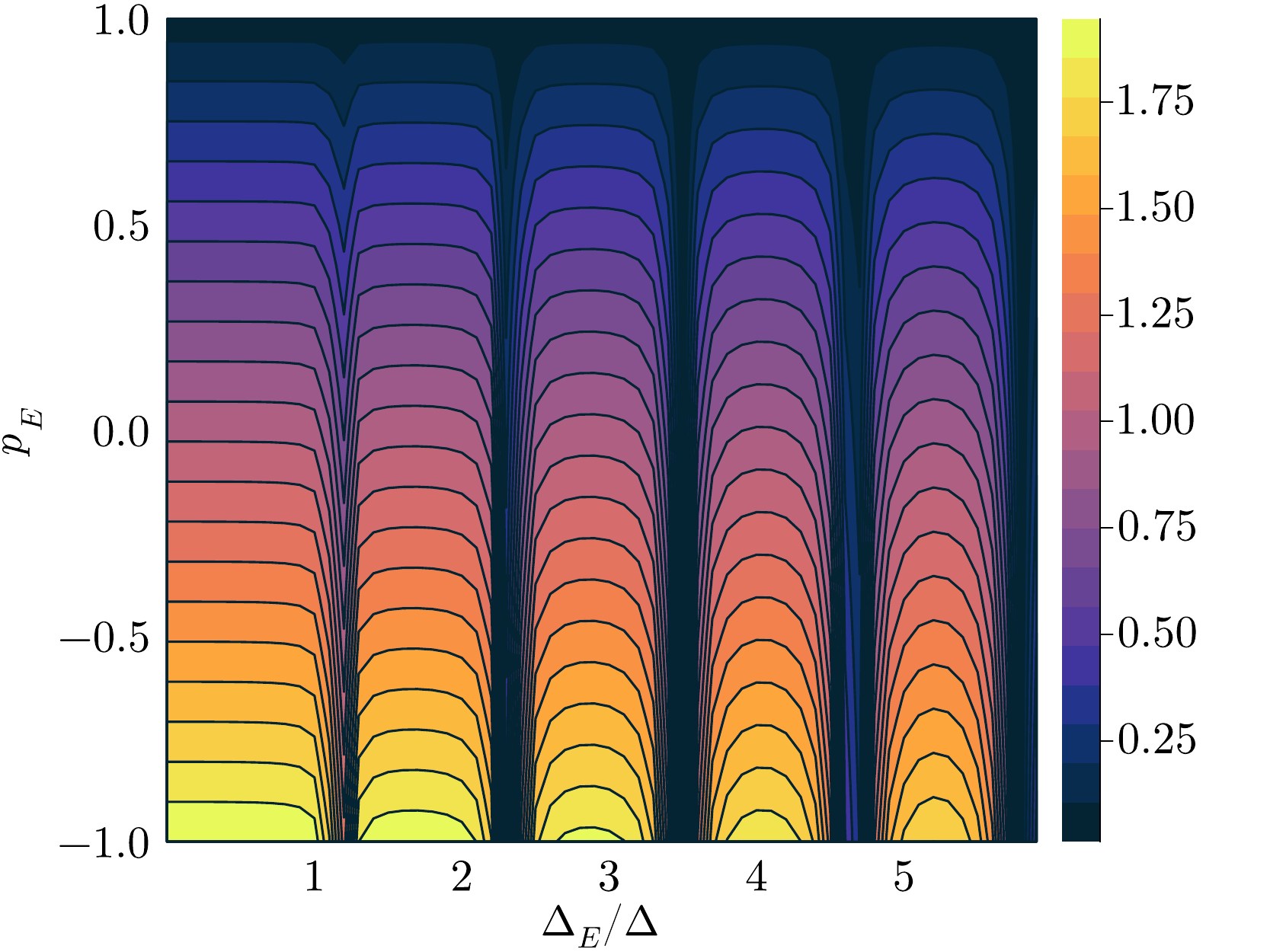}
    \caption{Same as \cref{fig:DeltaEsweep_cooling} but for DSP. DSP is more sensitive to the environment, with relative energies reaching much higher values.}
    \label{fig:DeltaEsweep_DSP}
\end{figure}

For weak noise ($\kappa' \ll g$), the increase in energy $E_{ss}$ in the steady state scales quadratically with the noise strength:
\begin{equation}
    E_{ss}^{\text{noisy}}-E_{ss}^{\text{noiseless}} \propto (\kappa')^2,
    \label{eq:energy_increase}
\end{equation}
consistent with a perturbative expansion of~\cref{eq:steady_state_energy_multibath}.

\begin{figure}[ht]
    \includegraphics[width=0.48\textwidth,height=\textheight,keepaspectratio]{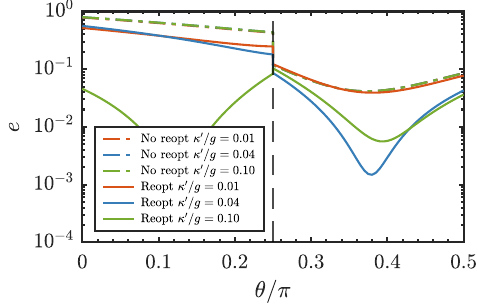}
    \caption{Comparison of DSP performance using noiseless and re-optimized phase-averaged parameters for $N=20$, $nn=1$, with the environment in its ground state ($p_E=1$) and $\Delta_E=10^{-5}\Delta$. The relative energy $e$ is plotted against $\theta/\pi$ for various noise strengths $\kappa'/g \in \{0.01, 0.04, 0.10\}$. Solid lines represent ideal parameters, while dashed lines show results using re-optimized parameters according to \cref{eq:finite_noise_optimization}. Re-optimization provides some improvement, but the benefits are less pronounced than in the cooling case.}
    \label{fig:noisy_reoptimization_stateprep}
\end{figure}
\begin{figure}[ht]
    \includegraphics[width=0.48\textwidth,height=\textheight,keepaspectratio]{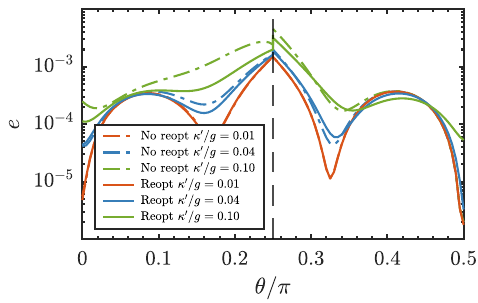}
    \caption{Same as \cref{fig:noisy_reoptimization_stateprep}, but for cooling. Re-optimization significantly improves cooling performance, especially for stronger noise, demonstrating the potential to mitigate environmental effects through parameter tuning.}
    \label{fig:noisy_reoptimization}
\end{figure}

For the dissipative state preparation (DSP) protocol, introducing environmental noise has a much more detrimental effect compared to the cooling algorithm.
Even with minimal coupling strength, the addition of an environment significantly increases the relative energy, occasionally driving the system close to its most excited state (see~\cref{fig:DeltaEsweep_DSP}).

Furthermore, the relative energy in the DSP case exhibits a cyclic behavior that persists beyond the regime of $\Delta_E \approx \Delta$, which can be attributed to higher-order contributions from the system's environment: terms of the form $(1 - \cos(x t))/x^2$ appear when expanding the equation for the energy density, creating the pattern seen in \cref{fig:DeltaEsweep_DSP}.

Quantitatively, the environment connected to the bath plays a small but non-negligible role in the case of DSP. This is evidenced by the fact that \cref{eq:steady_state_energy_multibath} does not accurately capture the change in energy density. The result after the re-optimization of parameters is also not very reliable (see \cref{fig:noisy_reoptimization_stateprep}): it depends heavily on noise strength, which is difficult to infer from the original energy density
While re-optimization can alleviate some effects of the environment, we still see here a significant advantage of cooling over DSP: even without re-optimization, cooling still performs better than re-optimized DSP in many instances. As shown in \cref{fig:noisy_reoptimization}, even for phase-averaged parameters the total energy density in the cooling protocol is kept under $10^{-3}$ for the whole phase up to noise strengths of $10\%$ of the coupling strength $g$.

\section{Scalability of the optimal parameters with system size}
\label{app:scalability}
\begin{figure}[ht]
    \includegraphics[width=0.48\textwidth,height=\textheight,keepaspectratio]{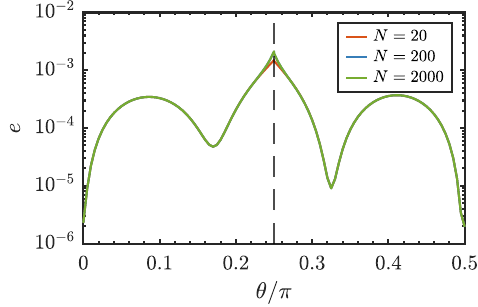}
    \caption{Scalability analysis of cooling using phase-averaged parameters optimized for $nn=2$ and $N=20$. The relative energy $e$ is plotted against $\theta/\pi$ for system sizes ranging from $N=20$ to $N=2000$. Cooling performance remains consistent as the system size increases, with relative energies showing no variation across most of the parameter space, demonstrating the scalability of the optimized protocol. A slight deviation is only observed near the critical point ($\theta \approx \pi/4$) for larger system sizes, where the relative energy increases with system size. This highlights the protocol's robustness across different system sizes and phases.}
    \label{fig:1bath_Nscaling_average}
\end{figure}

To assess the scalability of our cooling protocols, we apply the noiseless optimized coupling parameters obtained for $N=20$ to larger system sizes.
The relative energy remains our primary figure of merit, as it naturally accounts for intensive quantities.
Fig.~\ref{fig:1bath_Nscaling_average} illustrates the cooling performance for systems up to $N=2000$ using the phase-averaged parameters optimized for $nn=2$. The relative energy remains consistently lwo (below $10^{-3}$  for most $\theta$) across most of the parameter space as the system size increases, demonstrating the scalability and robustness of the optimized protocol.
A slight increase in relative energy is observed near the critical point ($\theta \approx \pi/4$) for larger system sizes. In this region, the value of $g$ starts to be similar to that of $\epsilon_k$ for some modes $k$ as the gap closes, which means the coupling term can no longer be treated as a perturbation.

\FloatBarrier

\bibliographystyle{apsrev4-1-title.bst}
\bibliography{library.bib}

\end{document}